\documentclass[10pt,twocolumn,letterpaper]{article}

\usepackage[pagenumbers]{wacv}  

\usepackage{graphicx}
\usepackage{tikz}
\usetikzlibrary{positioning, arrows.meta, calc, fit, backgrounds}
\usepackage{amsmath}
\usepackage{amssymb}
\usepackage{booktabs}
\usepackage{multirow}
\usepackage{subcaption}

\definecolor{wacvblue}{rgb}{0.21,0.49,0.74}
\usepackage[breaklinks,colorlinks,allcolors=wacvblue]{hyperref}

\usepackage{xcolor}

\def\wacvPaperID{1641}
\def\confName{WACV}
\def\confYear{2027}

\begin{document}

\title{Practical High-Fidelity Novel-View Synthesis of Mounted Lepidoptera}

\author{Kristof Overdulve \quad Lode Jorissen \quad Nick Michiels\\
Digital Future Lab, Flanders Make, Hasselt University\\
{\tt\small \{kristof.overdulve, lode.jorissen, nick.michiels\}@uhasselt.be}
}

\newlength{\teaserw}\setlength{\teaserw}{0.19\textwidth}
\newcommand{\timg}[1]{\raisebox{-0.5\height}{\includegraphics[trim=2 2 2 2, clip, width=\teaserw]{#1}}}
\newcommand{\teaserrow}[1]{\rotatebox[origin=c]{90}{\small\textbf{#1}}}

\makeatletter
\g@addto@macro\@maketitle{%
  \vspace{-12pt}
  \begin{center}
  \setlength{\tabcolsep}{2pt}%
  \renewcommand{\arraystretch}{1.0}%
  \begin{tabular}{@{}c@{\hspace{5pt}}c@{\hspace{0pt}}c@{\hspace{0pt}}c@{\hspace{0pt}}c@{\hspace{0pt}}c@{}}
    \teaserrow{Top (dorsal)} &
      \timg{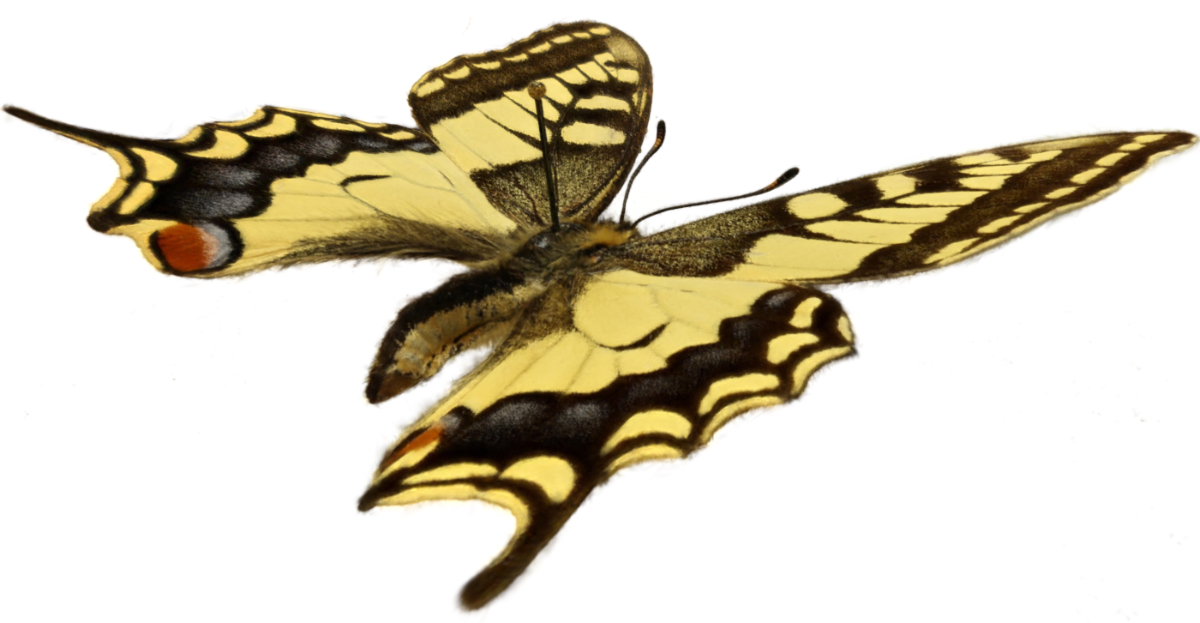} & \timg{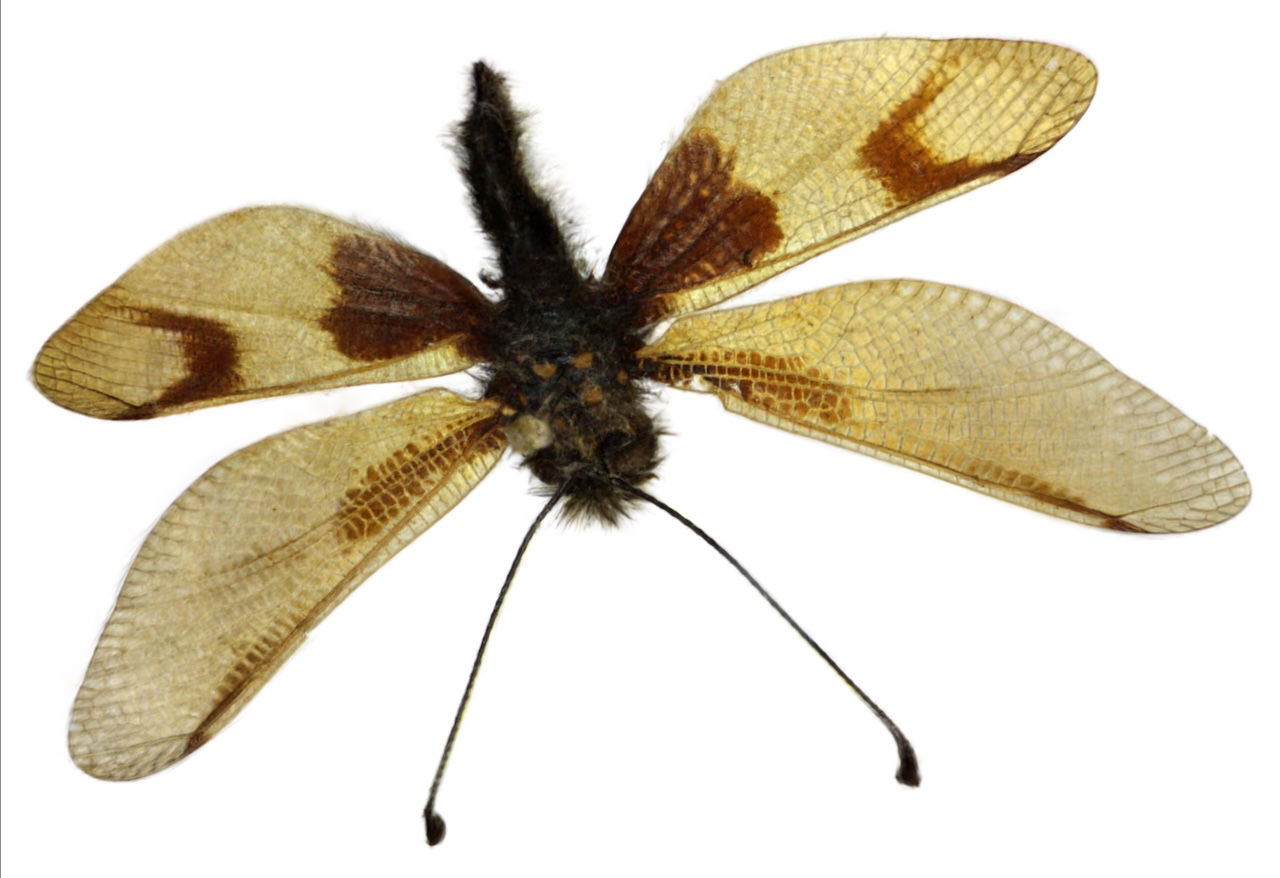} & \timg{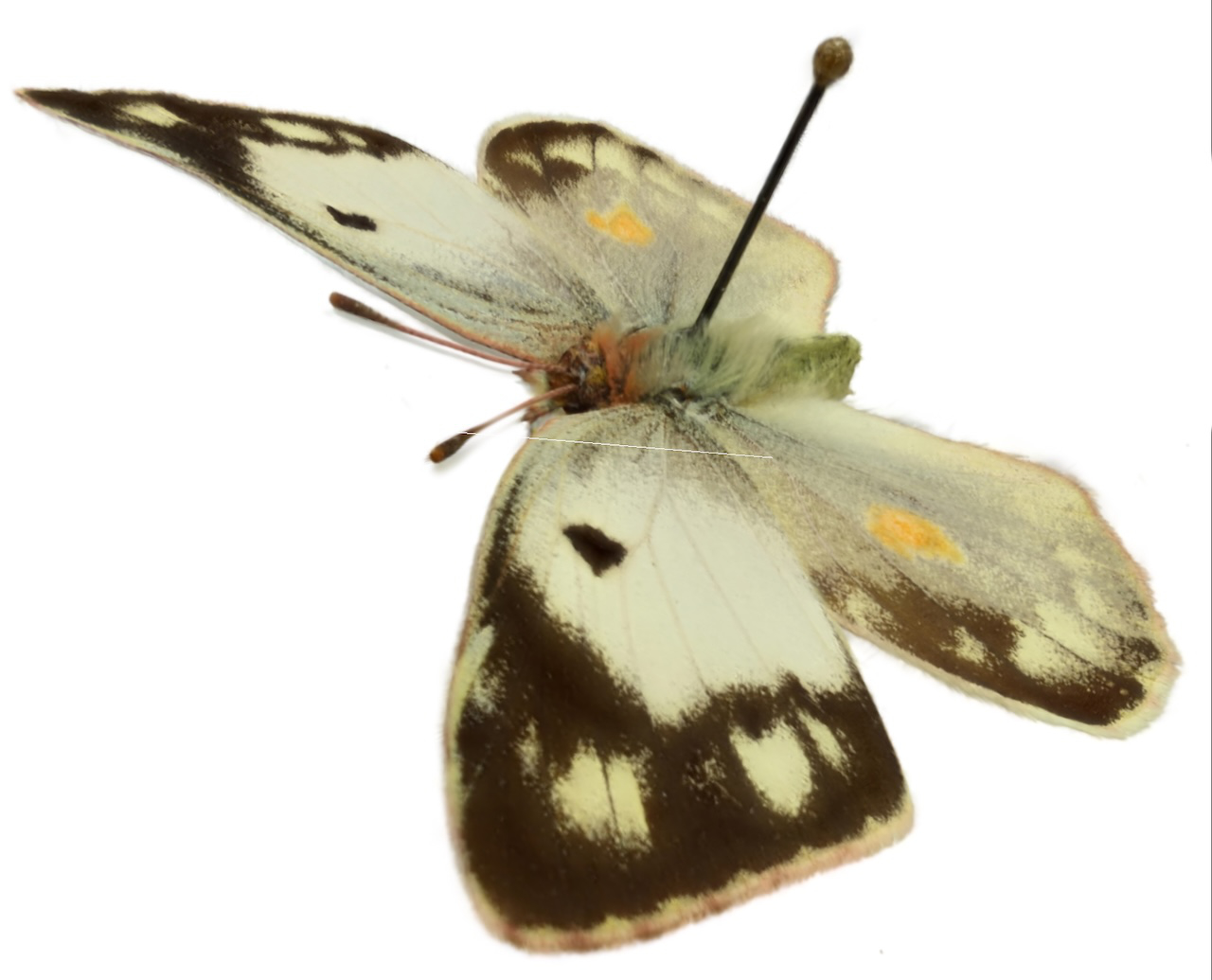} & \timg{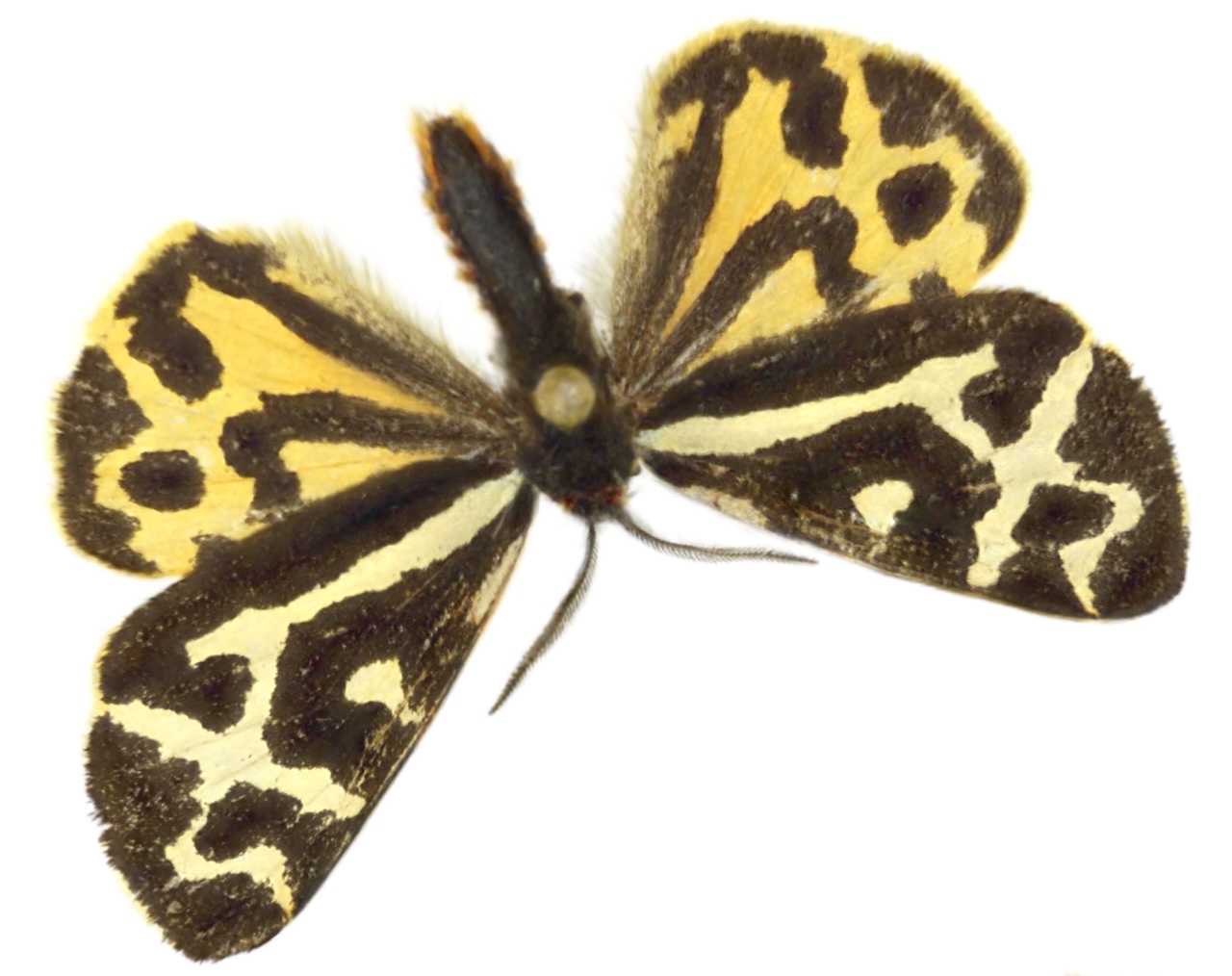} & \timg{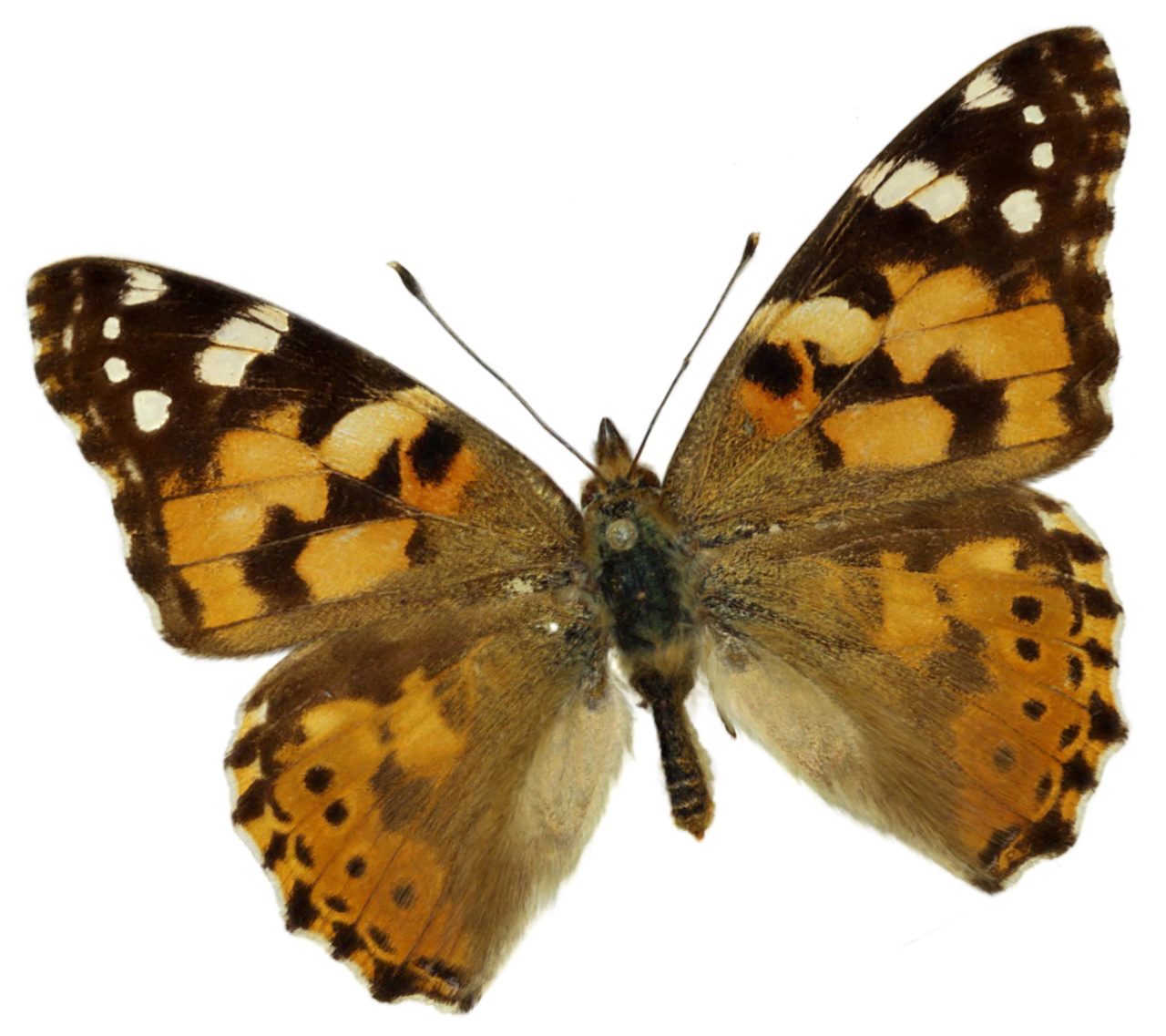} \\[2pt]
    \teaserrow{Bottom (ventral)} &
      \timg{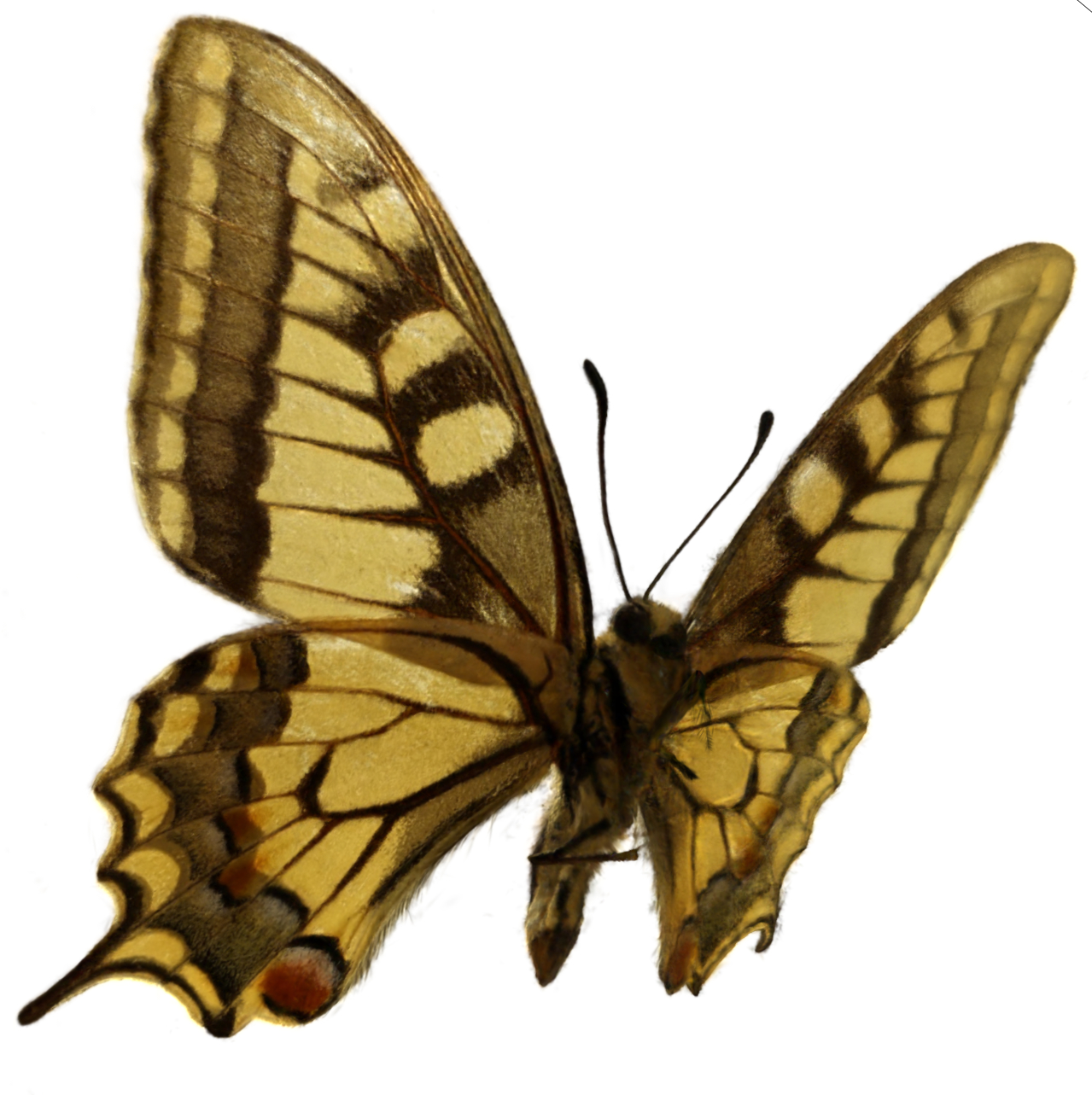} & \timg{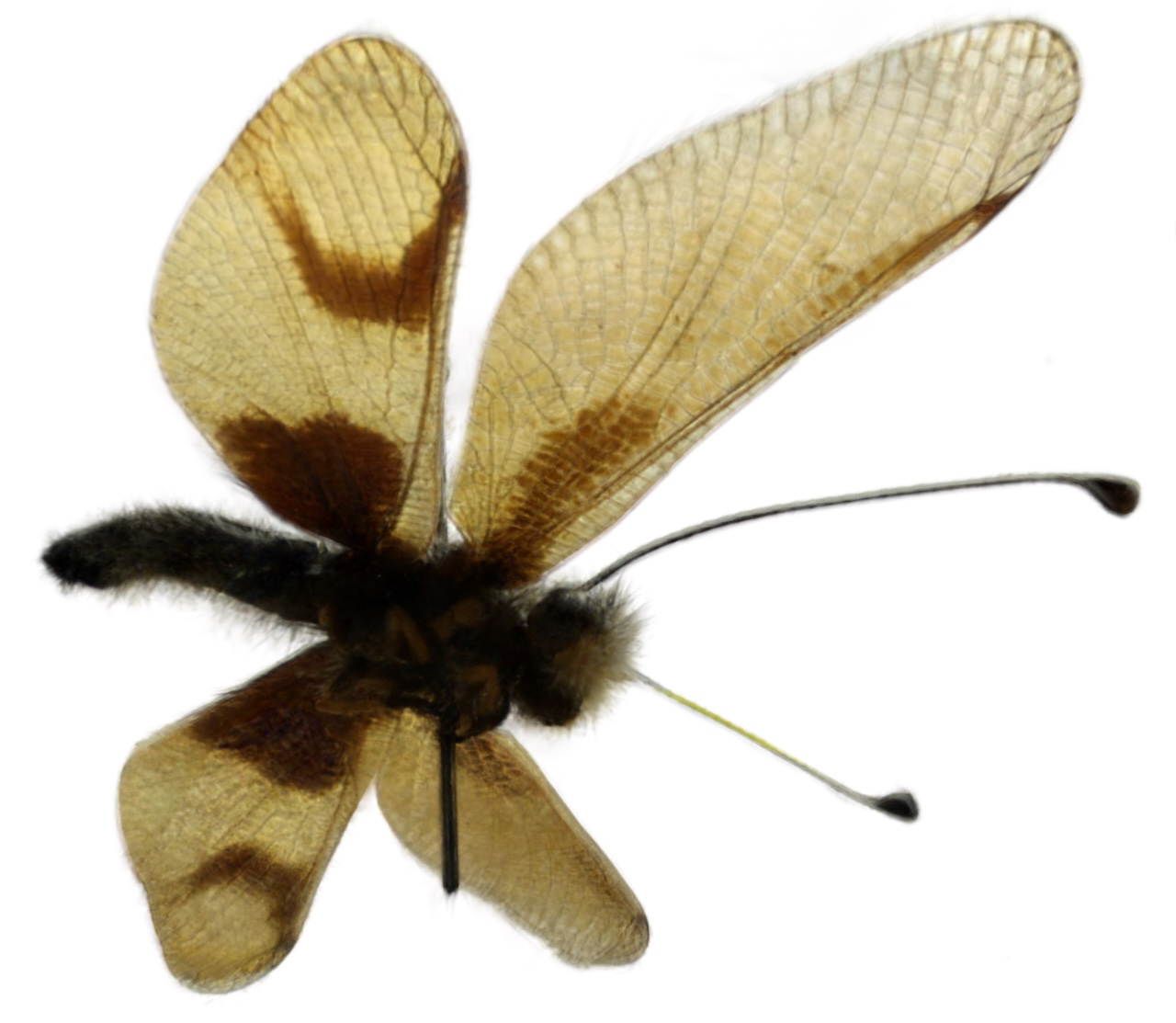} & \timg{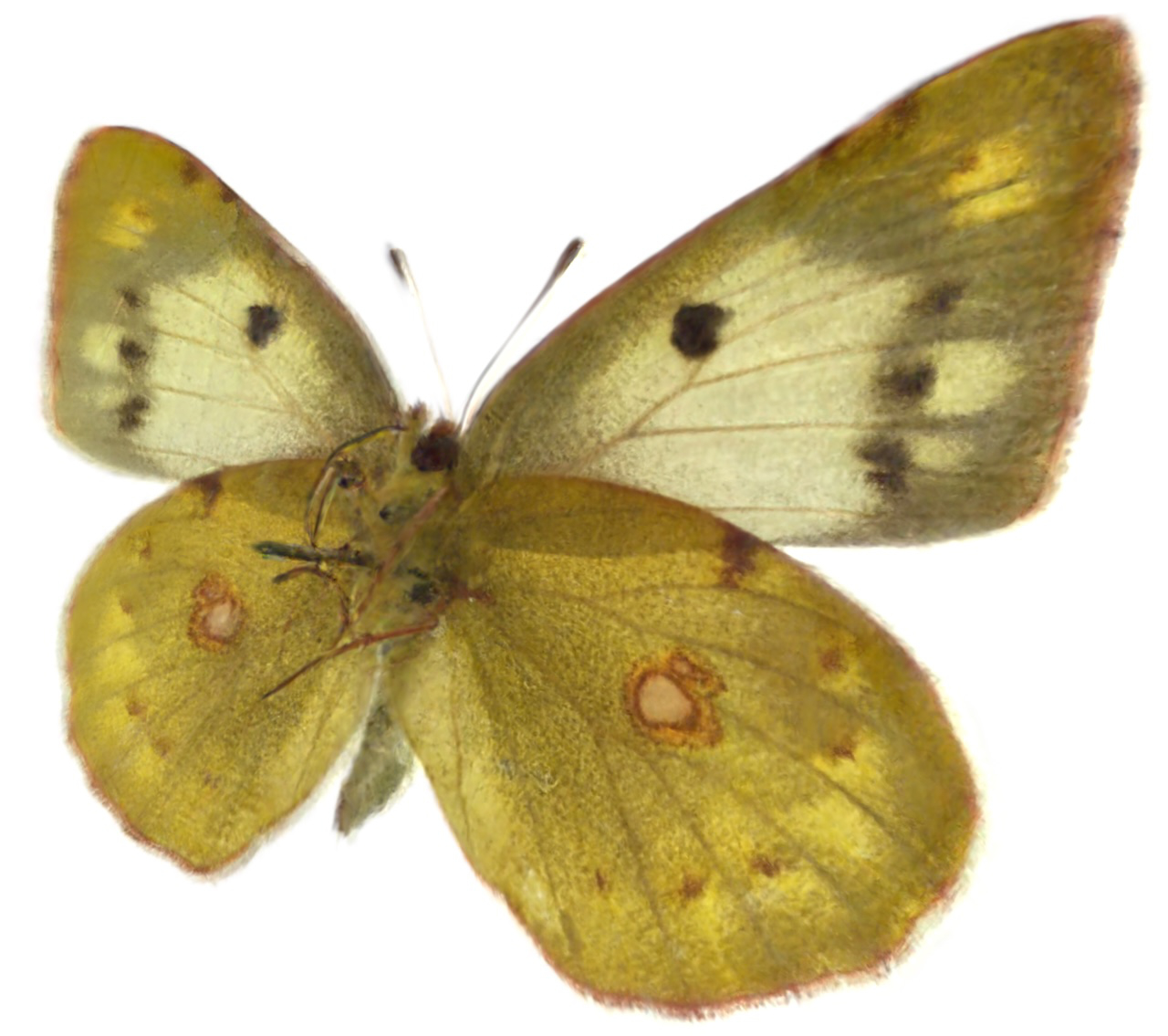} & \timg{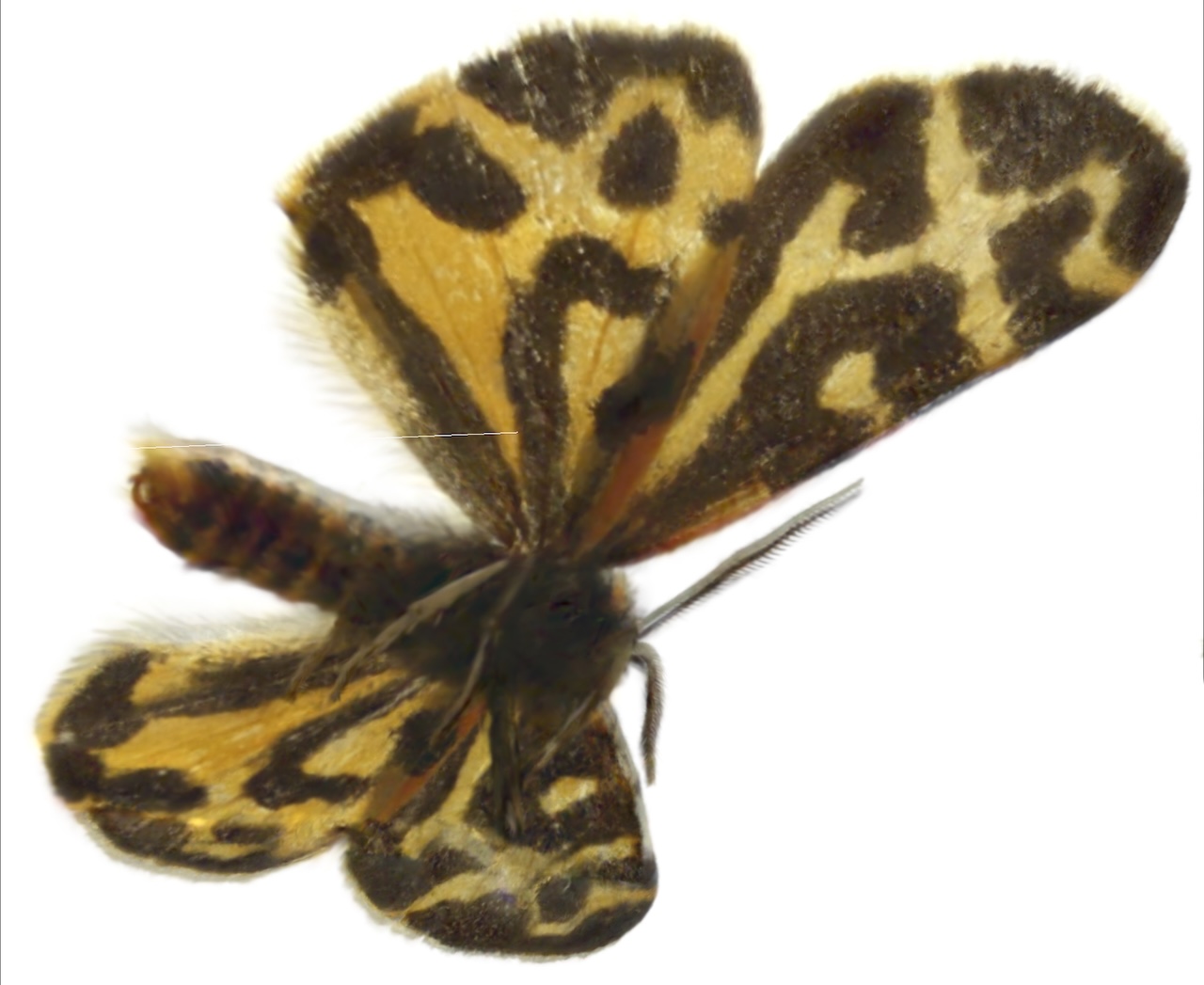} & \timg{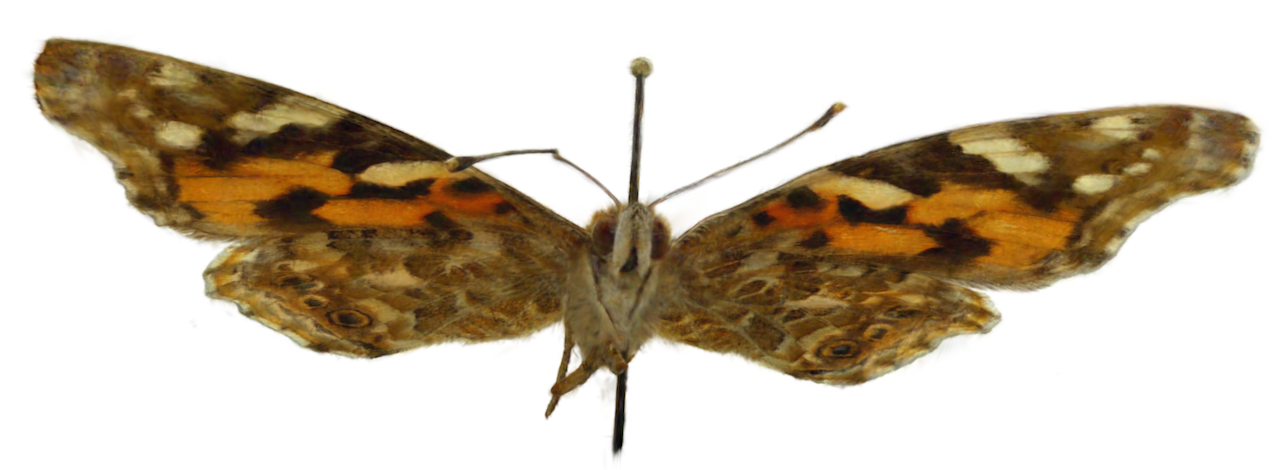} \\[2pt]
  \end{tabular}
  \captionof{figure}{%
    \textbf{Photorealistic 3D reconstruction of mounted Lepidoptera.} Dorsal (top) and ventral (bottom) renders of five specimens reconstructed with our pipeline; the ventral surface is only ever observed through a mirror. For display, each specimen is isolated from its background using a 3D bounding box crop.}
  \label{fig:teaser}
  \end{center}%
}
\makeatother
\maketitle

\begin{abstract}
Mounted butterflies are among the most striking objects in natural history collections. However, their beauty is notoriously hard to digitize in 3D: they are small and fragile, with microscopic hairs and vein structures. Capturing them in sufficient detail, therefore, requires a macro lens, which has a very limited depth of field (DoF). Moreover, a camera body cannot be maneuvered beneath a pinned specimen to photograph its ventral surface. We introduce an end-to-end pipeline that resolves these challenges, turning such specimens into photorealistic 3D models viewable from every direction. It combines three ingredients: handheld focus stacking for all-in-focus macro capture without a tripod, a non-contact first-surface mirror system that exposes the ventral surface without touching the specimen, and a segmentation-free, mirror-aware 3D Gaussian Splatting extension. We validate the reconstructions and design decisions on nine diverse specimens.
\end{abstract}

\section{Introduction}
\label{sec:intro}

Natural history collections worldwide hold millions of pinned insect specimens, representing an irreplaceable record of biodiversity over centuries. Lepidoptera (butterflies and moths) are among the most visually stunning of those specimens, prized aesthetically for the iridescent scales and patterns covering their wings. Digitizing such collections for virtual museums, educational platforms, and open-access scientific repositories is a priority, as reflected in initiatives such as iDigBio, DiSSCo~\cite{nelson2019dissco}, and SYNTHESYS+~\cite{smith2019synthesys+}. However, existing scanning workflows are designed for flat herbarium sheets or robust specimens that can be moved for professional imaging. 

Mounted Lepidoptera are uniquely challenging. Firstly, macro-lens photography is required to capture fine-scale microstructures (microscopic hairs, scale rows, and wing veins) at sufficient resolution, as they carry aesthetic value. Such macro lenses, however, have a depth of field (DoF) of only a few millimeters, so most of the specimen is out of focus in a single shot. Secondly, a mirrorless camera body cannot be maneuvered beneath a pinned butterfly to photograph the ventral surface (the underside of its wings). Removing a specimen from its case to hang it on nylon monofilament is standard practice for insects such as beetles. However, it poses a risk of irreparable damage, as mounted Lepidoptera are extraordinarily fragile. Curators describe handling mounted butterflies as working with ``dust''. Synthetically mirroring the dorsal (upper) surface to stand in for the ventral surface is not a viable shortcut, as the dorsal and ventral wing patterns frequently differ.

We address these challenges with a single, reproducible \emph{methodology} that combines established techniques into a workflow tailored to the reconstruction of mounted Lepidoptera (\cref{fig:pipeline}). Our contributions are:
\begin{itemize}
  \item A \textbf{non-contact, segmentation-free mirror approach} for imaging the ventral surface of Lepidoptera through first-surface mirrors (\cref{sec:mirrors}) and a mirror-aware extension of 3DGS (\cref{sec:mirrorgs}) to reconstruct a single, consistent specimen without the need for per-pixel segmentation masks.
  \item A \textbf{handheld focus stacking pipeline} with automatic image registration (\cref{sec:focusstack}) producing all-in-focus images from a handheld mirrorless camera equipped with a macro lens, eliminating the need for tripods or motorized focus rails.
\end{itemize}

Together, these components form a practical end-to-end pipeline that turns a fragile, pinned butterfly into a photorealistic 3D model viewable from any direction.
\section{Related Work}
\label{sec:related}

\subsection{Heritage and Specimen Digitization}
Early work on 3D insect digitization relied on classical reconstruction methods such as visual hulls, which struggle with concavities, fine structures, and view-dependent appearance~\cite{Nguyen2014NaturalColour3D}, thereby lacking photorealism. Later systems improved acquisition by using calibrated extended-depth-of-field imaging~\cite{Li2019PerspectiveConsistent, Xu2021InsectReconstruction} and automated hardware setups~\cite{Strobel2018DISC3D, Plum2021scAnt}. Although effective, these systems are expensive and difficult to transport. At the other end of the spectrum, single-view methods deform a template mesh to a photograph, e.g.\ to measure fish on longline vessels~\cite{mei2022dmr}, trading photorealism for a strong shape prior; we instead capture dense multi-view imagery and model photorealistic, view-dependent appearance without any template.

\subsection{Novel-View Synthesis}
NeRF~\cite{mildenhall2020nerf} introduced the radiance-field formulation of novel-view synthesis (NVS): a continuous volumetric scene function optimized by differentiable ray marching, which delivers photorealistic quality, fine structures, and view-dependent appearance, but trains and renders slowly. Explicit representations improved on this speed gap: voxel grids~\cite{fridovich2022plenoxels}, hash-grid encodings~\cite{muller2022instant}, and most notably 3D Gaussian Splatting (3DGS)~\cite{kerbl2023gaussian}, which rasterizes a set of anisotropic 3D Gaussians and pairs state-of-the-art quality with real-time rendering. Because radiance fields excel at fine structures and view-dependent effects, they are uniquely suited to insect digitization: they can capture microscopic hairs and, for butterflies, the strongly angle-dependent iridescence of the scales. Accordingly, Xu et al.\ evaluated learning-based reconstruction, including NeRF-style methods, under insect-specific failure modes~\cite{Xu2021InsectReconstruction}, and Amrani et al.\ adapted neural implicit surfaces to preserve thin structures through detail-aware sampling~\cite{Amrani2025DNASC}. Neither addresses mounted Lepidoptera, whose dimensions exacerbate macro defocus and whose ventral surface is unobservable while mounted.

\subsection{Macro Defocus}
Most NVS methods assume all-in-focus inputs, which a macro lens does not provide. Two remedies exist. The first models the lens blur: Deblur-NeRF~\cite{ma2022deblur} learns per-ray blur kernels, and DOF-GS~\cite{wang2024dofgs} and DoF-Gaussian~\cite{shen2025dof} extend 3DGS with a thin-lens camera and a differentiable circle of confusion to recover a sharp scene from defocused inputs. These methods target \emph{moderate} defocus, where every image still carries a largely recoverable sharp signal; at the macro scale, where most of the specimen is not in focus, they fail to recover fine textures (\cref{sec:eval_dof}). The second remedy removes the defocus \emph{before} reconstruction through focus stacking. Focus stacking composites a sequence of images focused at different depths into one all-in-focus image, using a per-pixel sharpness measure (gradient energy~\cite{nayar1994focus}, Laplacian variance, wavelet coefficients) to drive the fusion. However, it assumes static, tripod-mounted captures, which makes capture time-consuming when many viewpoints are needed. We therefore enable \emph{handheld} focus stacking by first aligning the frames in each stack to undo hand jitter (\cref{sec:registration}).

\subsection{Mirrors in Radiance Fields}
Without explicit handling, a radiance field reconstructs a mirror reflection as real geometry behind the mirror. MirrorGaussian~\cite{liu2024mirrorgaussian} and Mirror-3DGS~\cite{meng2024mirror} model the mirror explicitly, producing the virtual image either by reflecting the 3D Gaussians across the mirror plane~\cite{liu2024mirrorgaussian} or by rendering from a camera mirrored across it~\cite{meng2024mirror}. Both use \emph{ground-truth per-image mirror-segmentation masks} to label splats as `mirror or not' and for blending the reflection into the image. Our method follows the mirrored-camera principle but is \emph{segmentation-free}: it recovers the plane from the bilateral symmetry of the dense multi-view-stereo point cloud (\cref{sec:mirrorplane}) and replaces the annotated mask by a learned mirror-surface layer whose rendered opacity delimits where the reflection appears (\cref{sec:gaussianreflection}).

\begin{figure*}[htbp]
  \centering
  \begin{tikzpicture}[
      >={Latex}, thick,
      img/.style={inner sep=0pt},
      flow/.style={font=\footnotesize, align=center, inner sep=1pt},
    ]
      \node[img] (rt) {\includegraphics[width=2cm]{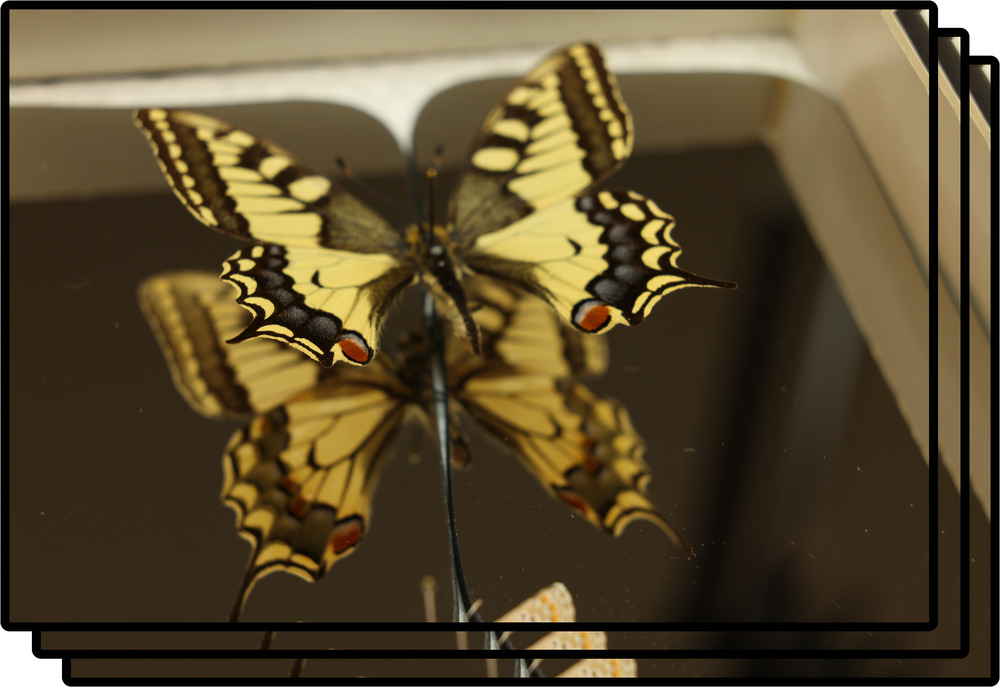}};
      \node[img, below=9mm of rt] (rb) {\includegraphics[width=2cm]{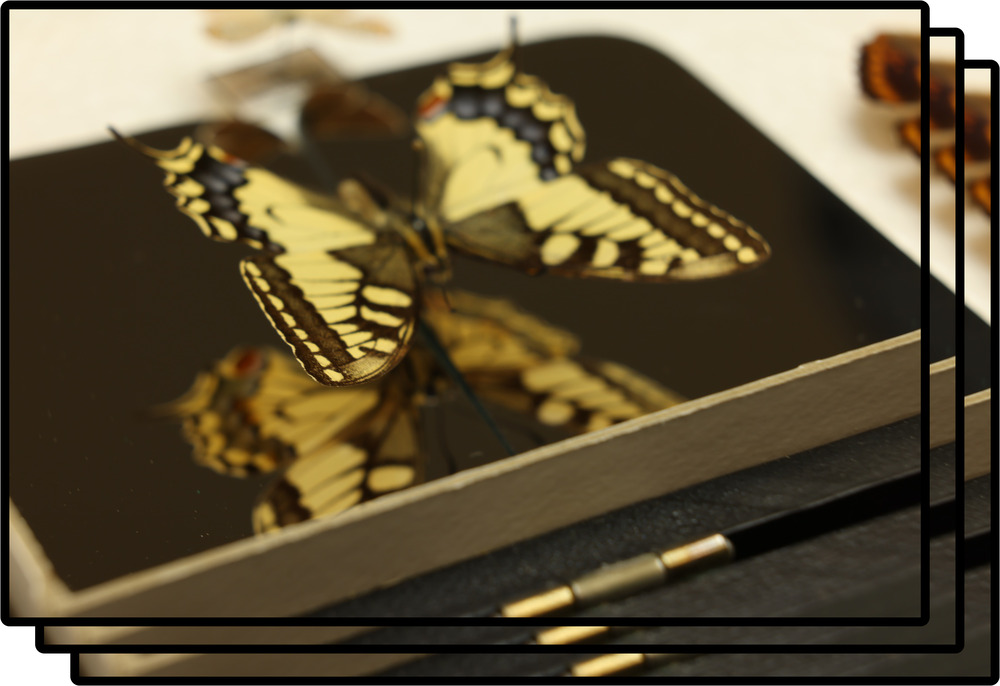}};
      \node at ($(rt)!0.5!(rb)$) {$\vdots$};
      \node[left=1mm of rt] {$\omega_1$};
      \node[left=1mm of rb] {$\omega_N$};
      \node[rotate=90, align=center] at ($(rt.west)!0.5!(rb.west) + (-12mm,0)$) {\textbf{Input:} Focus stack \\ per viewpoint $\omega_i$};
      \node[img, right=24mm of rt] (st) {\includegraphics[width=2cm]{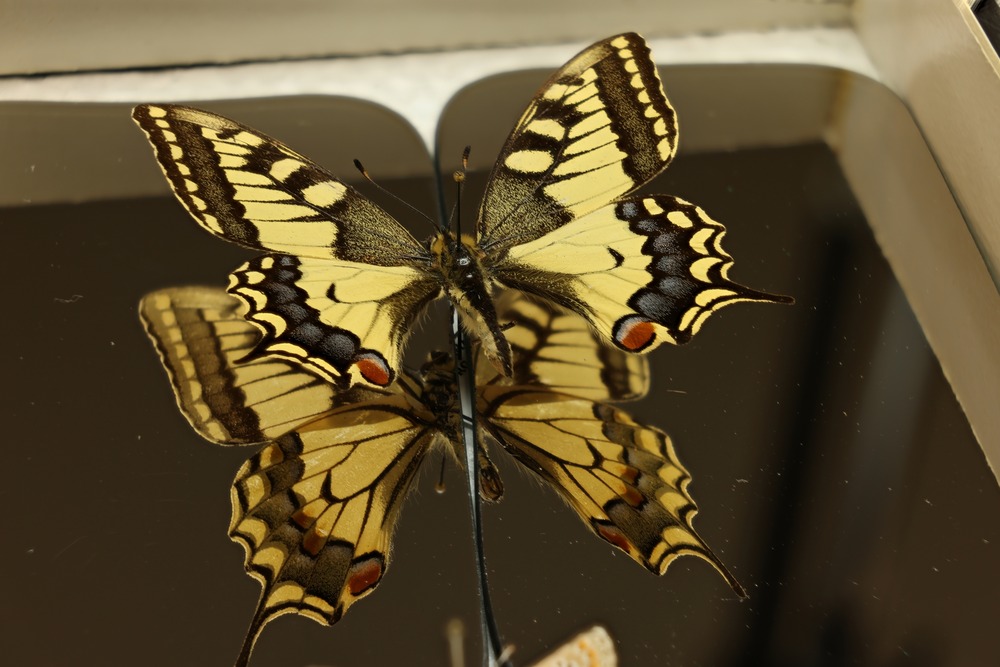}};
      \node[img, right=24mm of rb] (sb) {\includegraphics[width=2cm]{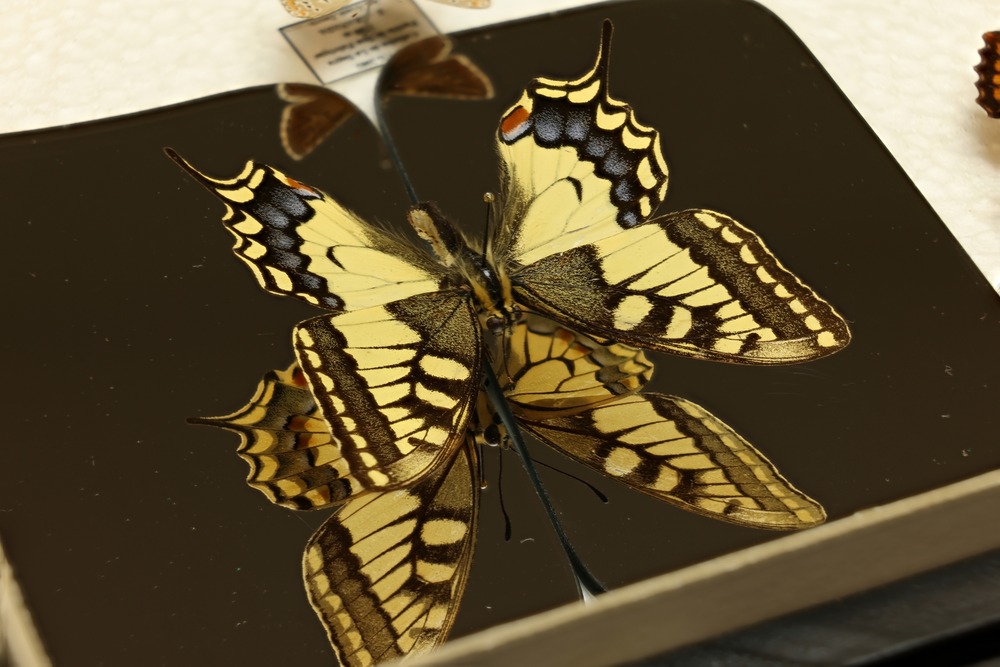}};
      \node[flow, anchor=south] at ($(rt.north east)!0.5!(st.north west)+(0,-5mm)$) {Image Registration \\ \& Pyramid Fusion};
      \coordinate (mid) at ($(st.east)!0.5!(sb.east)$);
      \node[img, right=9mm of mid] (sfm) {\includegraphics[height=2cm]{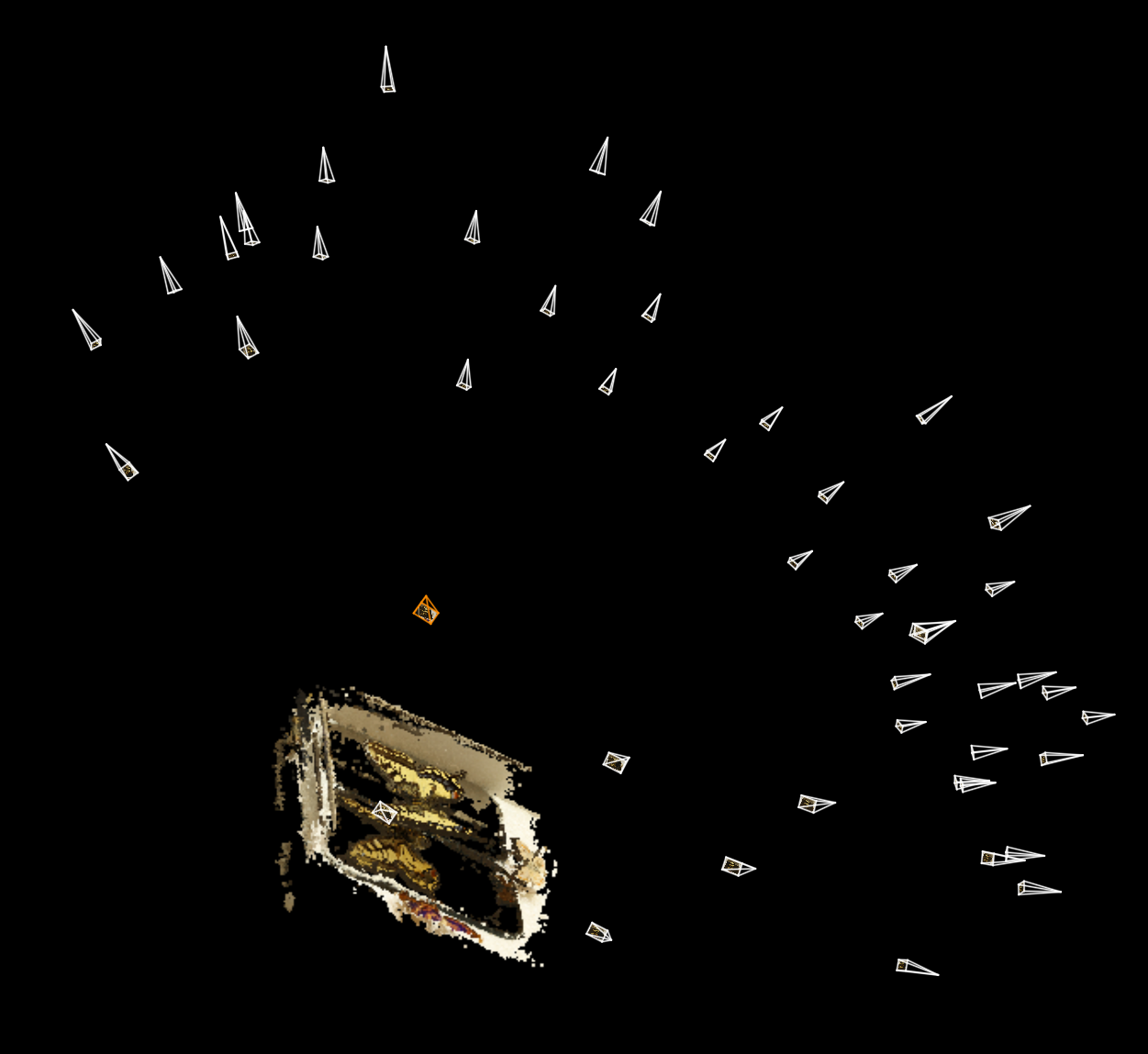}};
      \node[font=\small, rotate=90] at ($(sfm.east) + (3mm,0)$) {SfM \& MVS};
      \node[img, right=22mm of sfm, yshift=12mm] (mir) {\includegraphics[height=1.7cm]{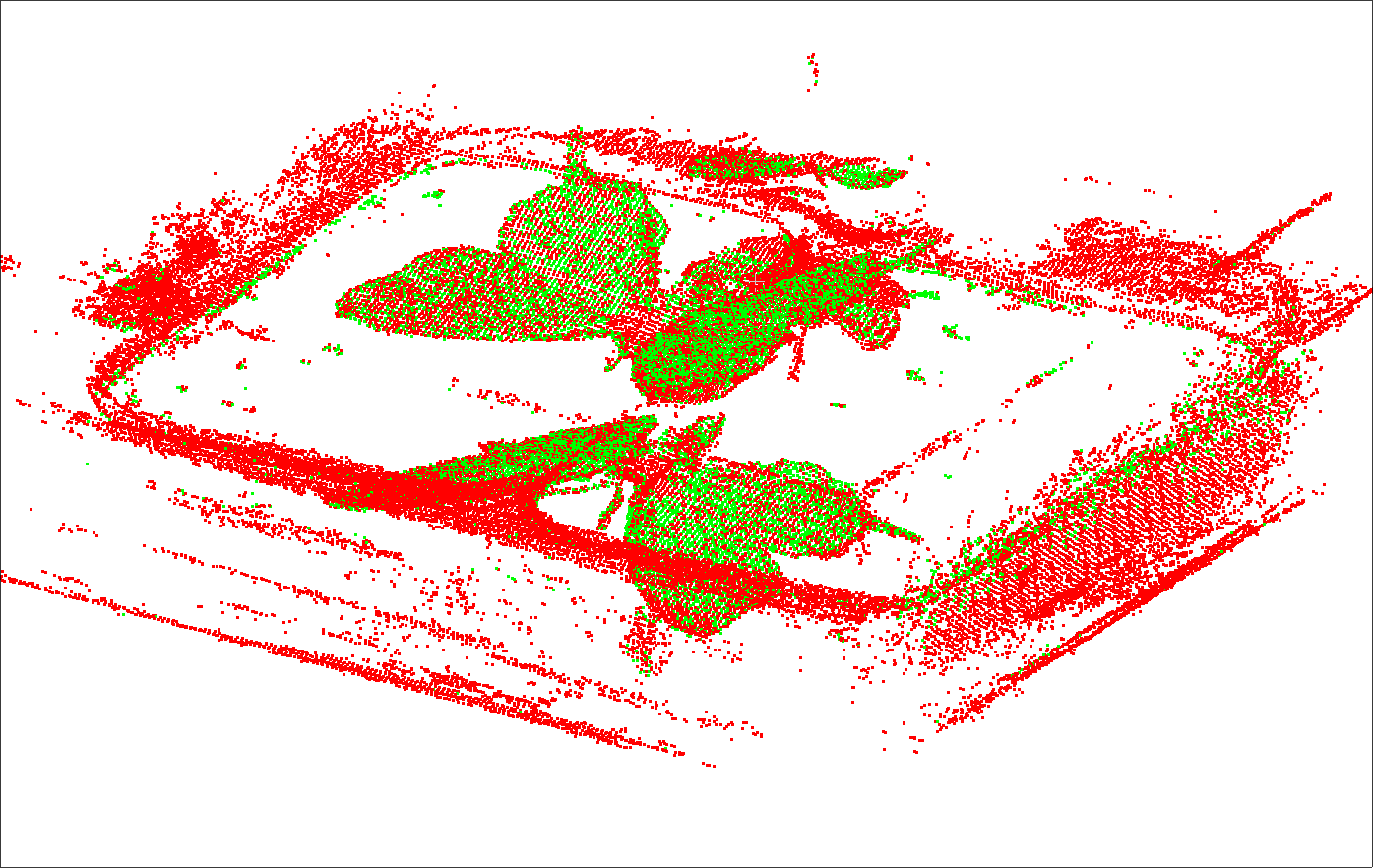}};
      \node[font=\small, above=0mm of mir] {Plane detection};
      \node[img, below=7mm of mir] (gs) {\includegraphics[height=1.7cm]{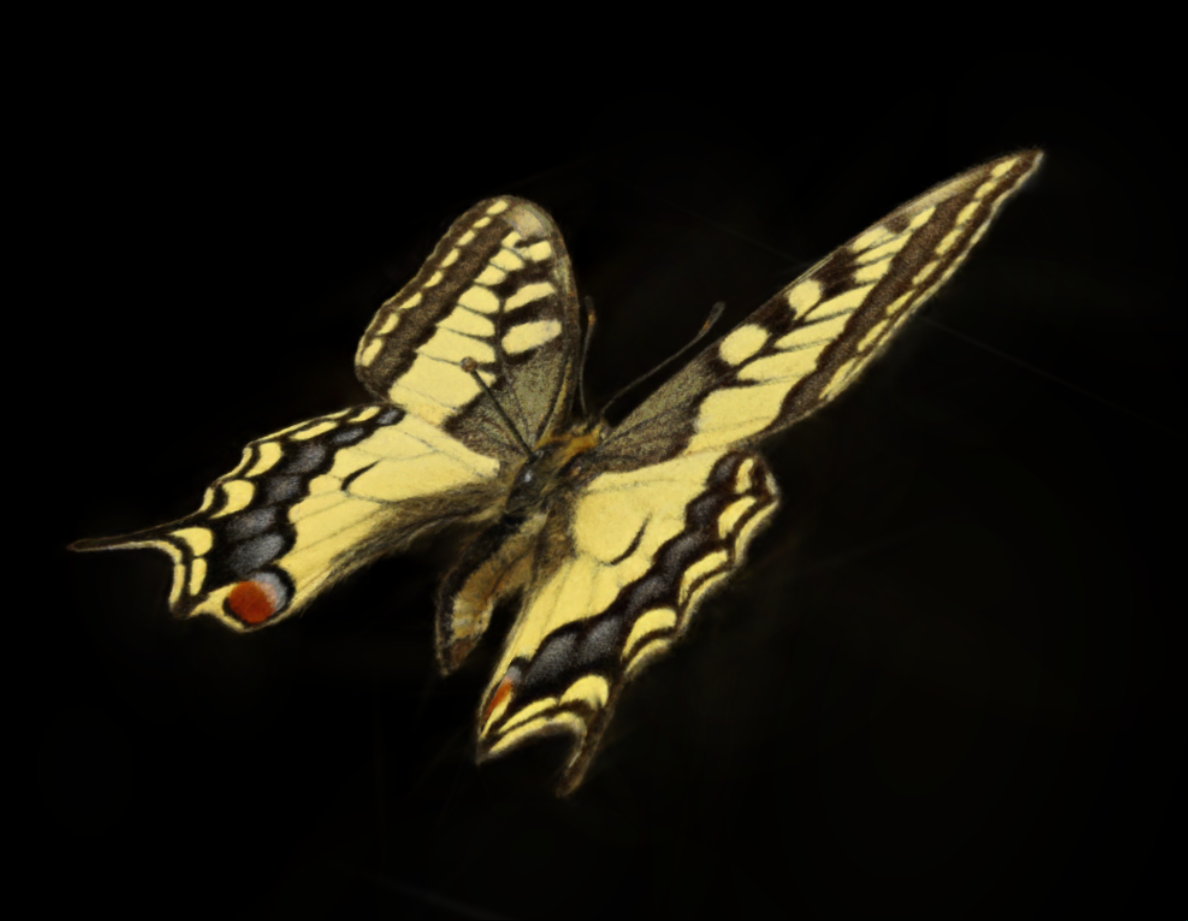}};
      \node[font=\small, below=0mm of gs] {Mirrored 3DGS};
      \draw[->] (rt) -- (st);
      \draw[->] (rb) -- (sb);
      \draw[->] (st) -- (sfm);
      \draw[->] (sb) -- (sfm);
      \coordinate (cmir) at (sfm.north |- mir.west);
      \coordinate (cgs)  at (sfm.south |- gs.west);
      \draw[->] (sfm.north) -- (cmir) -- (mir.west);
      \draw[->] (sfm.south) -- (cgs) -- (gs.west);
      \node[flow, anchor=south] at ($(cmir)!0.5!(mir.west) + (0,1mm)$) {Dense point cloud};
      \node[flow, anchor=north] at ($(cgs)!0.5!(gs.west) + (0,-1mm)$) {Poses + dense point cloud};
      \draw[->] (mir) -- (gs);
      \node[flow, anchor=east] at ($(mir.south)!0.5!(gs.north) + (-1mm,0)$) {Plane parameters};
    \end{tikzpicture}
  \caption{ \textbf{Pipeline overview.} Handheld focus stacks at $N$ viewpoints are registered and fused into all-in-focus images. SfM \& MVS recover camera poses and a dense point cloud, from which the mirror plane is detected. Poses, point cloud, and plane parameters feed our mirror-aware 3DGS. }
  \label{fig:pipeline}
\end{figure*}

\section{Method}
\label{sec:method}

\Cref{fig:pipeline} gives an overview of the pipeline. A photographer acquires $N$ focus stacks from different viewpoints $\omega_1$, \ldots, $\omega_N$ using a mirrorless camera with a macro lens. First-surface mirrors are positioned beneath the specimen's wings to expose the ventral surface. Each stack of $K$ images is registered and merged into one all-in-focus image. From the resulting image collection, we estimate camera poses with HLOC~\cite{sarlin2019hloc}, which runs COLMAP structure-from-motion on SuperPoint features matched with SuperGlue~\cite{sarlin2020superglue}. Our viewpoints are relatively sparse, and SIFT features, COLMAP's default, could not consistently match all images. We then recover a \emph{dense} point cloud by COLMAP multi-view stereo~\cite{schonberger2016colmap}. This dense point cloud initializes the Gaussians and is used to detect the mirror plane. The images, poses, and dense point cloud are fed into our mirror-aware 3DGS, yielding a single, consistent digital 3D representation that supports real-time rendering. 
\subsection{Non-Contact Mirror System}
\label{sec:mirrors}
A mounted butterfly rests with one side up (typically its dorsal side) on a spreading board with its pin driven through the thorax, leaving the ventral surface only a few centimeters above the case floor and thus unreachable by high-quality cameras. Removing the specimen to image it from below is inadvisable, as the wing-to-thorax attachment is held only by dried soft tissue and is easily damaged. We therefore expose the ventral surface with two \emph{first-surface mirrors} placed symmetrically around the pin (as shown in \cref{fig:dataset}). Unlike ordinary second-surface mirrors, first-surface mirrors have the reflective coating on the front face, so light never passes through the glass. This avoids the faint ghost reflections and color fringing that an ordinary mirror would show. We use inexpensive dental examination mirrors well suited to the $\sim$3--10\,cm scale of mounted specimens.
\subsection{Handheld Focus Stacking}
\label{sec:focusstack}

\subsubsection{Capture Protocol}

\begin{figure}[htbp]
  \centering
  \begin{subfigure}{0.48\linewidth}
    \includegraphics[width=\linewidth]{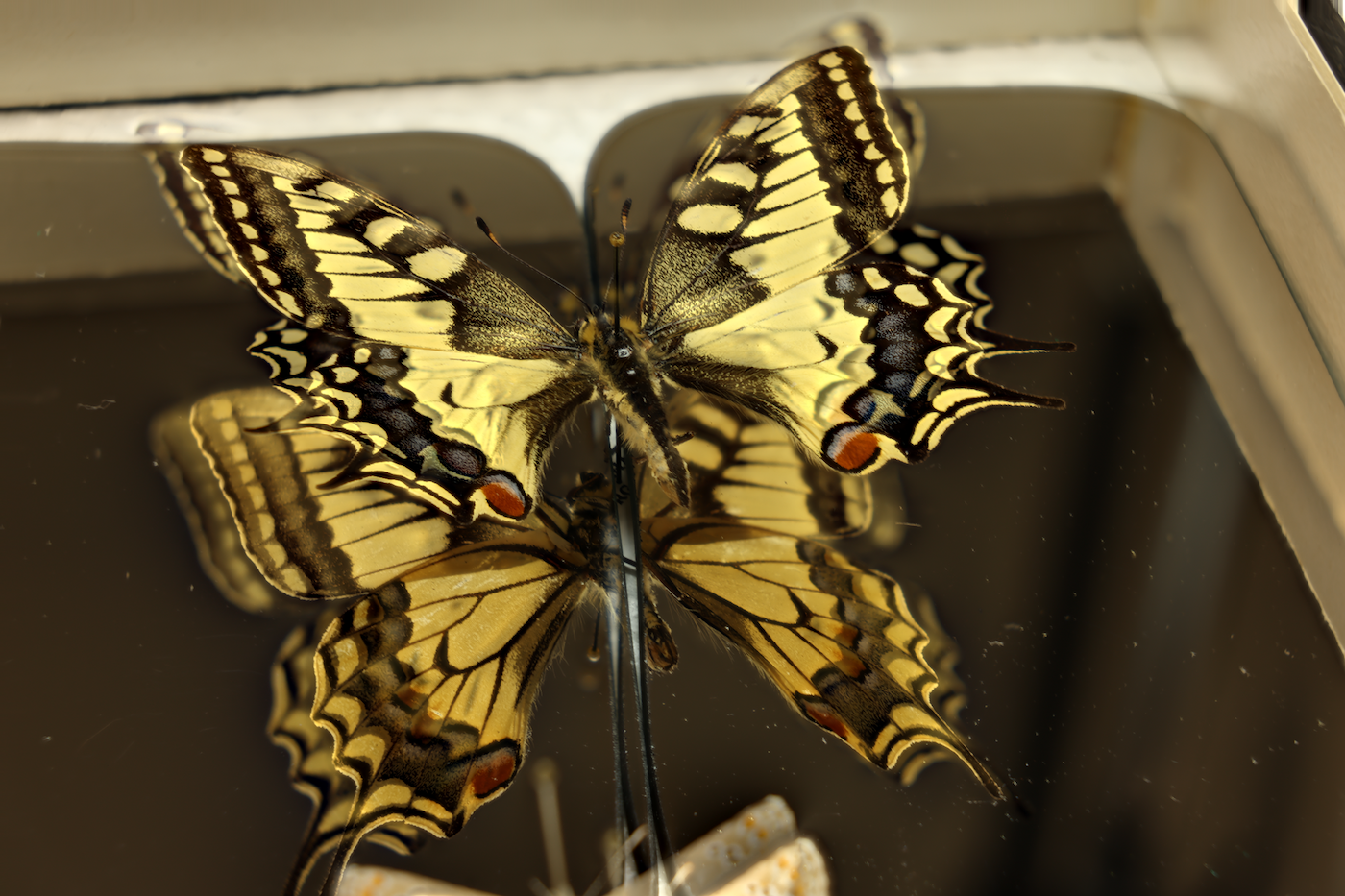}
    \caption{Fused, no alignment}
    \label{fig:stack_noalign}
  \end{subfigure}
  \hfill
  \begin{subfigure}{0.48\linewidth}
    \includegraphics[width=\linewidth]{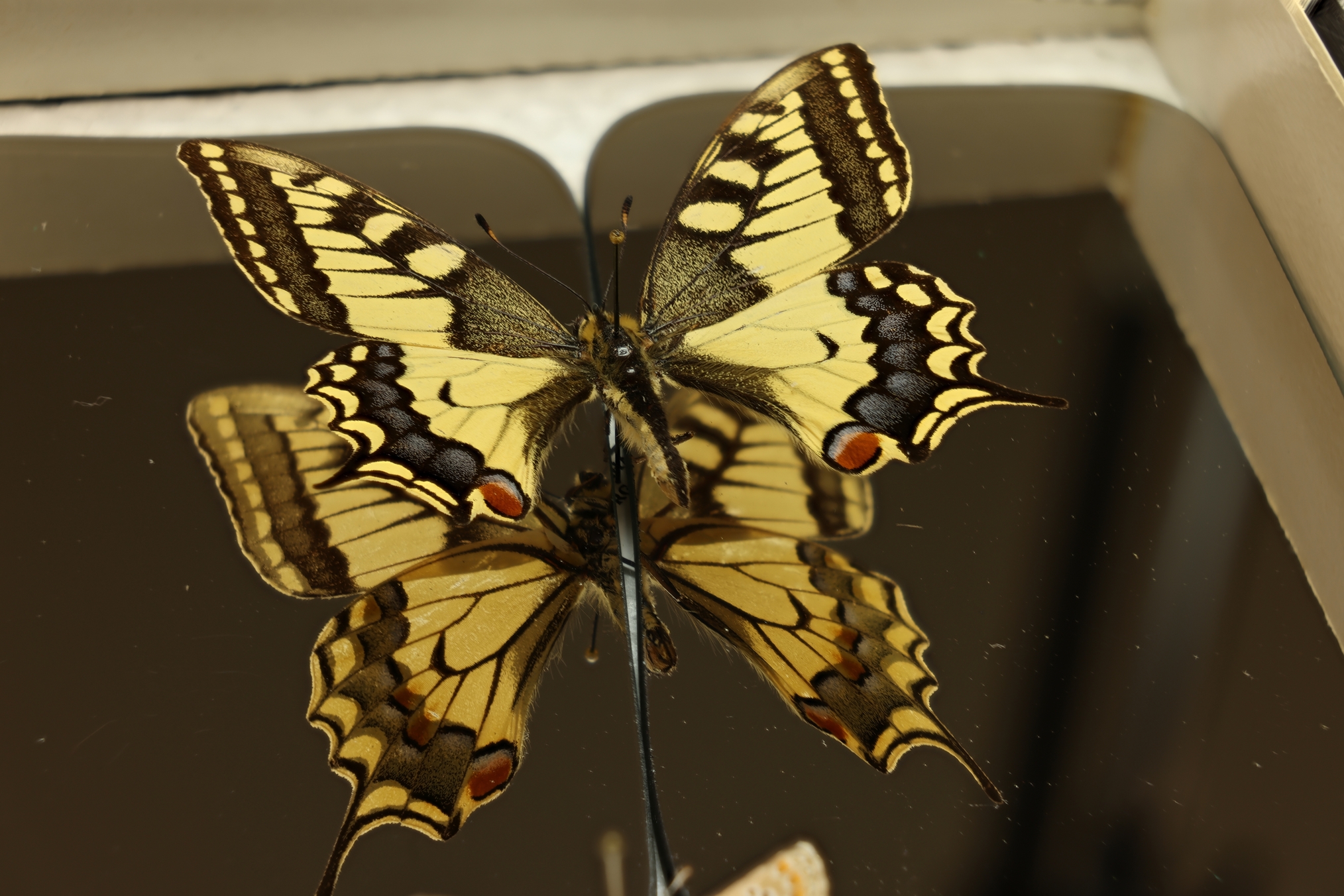}
    \caption{Fused, with alignment}
    \label{fig:stack_align}
  \end{subfigure}
  \caption{ \textbf{Handheld focus stacking.} Pyramid fusion without prior alignment~(a) produces ghost edges from handheld jitter; ECC-based sequential registration corrects these misalignments before fusion~(b). \emph{Best viewed zoomed in.} }
  \label{fig:focusstack}
\end{figure}

Images are captured with a Canon EOS R5C mirrorless camera fitted with a macro lens. At each viewpoint, we acquire a burst of $K$ frames (20--40) in continuous-drive mode, initially focusing on the nearest region of the specimen. Each frame captures a different depth slice of the butterfly in sharp focus. Because the camera is handheld, consecutive frames exhibit small shifts and rotations due to hand jitter. If uncorrected, these two effects combine to produce ghosting and double-edge artifacts in the fused output (\cref{fig:stack_noalign}).

\subsubsection{Image Registration}
\label{sec:registration}

We align the $K$ frames of each stack with the Enhanced Correlation Coefficient (ECC) algorithm~\cite{ecc2008}, which directly maximizes the correlation between a reference image and a warped source image in pixel space. ECC operates on single-channel images, so we project each frame onto the first principal component of the first frame's RGB values to preserve image contrast. We then align frames sequentially: for each consecutive pair, we estimate a 2D affine transform by maximizing the ECC criterion between the already-aligned reference frame and the warped source frame, with up to 50 Gauss--Newton iterations, stopping early once the parameter update norm falls below $10^{-3}$. We use an affine model rather than a full projective homography because parallax between consecutive frames is limited.

\subsubsection{Multi-Resolution Pyramid Fusion}
\label{sec:fusion}

After alignment, Helicon Focus~\cite{heliconfocus} merges the $K$ registered frames into one all-in-focus image with its \emph{pyramid stacking} method, a multi-resolution fusion in the spirit of Laplacian pyramid blending~\cite{burt1983pyramid}. Each frame is decomposed into a Laplacian pyramid, which separates fine detail from coarse color across scales. At every scale and pixel, the composite takes the detail from whichever frame is locally sharpest at that scale, and the pyramid is then collapsed back to full resolution. By selecting per-scale rather than per-pixel, the finest structure always comes from the sharpest frame, while coarse color and luminance blend smoothly across frames. This avoids the formation of halos typical of pixel-wise selection.
\subsection{Segmentation-Free Mirror-Aware 3DGS}
\label{sec:mirrorgs}
\subsubsection{Background: 3D/2D Gaussian Splatting}
3DGS~\cite{kerbl2023gaussian} represents a scene as $N$ anisotropic 3D Gaussians $\{\mathcal{G}_i\}_{i=1}^N$, each parameterized by a mean position $\boldsymbol{\mu}_i \in \mathbb{R}^3$, a covariance $\boldsymbol{\Sigma}_i$ stored as a scale vector $\mathbf{s}_i$ and unit quaternion $\mathbf{q}_i$, an opacity $\alpha_i$, and view-dependent color encoded as spherical harmonic (SH) coefficients. Given a camera viewpoint, Gaussians are projected to 2D, sorted by depth, and alpha-composited front-to-back. The representation is photometrically optimized with respect to the training images. 2D Gaussian Splatting (2DGS)~\cite{huang2024_2dgs} replaces the volumetric Gaussians by flat, oriented surfels that are intersected with the viewing ray in 3D rather than projected, which improves surface alignment. It also introduces a depth-distortion regularizer that concentrates each ray's weight onto a single surface. Our pipeline supports both primitives (\cref{sec:eval_dof}) and adopts 2DGS for its improved surface adherence.

\subsubsection{Mirror-Plane Detection}
\label{sec:mirrorplane}

A camera facing a mirror sees the specimen twice: directly, and reflected in the mirror. Multi-view stereo, unaware of the mirror, reconstructs both as real geometry, so the dense point cloud contains the specimen \emph{and} its mirrored copy. This duplicated geometry is an approximate \emph{bilateral symmetry} of the point cloud, with the mirror plane as the plane of symmetry. Detecting that symmetry therefore recovers the plane without any segmentation or depth supervision. We use the dense multi-view-stereo cloud rather than the sparse SfM keypoints because its density yields more reliable local geometric descriptors.

We detect the plane using the Mirror-Symmetry-via-Registration (MSR) principle~\cite{cicconet2017mirror}: reflect the point cloud across an \emph{arbitrary} plane, rigidly register the reflected copy onto the original, and compose the two transforms. For a symmetric cloud, this composition is the reflection across the \emph{true} symmetry plane, whose parameters follow in closed form (Steps 1--5). We repeat the registration and keep the most confident result (Step 6), adding a camera-side test to resolve the ambiguity introduced by the specimen's own bilateral symmetry. In more detail:

\noindent\textbf{Step 1: Artificial self-mirror.} We create an artificial mirror clone $\mathcal{P}'$ of the dense point cloud $\mathcal{P}$ by reflecting all points and normals across the world-coordinate plane $X{=}0$, $\mathbf{p}' = \mathbf{M}_X \mathbf{p}$ with $\mathbf{M}_X = \mathrm{diag}(-1,1,1)$. After this flip, the bilateral symmetry of the scene means that $\mathcal{P}'$ and $\mathcal{P}$ are related by a rigid transform.

\noindent\textbf{Steps 2--4: Registration.} Both clouds are voxel-downsampled and described by FPFH descriptors~\cite{rusu2009fpfh}; feature-matching RANSAC~\cite{fischler1981ransac} yields a coarse rigid transform that maps the clone back toward the original, which point-to-plane ICP~\cite{chen1992pointtoplane} refines to $\mathbf{T}_\text{icp}$.

\noindent\textbf{Step 5: Plane parameter extraction.} Since $\mathbf{T}_\text{icp}$ maps $\mathcal{P}'$ back onto $\mathcal{P}$, the composition $\mathbf{S} = \mathbf{T}_\text{icp}\,\mathbf{M}_X$ is itself a reflection across the recovered mirror plane. Its $3\times3$ part has eigenvalues $\{+1,+1,-1\}$, and the eigenvector of $-1$ is the plane normal $\mathbf{n}$; with $\mathbf{t}_s$ the translation part of $\mathbf{S}$, the midpoint $\mathbf{t}_s/2$ lies on the plane by construction, so the offset is $d = -\mathbf{n}^\top\mathbf{t}_s/2$.

\noindent\textbf{Step 6: Robust hypothesis selection.} A butterfly is itself strongly bilaterally symmetric, so the registration of Steps 2--5 is ambiguous: depending on the stochastic RANSAC sampling, it may converge to the true mirror plane or to the specimen's own plane of symmetry. Following MSR~\cite{cicconet2017mirror}, we repeat Steps 2--5 $T=10$ times to obtain a set of candidate planes and merge near-duplicates. We add a camera-side test: the camera viewpoints form a hemisphere entirely on the real side of the mirror, so they must all lie on the same side of the true plane. Among the candidates that pass this test, we keep the one with the highest ICP fitness.

\subsubsection{Mirror-Aware Rendering}
\label{sec:gaussianreflection}

\newlength{\psw}\setlength{\psw}{0.33\linewidth}
\begin{figure}[htbp]
  \centering
  \setlength{\tabcolsep}{0.5pt}%
  \begin{tabular}{@{}cccc@{}}
    \includegraphics[width=\psw]{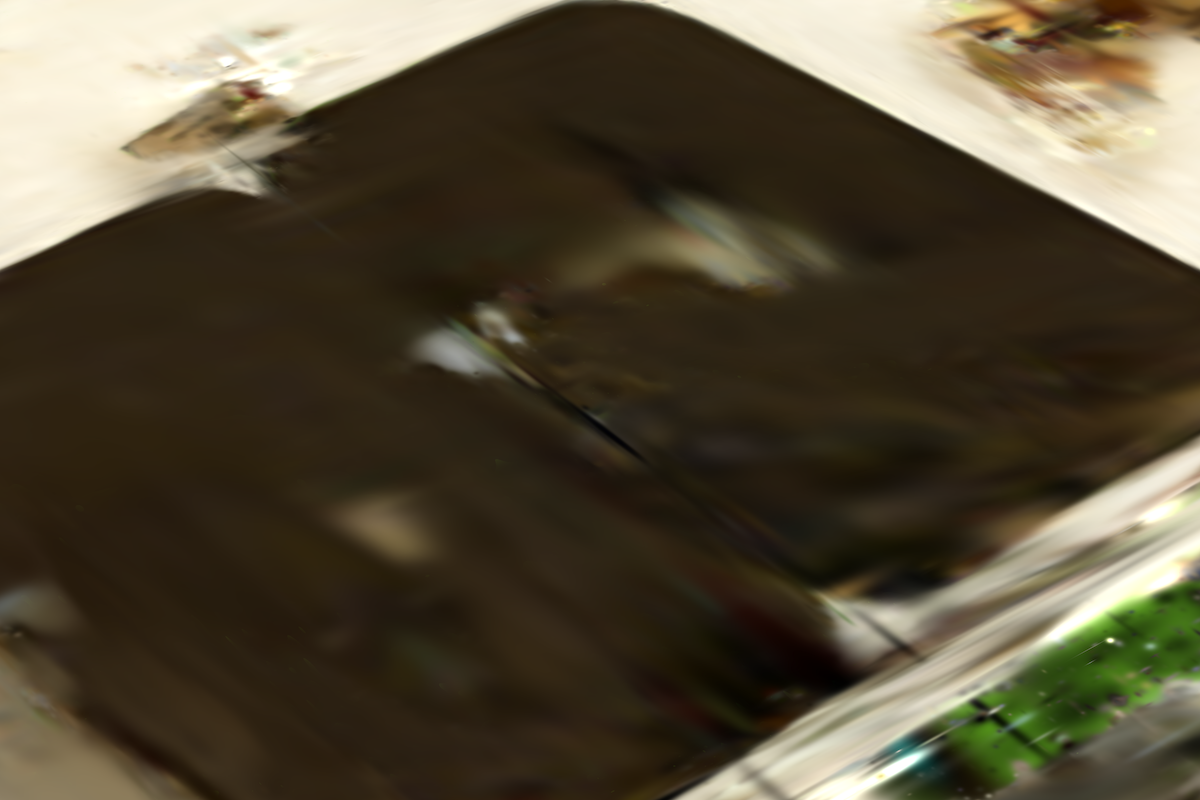} & \includegraphics[width=\psw]{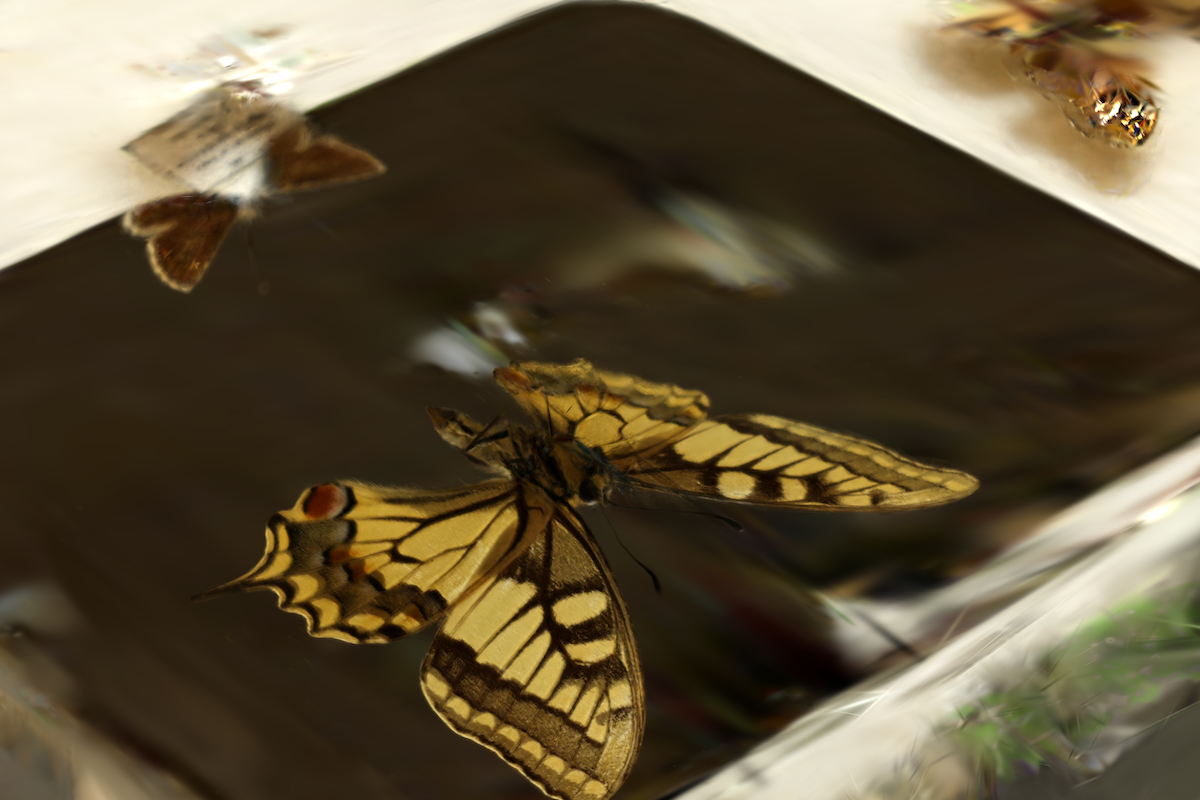} & \includegraphics[width=\psw]{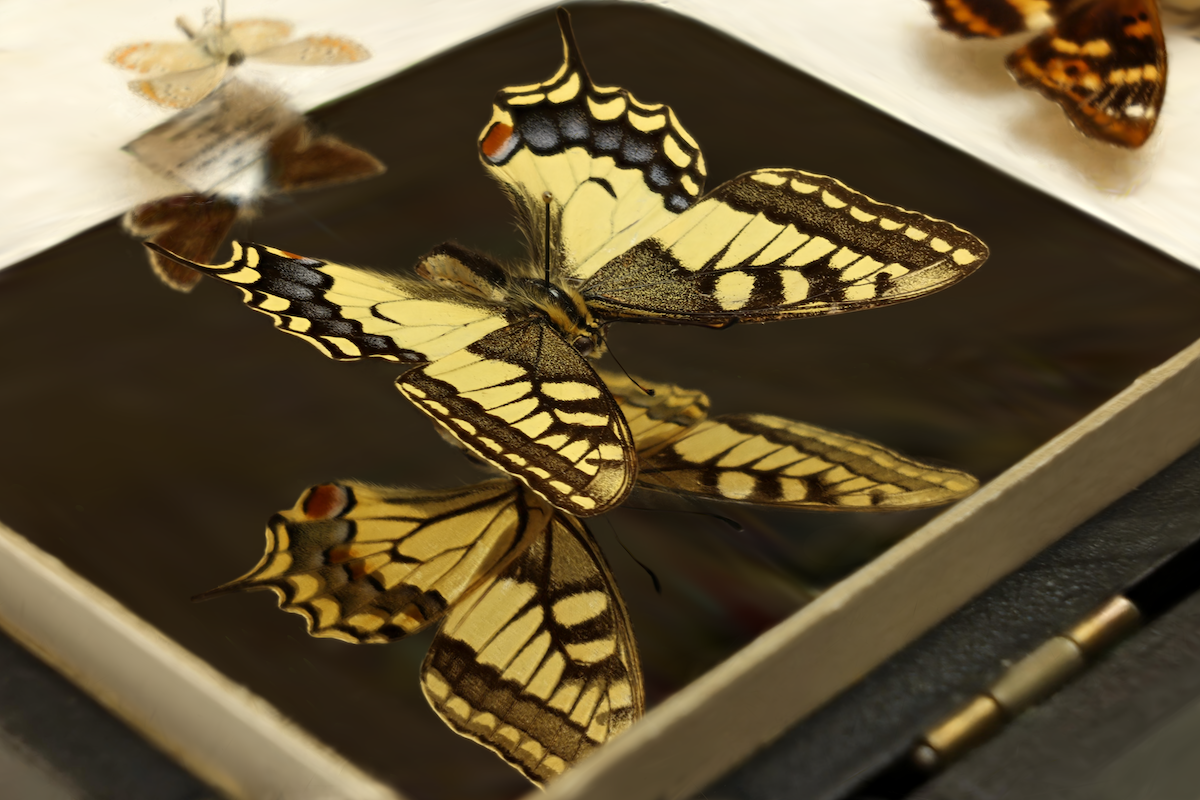} \\
    \small Mirror & \small $+$ Reflected & \small $+$ Direct \\
  \end{tabular}
  \caption{\textbf{Rendering passes}, accumulated back-to-front as in \cref{eq:composite_render}: the mirror Gaussians, the reflected specimen Gaussians confined to the mirror opacity, and the direct specimen Gaussians on top (final render).}
  \label{fig:passes}
\end{figure}

We represent the scene with two sets of Gaussians: the \emph{specimen Gaussians} $\mathcal{G} = \{\mathcal{G}_i\}_{i=1}^N$, a \emph{single} set that models the butterfly, and the \emph{mirror Gaussians} $\mathcal{H} = \{\mathcal{H}_j\}_{j=1}^M$, a separate set that models the mirror surface. The specimen and its reflection explain the butterfly pixels, but not the mirror surface itself: a large, highly reflective, weakly textured region surrounding the specimen. Without an explicit model for it, the only way the optimizer can explain those pixels is by scattering low-opacity floaters in the space around the specimen, which leave a halo that must be removed manually. The mirror Gaussians give the optimizer a dedicated explanation instead: they are initialized as a flat grid on the detected plane $(\mathbf{n}, d)$, sized to the specimen's projection onto it, and kept on the plane by projecting them back onto it after every optimization step. Every training view is rendered in three passes and alpha-composited. The three passes are only needed during training; after training, only the specimen Gaussians are retained.

The \emph{direct pass} rasterizes the specimen $\mathcal{G}$ from the real camera pose $\mathbf{P}$, giving the directly observed (dorsal) view with premultiplied color $\mathbf{C}_r$ and opacity $A_r$. The \emph{reflected pass} produces the virtual image inside the mirror, which is what the specimen looks like from the viewpoint mirrored across the plane: we rasterize the \emph{same} specimen Gaussians from the mirrored camera $\mathbf{P}_m = \mathbf{S}\,\mathbf{P}$, with $\mathbf{S}$ the reflection of \cref{sec:mirrorplane}, yielding the reflected (ventral) view $\mathbf{C}_m, A_m$. The rasterizer evaluates the view-dependent SH color from the mirrored camera center, so the reflected appearance is correct by construction and no transform of the SH coefficients is required. The \emph{mirror pass} rasterizes the mirror Gaussians $\mathcal{H}$ from $\mathbf{P}$, giving the mirror surface $\mathbf{C}_p, A_p$; its opacity $A_p$ delimits the mirror region and confines the reflected pass to it. Over a background color $\mathbf{b}$, the three passes combine back-to-front as
\begin{equation}
  \resizebox{0.9\columnwidth}{!}{$\displaystyle \mathbf{C} = \mathbf{C}_r + (1{-}A_r)\big[A_p\mathbf{C}_m + (1{-}A_pA_m)(\mathbf{C}_p + (1{-}A_p)\mathbf{b})\big]$}.
  \label{eq:composite_render}
\end{equation}
As the reflection is composited in front of the mirror surface, the mirror Gaussians only explain residual mirror pixels. They cannot paint a copy of the specimen onto the plane. \Cref{fig:passes} shows the contribution of each pass.

\subsubsection{Regularization for a Clean Cut-out}
\label{sec:regularizers}

Semi-transparent Gaussians can still surround the specimen to explain background content. We add two regularizers to the specimen Gaussians that reduce these floaters. The first is an \emph{alpha-entropy} loss on the direct pass. With $A_r(\mathbf{x})$ the accumulated specimen opacity at pixel $\mathbf{x}$,
\begin{equation}
  \mathcal{L}_\alpha = -\frac{1}{|\Omega|}\sum_{\mathbf{x}\in\Omega}\Big[A_r\log A_r + (1{-}A_r)\log(1{-}A_r)\Big]
  \label{eq:alphaentropy}
\end{equation}
pushes every pixel toward being fully opaque or fully empty, so the specimen renders as a crisp cut-out rather than a soft cloud; it is enabled once the geometry has settled (from iteration $5$k, weight $0.1$). The second is 2DGS's \emph{depth-distortion} loss~\cite{huang2024_2dgs}, which concentrates per-ray weights into a single depth value to encourage surface alignment. We apply it to the reflected pass (from iteration $3$k, weight $1.0$), where floaters accumulate the most. 

\subsubsection{Implementation}
\label{sec:mirror_implementation}

We build on gsplat~\cite{ye2025gsplat}, an open-source PyTorch library for 3D Gaussian Splatting with differentiable rasterizers for both 3D Gaussians and 2D surfels; the three passes and \cref{eq:composite_render} are implemented directly in its training loop, and the reflection adds no trainable parameters. The specimen Gaussians are initialized from the full dense MVS cloud, pre-folded across the detected plane onto the camera side.
\section{Experiments}
\label{sec:experiments}

\newlength{\qphw}\setlength{\qphw}{0.245\textwidth}
\newcommand{\qimg}[1]{\raisebox{-0.5\height}{\includegraphics[width=\qphw,height=\qphw, keepaspectratio]{#1}}}
\newcommand{\qrow}[1]{\rotatebox[origin=c]{90}{\small #1}}

No prior end-to-end system targets our task, so there is no complete system to benchmark against. We therefore evaluate the two halves of the pipeline separately. \Cref{sec:eval_dof} treats the captures as an ordinary novel-view-synthesis problem, ignoring the mirror, and compares macro defocus extensions of 3DGS against focus stacking. \Cref{sec:eval_mirror} then evaluates the mirror-aware reconstruction on the eight flat specimens: the accuracy of the detected plane, whether mirror-awareness costs novel-view quality on the dorsal side, and finally the fidelity of the ventral surface, which the model only sees reflected.

\subsection{Dataset}
\label{sec:dataset}

We evaluate on \textbf{nine mounted Lepidoptera specimens}, chosen to span a wide diversity of iridescent color effects, physical sizes, dorsal and ventral wing patterns, and mounting types. For each specimen, we acquire 40--90 focus-stacked viewpoints at 8K resolution. Digitizing one specimen takes roughly half an hour of hands-on photography and about $1.5$ hours of largely unattended compute (calibration, plane detection, training) on a single RTX~4090. All nine specimens are used for the macro-defocus comparison (\cref{sec:eval_dof}); one specimen is not mounted flat and therefore does not require mirrors, so the mirror-aware evaluation (\cref{sec:eval_mirror}) uses the remaining eight. Specimens $7$ and $8$ were photographed together (\cref{fig:dataset}, right) and share a single capture and reconstruction; each is evaluated with its own 2D masks. \Cref{fig:dataset} shows two example captured images; photographs and descriptions of all specimens are given in the supplementary material (Sec.~A).

\newlength{\dsw}\setlength{\dsw}{0.47\linewidth}
\begin{figure}[htbp]
  \centering
  \includegraphics[width=\dsw]{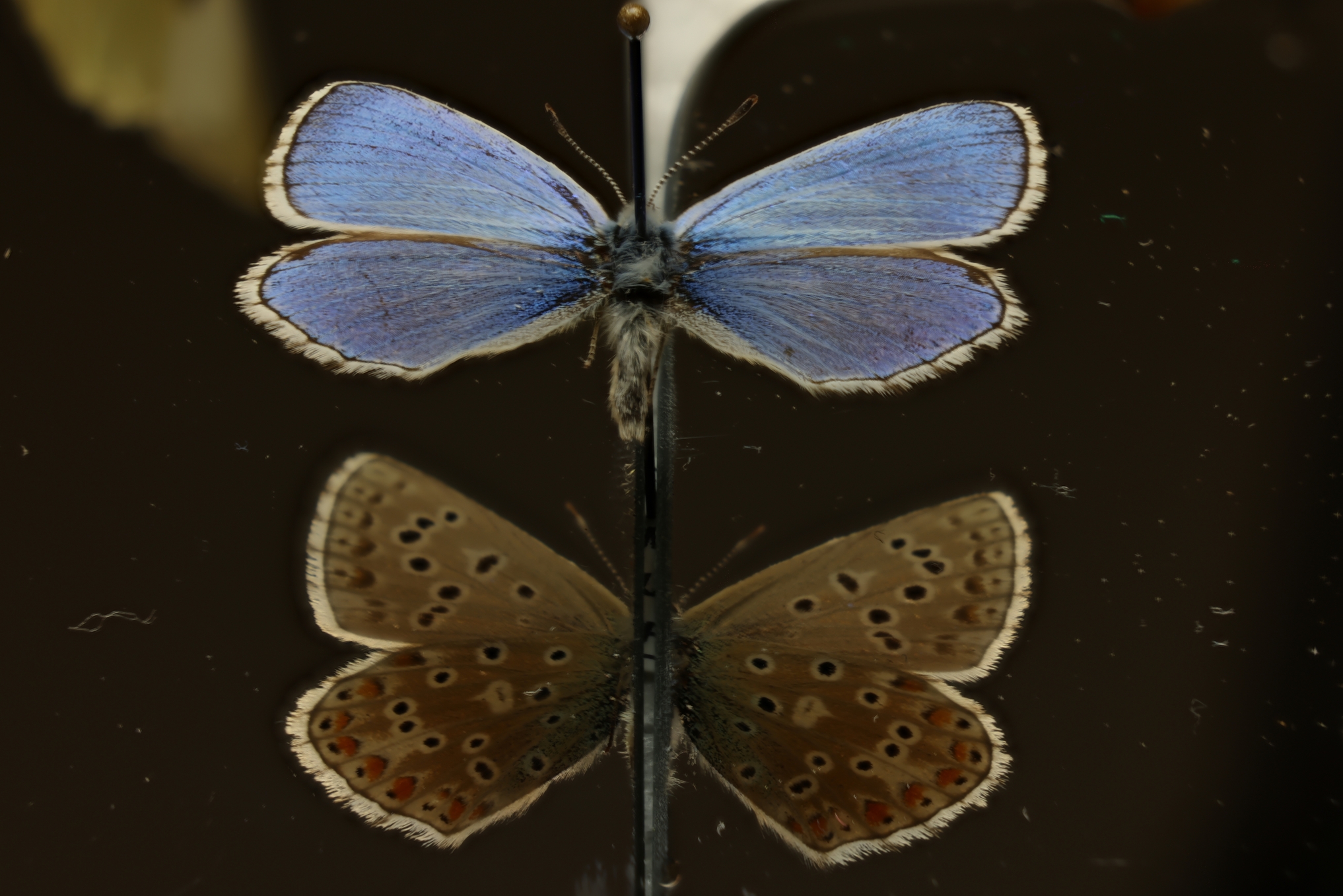}\hfill\includegraphics[width=\dsw]{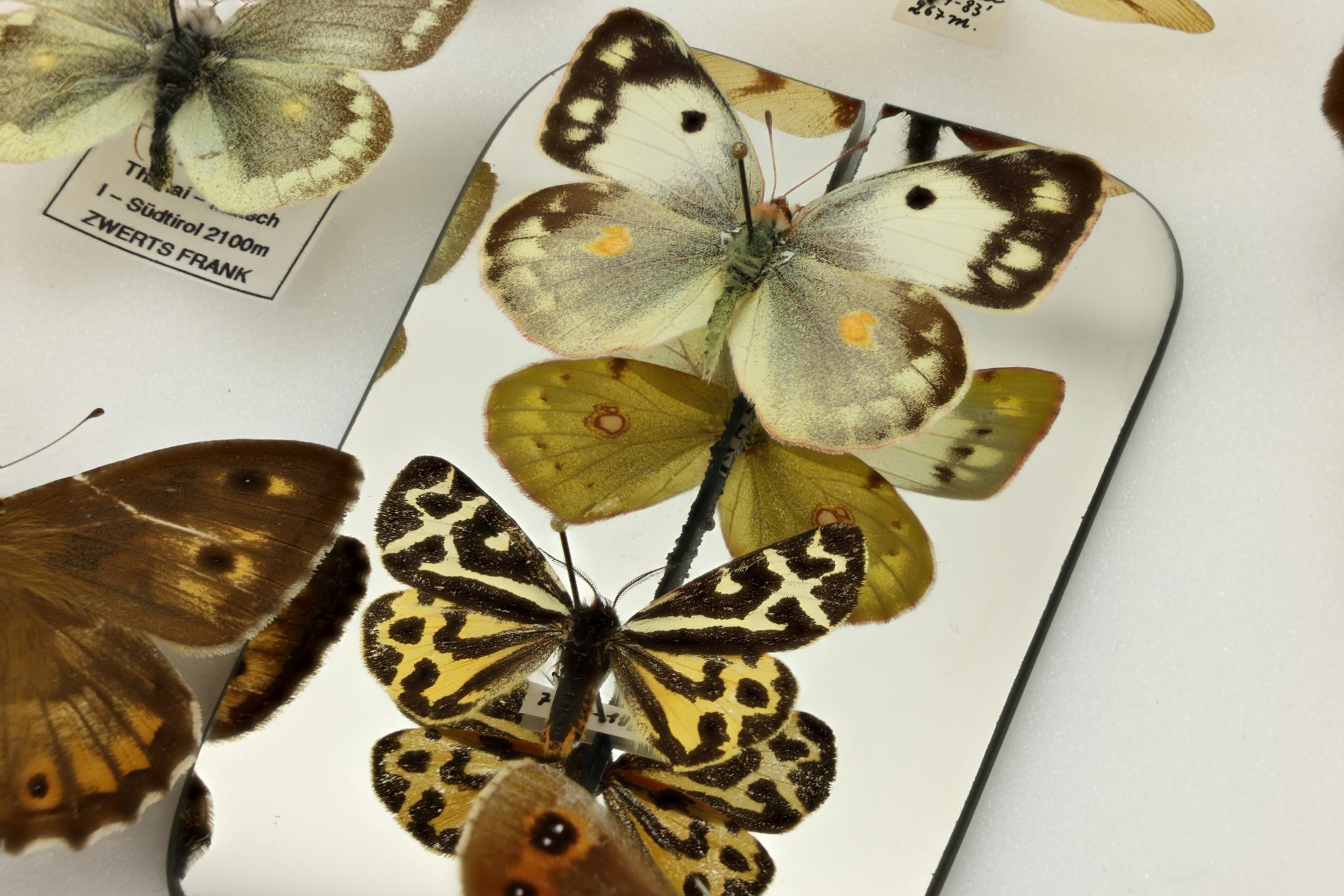}
  \caption{ \textbf{Evaluation dataset.} Three example mounted Lepidoptera specimens, illustrating their visual diversity and the mirror setup. All specimens are shown in the supplementary material (Sec.~A).}
  \label{fig:dataset}
\end{figure}

\subsection{Evaluation Protocol}
\label{sec:protocol}

We report PSNR, SSIM~\cite{wang2004ssim}, and LPIPS~\cite{zhang2018lpips} on held-out views, with images downscaled by a factor of four. Because the captured viewpoints per specimen are relatively sparse, we do not use the 7/8-for-training, 1/8-for-testing split common in the NVS literature. Instead, for each specimen, we train five separate models, each holding out a different view, and report the mean over all held-out views. In all dorsal-view experiments, render and ground truth (GT) are both composited to white outside a manually drawn 2D mask of the specimen and its reflection. Masking the ground truth scores only the butterfly, and masking the render prevents unrelated content in the reconstruction (e.g., the butterfly case or neighboring specimens) from being counted as error, since we do not crop the reconstruction for these experiments. The masks are used only for evaluation. Per-specimen and full-frame results of all experiments are in the supplementary material (Secs.~F--H).

\subsection{Without Mirrors: Macro Defocus \& Primitives}
\label{sec:eval_dof}

We first ignore the mirror entirely, training plain 3DGS and 2DGS without our mirror extension. Two strategies for handling macro defocus exist: handheld focus stacking before training, or training a \emph{depth-of-field-aware} 3DGS on the raw shallow-DoF frames. We compare DoF-Gaussian~\cite{shen2025dof}, which fits a lens-based imaging model with a differentiable circle of confusion during training, with our focus-stacking preprocessing. DoF-Gaussian receives one randomly chosen frame per viewpoint, at the same resolution and with the same camera poses as our fused images.

\begin{table}[htbp]
  \centering
  \caption{ \textbf{Macro-defocus handling.} Novel-view-synthesis quality of handheld focus stacking (ours, under both splatting primitives) versus DoF-Gaussian~\cite{shen2025dof}, a depth-of-field-aware 3DGS.}
  \label{tab:dof}
  \small
  \setlength{\tabcolsep}{3.5pt}
  \begin{tabular}{@{}lccc@{}}
    \toprule
    Method & PSNR$\uparrow$ & SSIM$\uparrow$ & LPIPS$\downarrow$ \\
    \midrule
    DoF-Gaussian~\cite{shen2025dof} & 26.06 & 0.926 & 0.0885 \\
    \midrule
    Focus stacking (3DGS) & 29.71 & 0.947 & 0.0438 \\
    Focus stacking (2DGS) & \textbf{30.30} & \textbf{0.949} & \textbf{0.0408} \\
    \bottomrule
  \end{tabular}
\end{table}

\begin{figure}[htbp]
  \centering
  \setlength{\qphw}{0.45\linewidth}%
  \setlength{\tabcolsep}{1.5pt}%
  \renewcommand{\arraystretch}{1.1}%
  \begin{tabular}{@{}c@{\hspace{3pt}}cc}
    \qrow{Ours} & \qimg{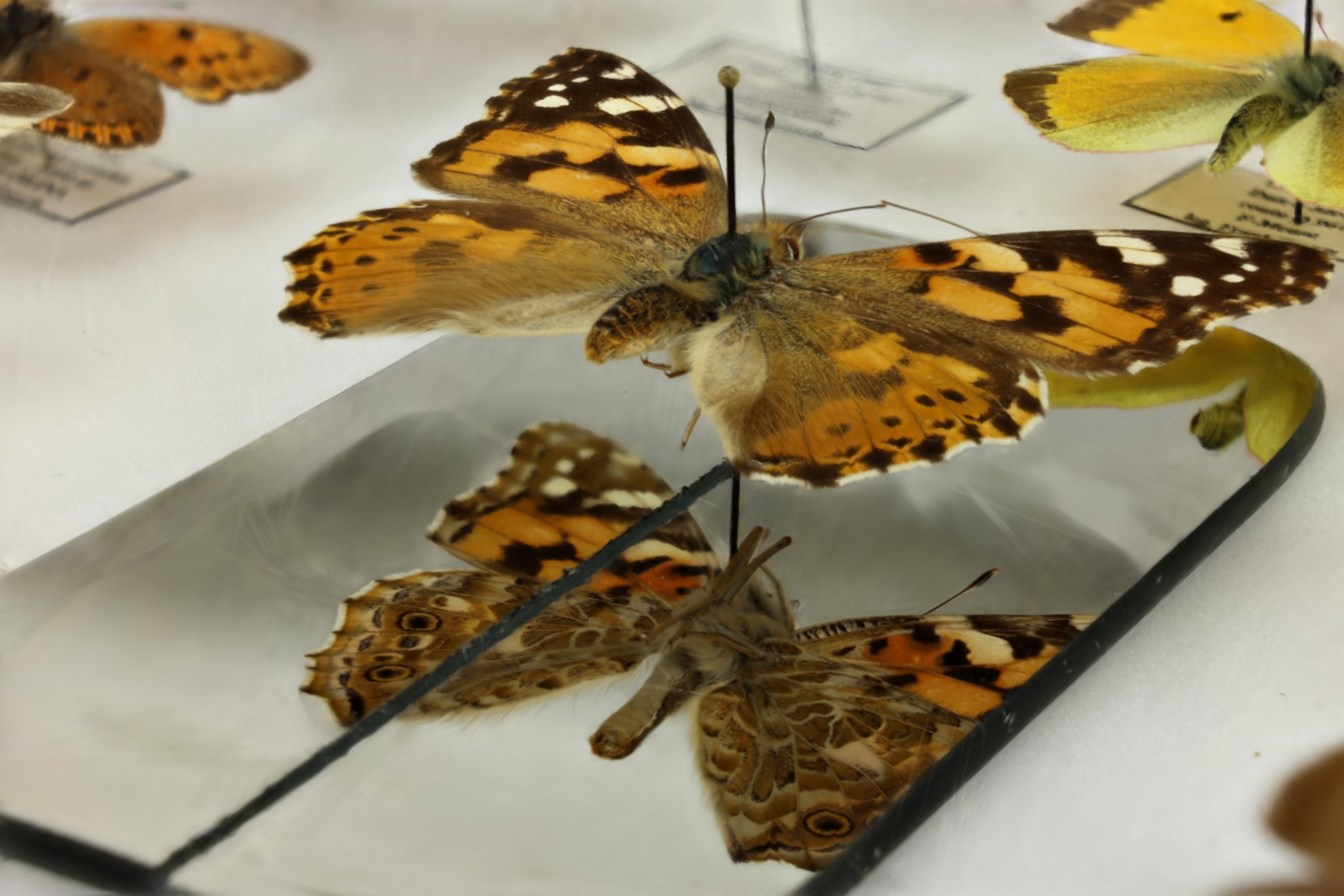} & \qimg{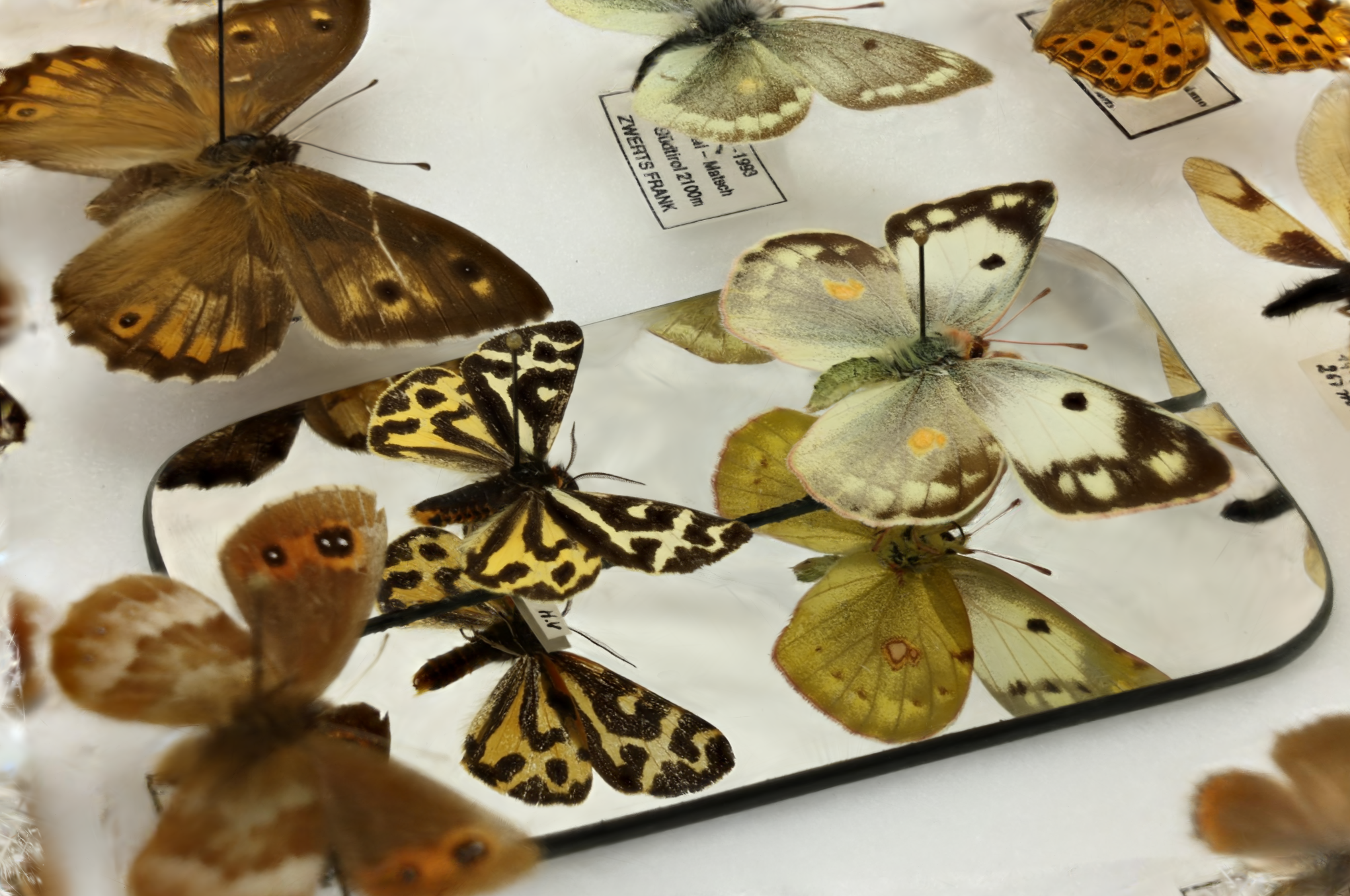} \\
    \qrow{DoF-Gaussian} & \qimg{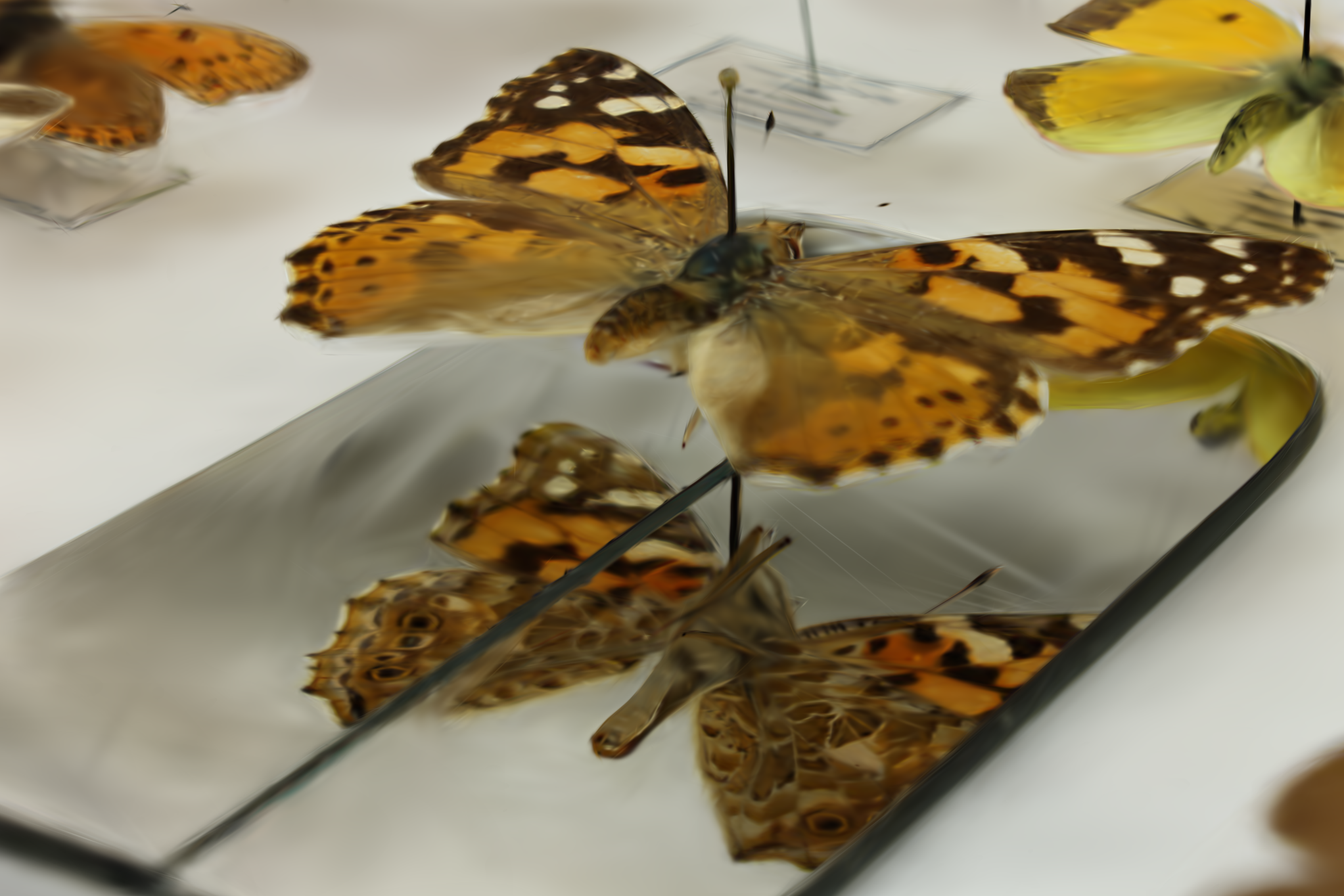} & \qimg{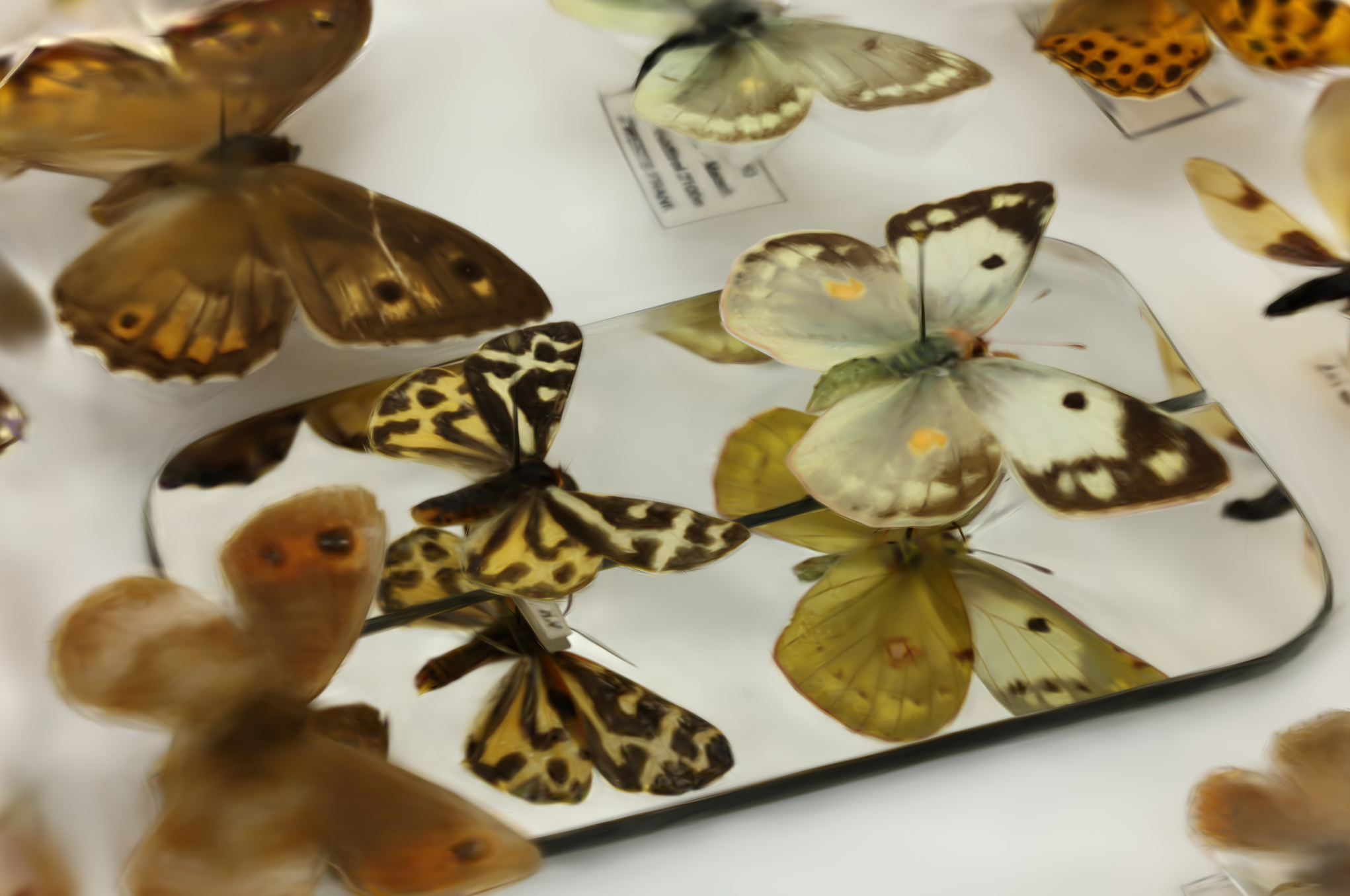} \\
  \end{tabular}
  \caption{\textbf{Qualitative comparison of macro-defocus handling.} Two held-out views; top: handheld focus stacking with 2DGS (ours), bottom: DoF-Gaussian~\cite{shen2025dof}. Focus stacking recovers sharp scales, wing veins, and patterns that DoF-Gaussian does not.}
  \label{fig:qual_dof}
\end{figure}

\Cref{tab:dof} reports the quantitative comparison and \cref{fig:qual_dof} the qualitative one. On our data, DoF-Gaussian fails to recover high-frequency scale and vein detail, whereas focus stacking consistently delivers sharp, all-in-focus renders. It improves masked PSNR by $3.6$--$4.2$\,dB and more than halves the perceptual (LPIPS) error. 2DGS is marginally ahead of 3DGS on every metric ($+0.6$\,dB masked, $+0.5$\,dB full frame) at essentially the same Gaussian count ($1.33$M vs.\ $1.37$M). We therefore adopt 2DGS in further experiments.

\subsection{Mirror-Aware Reconstruction}
\label{sec:eval_mirror}

\subsubsection{Mirror-Plane Accuracy}
\label{sec:eval_plane}

We quantify the accuracy of the automatic mirror-plane detection of \cref{sec:mirrorplane} against a manual reference plane, fitted robustly to points hand-picked along the visible mirror edges in the dense point cloud (see the supplementary material, Sec.~C). Averaged over the eight mirror specimens, the detected plane agrees with the reference to within $\Delta\theta = 0.64^\circ \pm 0.34^\circ$ in normal direction and $\Delta d = 0.40\% \pm 0.34\%$ of the specimen extent in offset, and the hand-picked mirror points lie on average within $0.57\%$ of the specimen extent from the detected plane. Manual point picking is itself imprecise, so part of this discrepancy is attributable to inaccuracies in the ground truth rather than to the detector.

\subsubsection{Dorsal-Side Novel-View Quality}
\label{sec:eval_dorsal}

We compare our mirror-aware 2DGS with regular 2DGS trained on the same images. The held-out views are dorsal-side photographs, in which both the butterfly and its reflection are scored. Ventral-side evaluation is done separately in \cref{sec:eval_ventral}.

\noindent\textbf{Densification.} On the mirror-unaware baseline, we first observed that adaptive density control (ADC)~\cite{kerbl2023gaussian} \emph{lowers} held-out quality: regular 2DGS loses $1.4$\,dB PSNR with ADC while using $1.5\times$ more Gaussians (\cref{tab:reflect}). Mounted Lepidoptera are so richly textured that the dense multi-view-stereo cloud we initialize from (\cref{sec:mirror_implementation}) already covers the specimen sufficiently. Densification predominantly adds floaters to account for the background and the mirror, which, seen from a held-out viewpoint, can drift in front of the butterfly, lowering the metrics. Therefore, we turn off densification for both the baseline and our model.

\noindent\textbf{Ours vs.\ regular 2DGS.} Our problem is technically harder than the baseline's. A mirror-unaware model is free to explain each view as two independent objects, the dorsal surface and a separate ``virtual'' butterfly behind the mirror, each fitted to the images without requiring that they describe the same specimen. In contrast, our model must reconstruct a single butterfly whose direct view and reflection are consistent with each other. Despite this, our model is on par with regular 2DGS on every metric (\cref{tab:reflect}; $-0.11$\,dB PSNR, $+0.002$ SSIM, $+0.001$ LPIPS) with $2.4\times$ fewer Gaussians ($0.38$M vs.\ $0.91$M), and it delivers a single specimen that is accurate from any viewpoint.

\noindent\textbf{Ablation.} \Cref{tab:reflect} ablates the three design choices of \cref{sec:gaussianreflection}: producing the reflection by reflecting the camera~\cite{meng2024mirror} or the splats~\cite{liu2024mirrorgaussian}; modeling the mirror surface with explicit mirror Gaussians or not (without them, our model is the segmentation-free counterpart of Mirror-3DGS~\cite{meng2024mirror}); and densifying the specimen Gaussians with ADC~\cite{kerbl2023gaussian} or not at all.

\begin{figure}[b]
  \centering
  \setlength{\tabcolsep}{1.5pt}%
  \renewcommand{\arraystretch}{1.2}%
  \begin{tabular}{@{}c@{\hspace{2pt}}cc@{}}
    \small $\alpha$-entropy off & $\alpha$-entropy on \\
    \includegraphics[width=2.5cm]{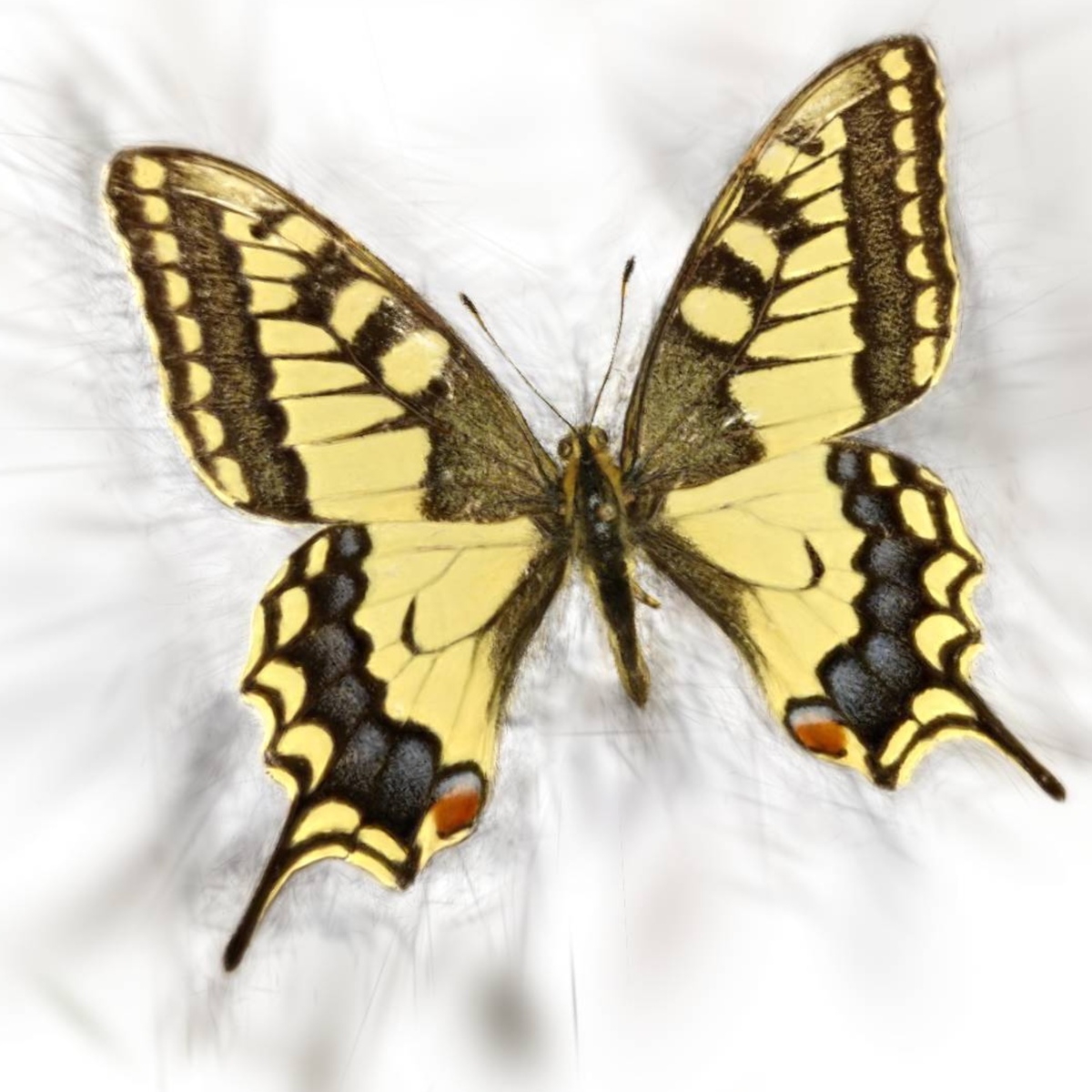} & \includegraphics[width=2.5cm]{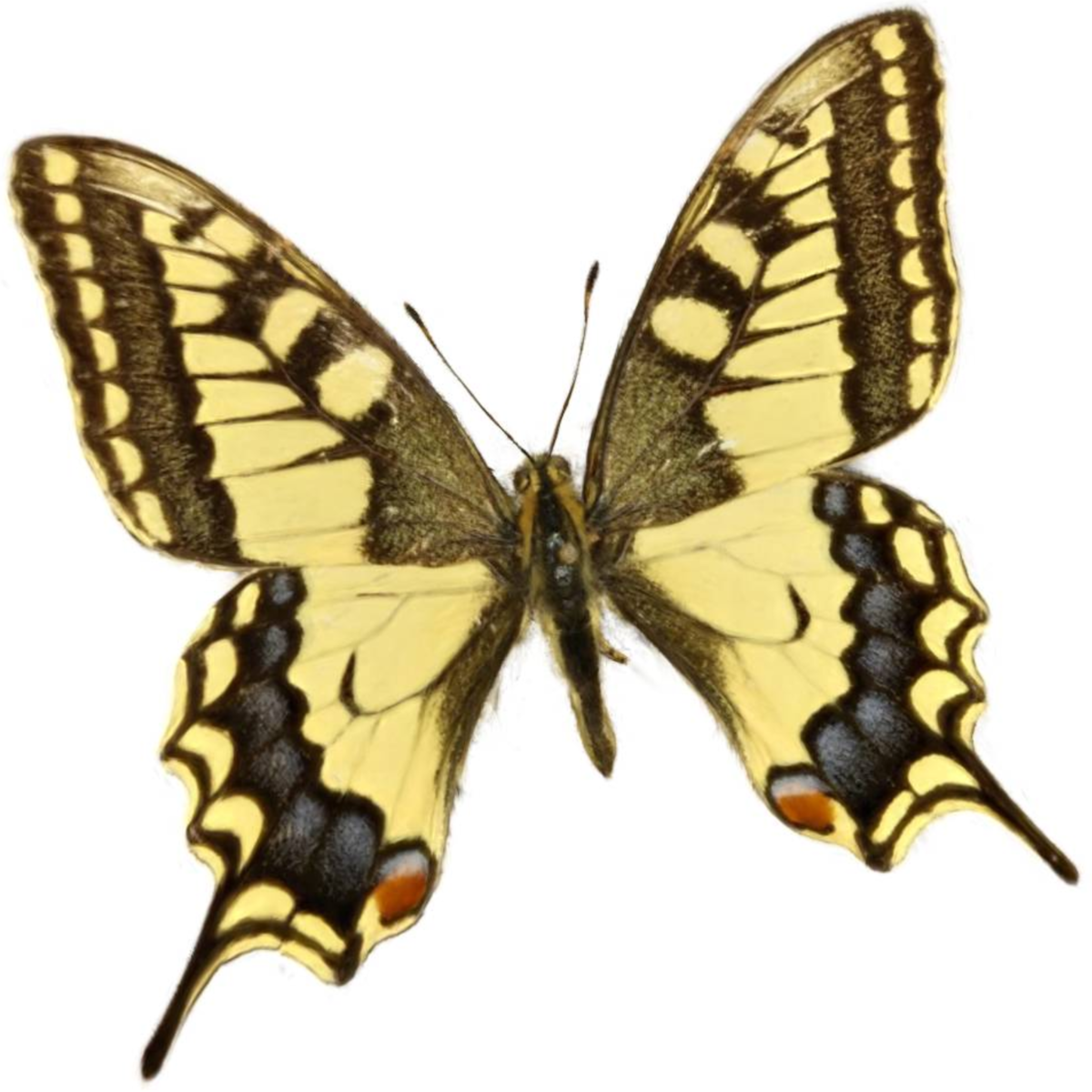} \\[2pt]
    \small distortion loss off & \small distortion loss on \\
    \includegraphics[width=2.5cm]{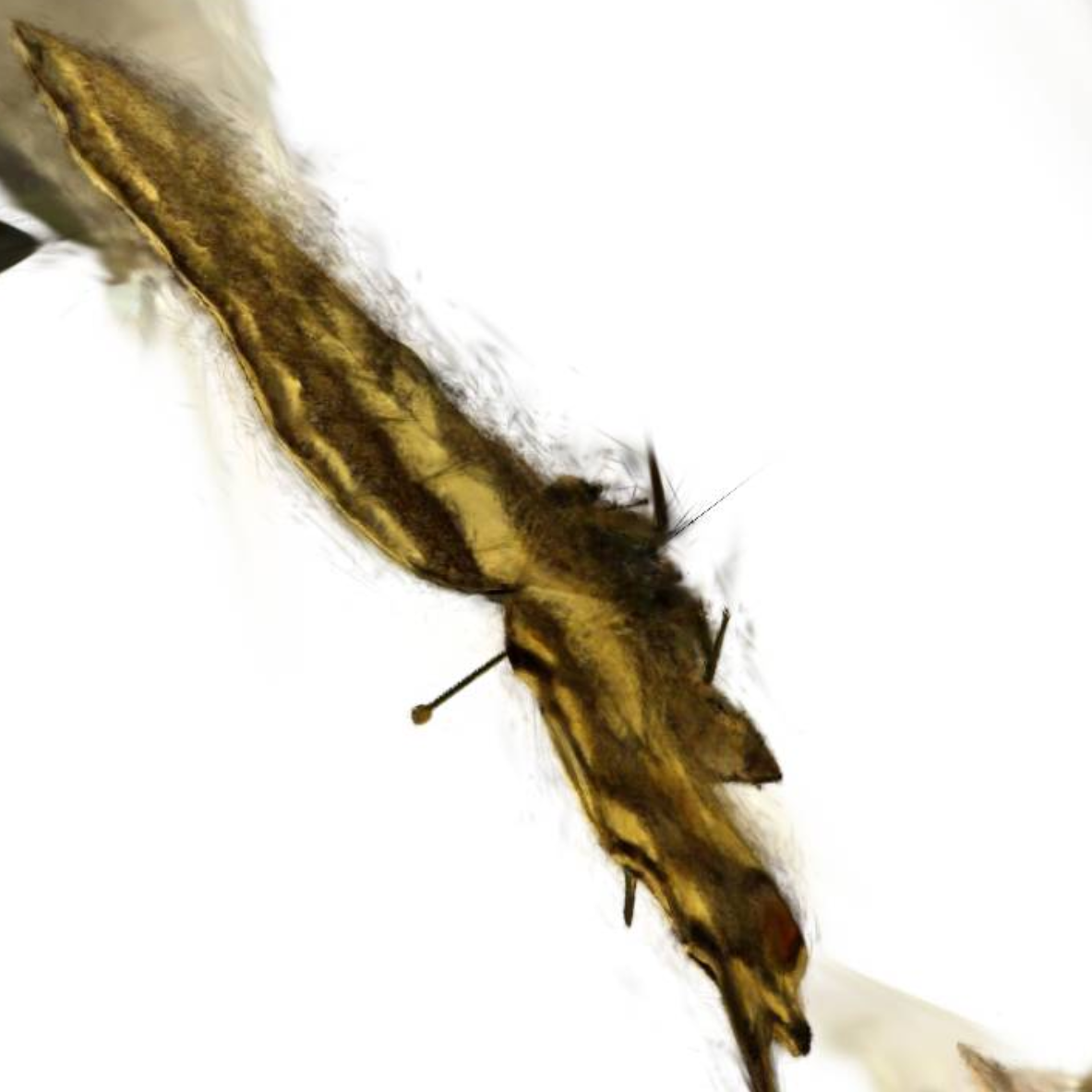} & \includegraphics[width=2.5cm]{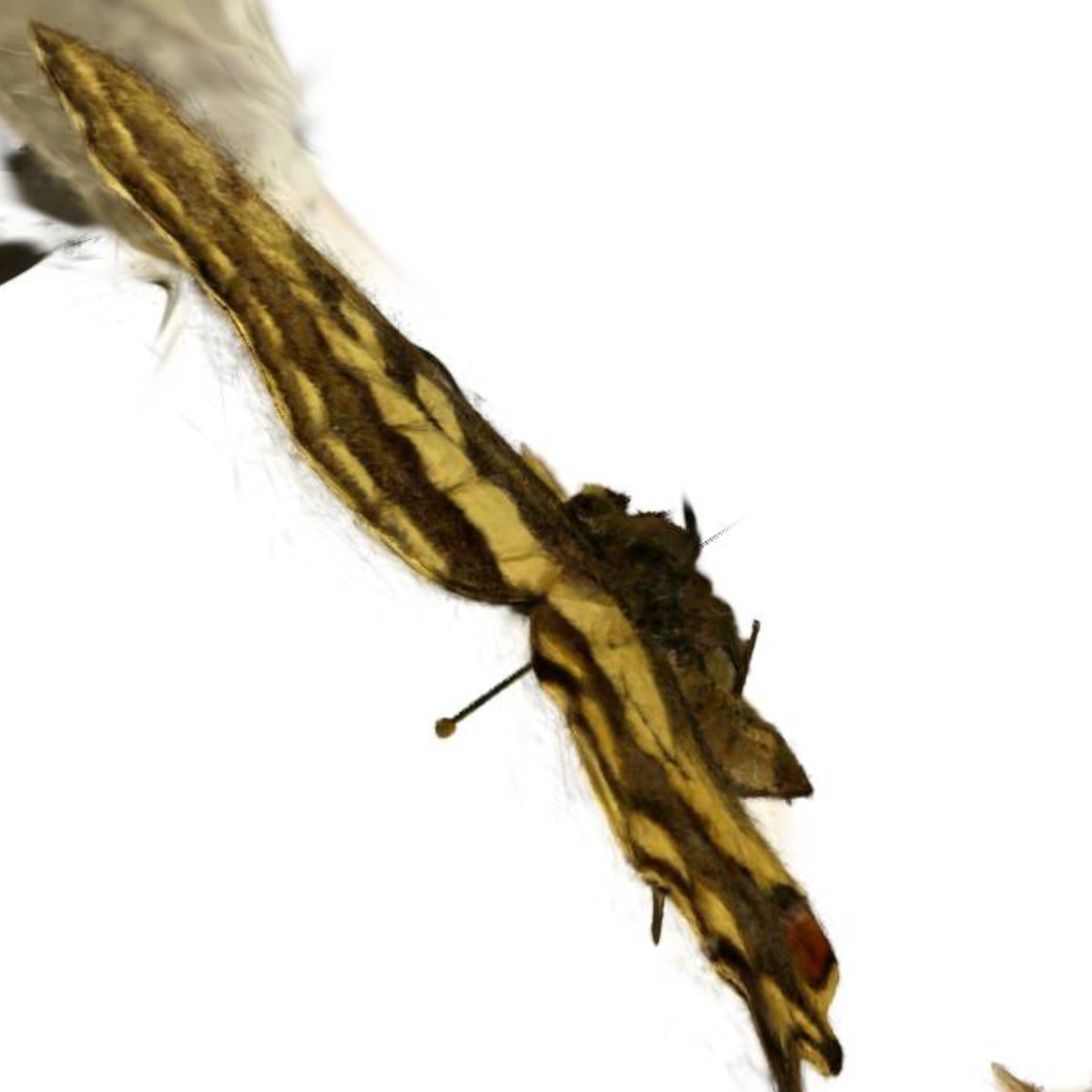} \\
  \end{tabular}
  \caption{ \textbf{Effect of the regularizers of \cref{sec:regularizers}.} Top: alpha-entropy loss off/on; bottom: distortion loss off/on. }
  \label{fig:ablation_reg}
\end{figure}

Reflecting the camera with explicit mirror Gaussians and \emph{no} densification is best mirror-aware configuration across all metrics while using the fewest primitives ($0.38$M vs.\ $0.42$--$1.17$M). Densification costs $3.7$\,dB under ADC, for the reason given above. Reflecting the \emph{splats} instead of the camera performs on par at the densification-free setting ($28.35$ vs.\ $28.25$\,dB, both without a plane) but doubles the primitive count. It suffers most when ADC is enabled ($-4.7$\,dB). Removing the mirror Gaussians under the same densification-free setting incurs a $2.9$\,dB penalty, as the algorithm attempts to explain the mirror, creating harmful floater artifacts in the process (\cref{sec:gaussianreflection}). On top of the chosen configuration, the final model adds the two regularizers of \cref{sec:regularizers} (the last row), whose purpose is to reduce post-processing rather than to improve novel-view quality: they leave masked quality essentially unchanged ($-0.07$\,dB) while yielding a crisp cut-out and a flat, floater-free surface (\cref{fig:ablation_reg}).

\begin{table}[htbp]
  \centering
  \caption{ \textbf{Mirror-aware reconstruction and ablation.} Masked quality over the held-out dorsal views of the eight mirror specimens. \emph{Reflect}: reflect the camera (Cam) or the splats;  \emph{Plane}: explicit mirror model; \emph{Densify}: ADC or none; \emph{Reg.}: regularizers of \cref{sec:regularizers}; \#GS: mean number of Gaussians. Best per metric in \textbf{bold}; the last row is our full model. The first two rows are the mirror-unaware 2DGS baselines.}
  \label{tab:reflect}
  \small
  \resizebox{\columnwidth}{!}{%
  \begin{tabular}{@{}llllcccc@{}}
    \toprule
    Reflect & Plane & Densify & Reg. & PSNR$\uparrow$ & SSIM$\uparrow$ & LPIPS$\downarrow$ & \#GS\,(M)$\downarrow$ \\
    \midrule
    No & no & ADC  & no & 29.77 & 0.945 & 0.0421 & 1.35 \\
    No & no & none & no & \textbf{31.16} & 0.947 & 0.0400 & 0.91 \\
    \midrule
    Cam & yes & none & no & 31.12 & \textbf{0.950} & \textbf{0.0392} & \textbf{0.38} \\
    Cam & yes & ADC  & no & 27.43 & 0.940 & 0.0566 & 0.98 \\
    Cam & no  & none & no & 28.25 & 0.940 & 0.0522 & 0.42 \\
    Splats & no & ADC  & no & 23.69 & 0.924 & 0.0781 & 1.17 \\
    Splats & no & none & no & 28.35 & 0.941 & 0.0513 & 0.84 \\
    \midrule
    Cam & yes & none & yes & 31.05 & 0.949 & 0.0412 & 0.38 \\
    \bottomrule
  \end{tabular}}
\end{table}

\subsubsection{Ventral-Surface Validation}
\label{sec:eval_ventral}

\begin{figure*}[htbp]
  \centering
  \setlength{\qphw}{0.135\textwidth}%
  \setlength{\tabcolsep}{1.5pt}%
  \renewcommand{\arraystretch}{1.1}%
  \begin{tabular}{@{}c@{\hspace{3pt}}cccccc@{}}
    \qrow{GT} & \qimg{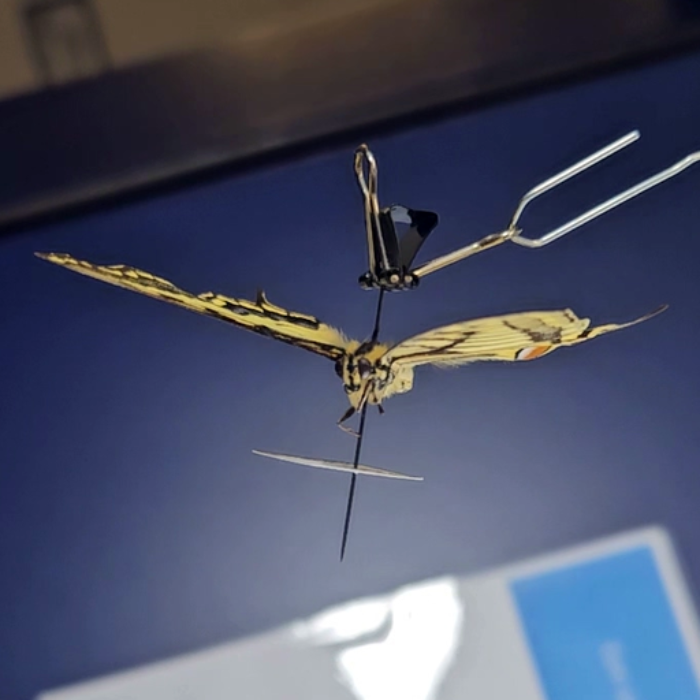} & \qimg{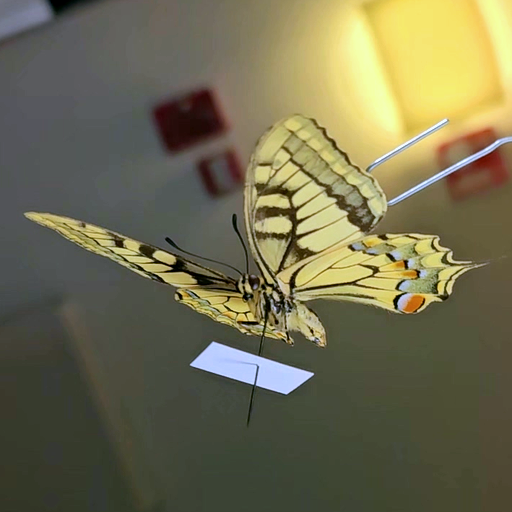} & \qimg{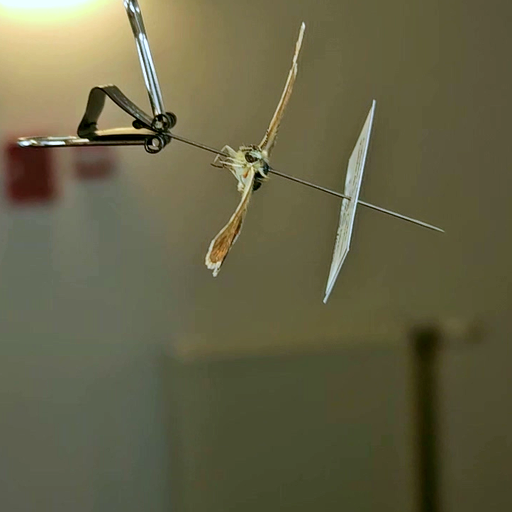} & \qimg{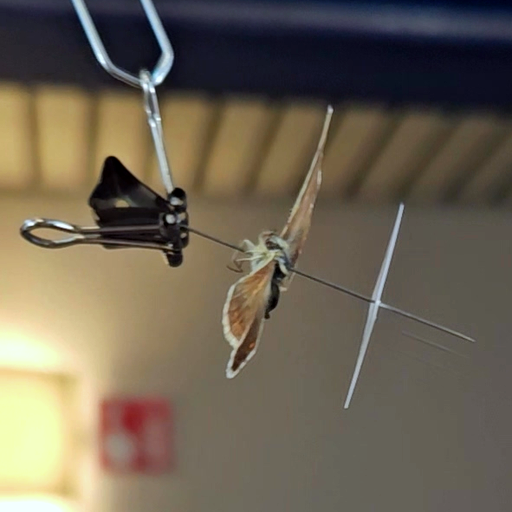} & \qimg{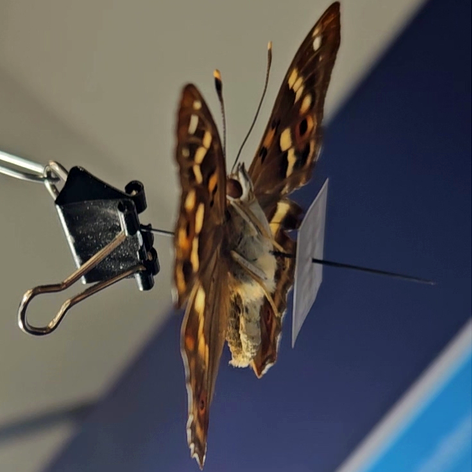} & \qimg{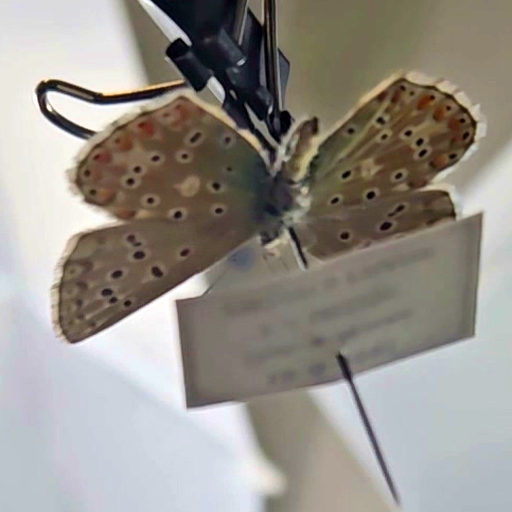} \\
    \qrow{Ours} & \qimg{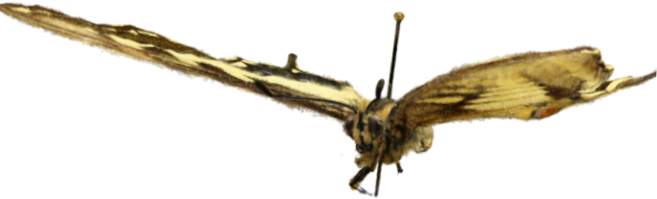} & \qimg{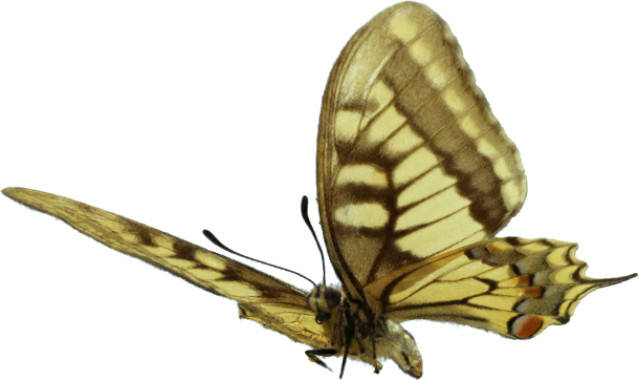} & \qimg{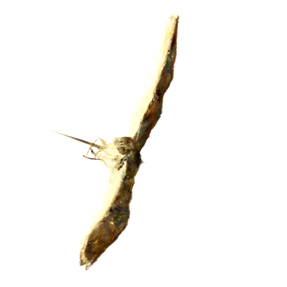} & \qimg{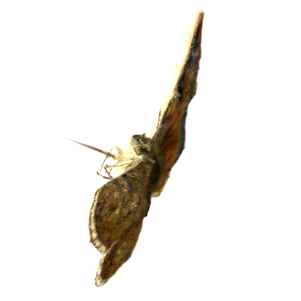} & \qimg{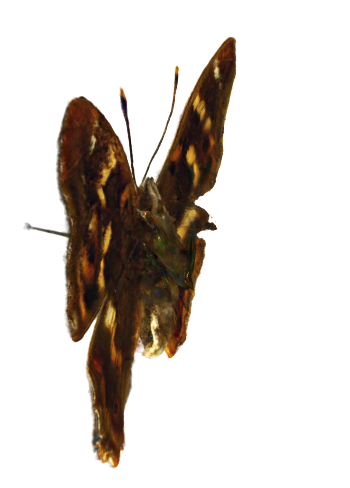} & \qimg{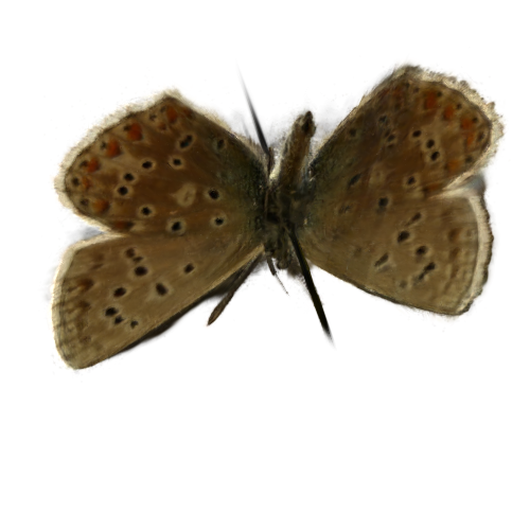} \\
    \qrow{No mirroring} & \qimg{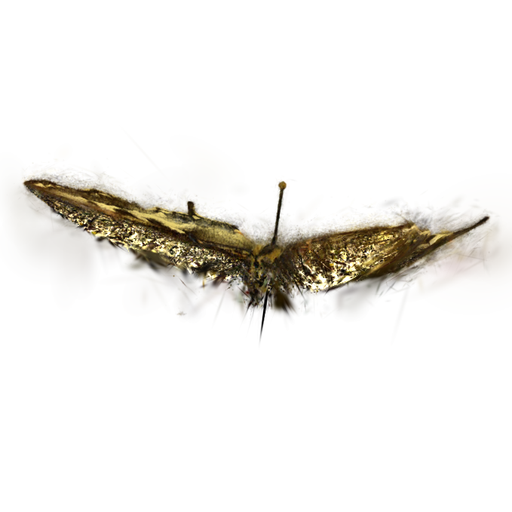} & \qimg{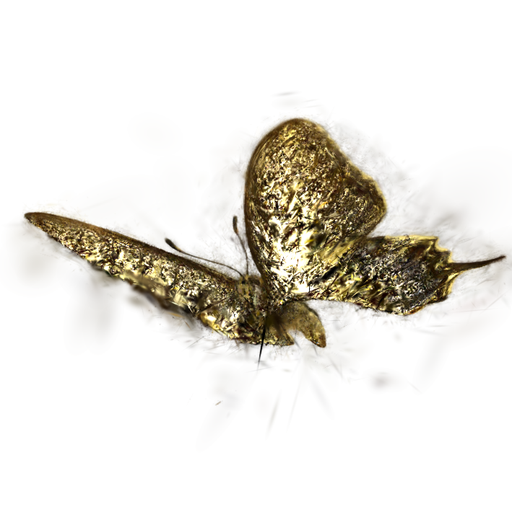} & \qimg{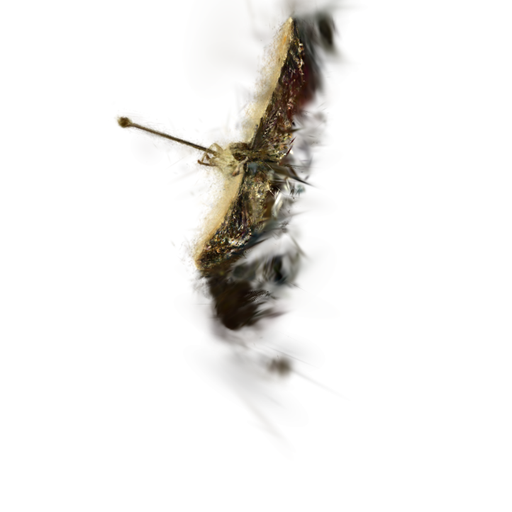} & \qimg{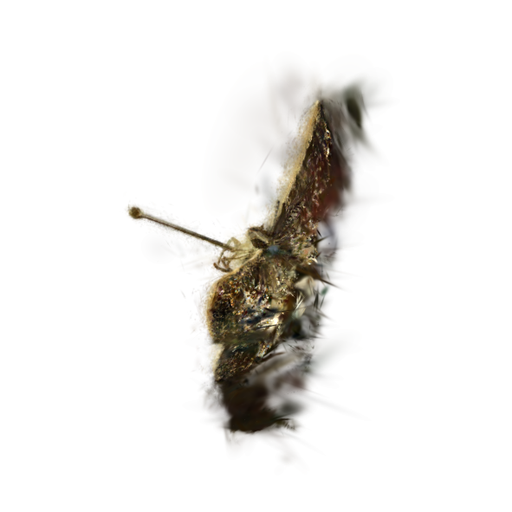} & \qimg{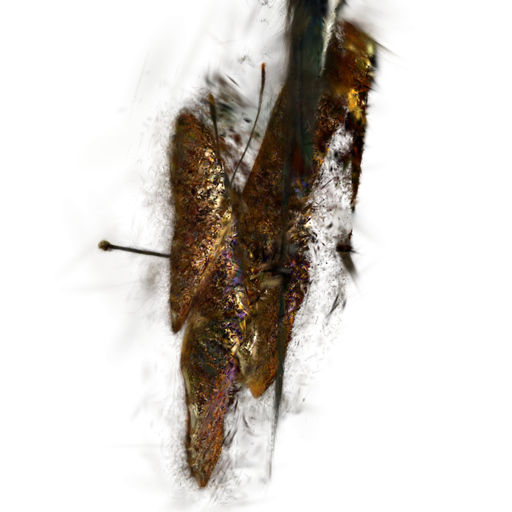} & \qimg{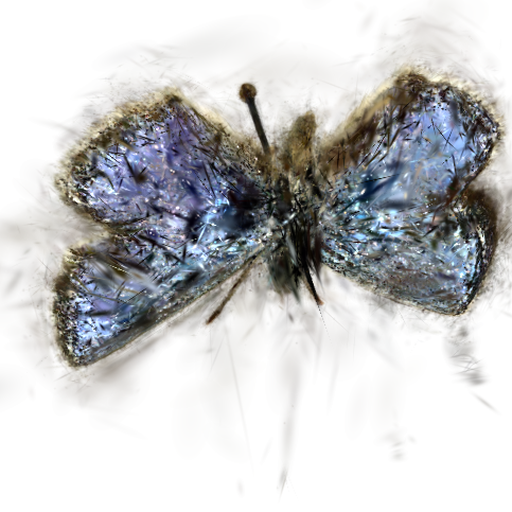} \\
  \end{tabular}
  \caption{ \textbf{Ventral-surface reconstruction.} Six held-out views from the separately acquired ventral photographs; top to bottom: ground truth, ours, and the mirror-unaware baseline.}
  \label{fig:qual_mirror}
\end{figure*}

The mirror allows our method to reconstruct the specimen's \emph{ventral} surface, even though it is observed only as a reflection. To validate that reconstruction against ground truth, we separately photograph the ventral surface of each specimen in a fresh acquisition, run an independent COLMAP reconstruction of these views, and treat them as held-out novel views of the ventral surface. Acquiring this ground truth requires exactly the handling our pipeline is designed to avoid (taking the specimen out of its case and hanging it on nylon monofilament). Despite our best care, two specimens lost their antennae during processing, so we limited this validation to four specimens.

The two acquisitions live in independent COLMAP frames, which we align with a manual similarity transform refined by ICP. Because this alignment is only coarse, we follow the in-the-wild protocol of NeRS~\cite{zhang2021ners}: the reconstruction is frozen, and each held-out camera is refined against its own view (details in the supplementary material, Sec.~E). We report PSNR, SSIM, LPIPS, and FLIP~\cite{andersson2020flip} on $256{\times}256$ object crops, masking only the ground truth. Because the specimen was moved for this acquisition, we isolate each reconstruction from its original background through a manually placed 3D bounding box. As a result, the render contains no background, and we can score the renders unmasked, which penalizes floaters around the specimen. The held-out ventral photographs were taken with a different camera and a standard lens under different lighting conditions, so the absolute color and exposure differ from the renders even when the geometry is exact, leading to lower PSNR scores. The perceptual metrics (LPIPS and FLIP) are therefore more meaningful here. \Cref{tab:ventral} and \cref{fig:qual_mirror} compare against the mirror-unaware baseline of \cref{sec:eval_dorsal}. Our method wins on every metric, most clearly on the perceptual ones. To separate the effects of camera and lighting changes from the quality discrepancy between dorsal and ventral surfaces, the last row of \cref{tab:ventral} serves as a reference: it scores the dorsal views from the same acquisition using our model under the same protocol. The ventral surface trails the dorsal view in PSNR by $1.3$\,dB, and in LPIPS by $0.03$, with SSIM and FLIP on par.

\begin{table}[htbp]
  \centering
  \caption{ \textbf{Ventral-surface validation.} $256{\times}256$ object crops, pose-refined; render unmasked, ground truth masked. \emph{Last row:} dorsal reference from the same acquisition.}
  \label{tab:ventral}
  \small
  \setlength{\tabcolsep}{4pt}
  \begin{tabular}{@{}lcccc@{}}
    \toprule
    Method & PSNR$\uparrow$ & SSIM$\uparrow$ & LPIPS$\downarrow$ & FLIP$\downarrow$ \\
    \midrule
    Ours (mirror) & \textbf{17.27} & \textbf{0.861} & \textbf{0.184} & \textbf{0.121} \\
    No mirroring  & 14.73 & 0.768 & 0.278 & 0.205 \\
    \midrule
    \textit{Ours, dorsal (reference)} & \textit{18.56} & \textit{0.862} & \textit{0.153} & \textit{0.126} \\
    \bottomrule
  \end{tabular}
\end{table}
\section{Limitations and Future Work}
\label{sec:limitations}

\noindent\textbf{Residual floaters under poor capture illumination.}
Our mirror Gaussians model the mirror surface as regular splats with a view-dependent radiance, whereas a real mirror specularly reflects the environment---in particular, the studio lights, whose high-frequency reflections move across the mirror as the viewpoint changes and can only be approximated by a radiance-only surface. How much this matters depends on the capture illumination. In early captures, our specimens were recorded under bright point lights over a dark background: there, the reconstruction absorbs the discrepancy as a halo of \emph{floaters} that cling to the specimen, which the alpha-entropy and distortion regularizers (\cref{sec:regularizers}) markedly reduce but do not fully remove, leaving dark borders. In later recordings against a light background with even, diffuse lighting, the mirror surface is modeled considerably better, and few floaters remain.

\noindent\textbf{Lower fidelity on the ventral surface.} The ventral surface is reconstructed at slightly lower fidelity than the dorsal side, because it is seen from fewer viewpoints and only indirectly through the mirror. Capturing additional oblique viewpoints would provide more direct evidence of the ventral surface and help close this gap. The ventral surface also appears slightly darker than in reality, as it was less brightly illuminated than the dorsal side during capture.
\section{Conclusion}
\label{sec:conclusion}
We have presented a complete acquisition and reconstruction pipeline for high-fidelity novel-view synthesis of mounted Lepidoptera. By combining a non-contact first-surface mirror system, handheld focus stacking with automatic image registration, and a segmentation-free mirror-aware extension of 3D Gaussian Splatting, our system overcomes the challenges encountered when digitizing this class of fragile, detail-rich objects in 3D. The resulting models will be deployed in a virtual museum experience.
\section*{Acknowledgments}
This research was funded by VLAIO under the TETRA program (grant agreement No. HBC.2025.0131). We gratefully acknowledge imec IDLab for computational resources via the iLab.t infrastructure, Pieter Op De Beeck of AlterEyes for providing the use case, and Frank Zwerts for providing the butterfly specimens.

{\small
\bibliographystyle{ieeenat_fullname}
\bibliography{egbib}

@string{CVPR      = {IEEE/CVF Conference on Computer Vision and Pattern Recognition (CVPR)}}

@string{ECCV      = {European Conference on Computer Vision (ECCV)}}

@string{ICCVW     = {IEEE International Conference on Computer Vision Workshops (ICCVW)}}

@string{NeurIPS   = {Advances in Neural Information Processing Systems (NeurIPS)}}

@string{ICRA      = {IEEE International Conference on Robotics and Automation (ICRA)}}

@string{SIGGRAPH  = {ACM SIGGRAPH Conference Papers}}

@string{DICTA     = {Digital Image Computing: Techniques and Applications (DICTA)}}

@string{ThreeDV   = {International Conference on 3D Vision (3DV)}}

@string{VCIP      = {IEEE International Conference on Visual Communications and Image Processing (VCIP)}}

@string{ICME      = {IEEE International Conference on Multimedia and Expo (ICME)}}

@string{TOG       = {ACM Transactions on Graphics}}

@string{TPAMI     = {IEEE Transactions on Pattern Analysis and Machine Intelligence}}

@string{TIP       = {IEEE Transactions on Image Processing}}

@string{CACM      = {Communications of the ACM}}

@string{IVC       = {Image and Vision Computing}}

@string{JMLR      = {Journal of Machine Learning Research}}

@string{PACMCGIT  = {Proceedings of the ACM on Computer Graphics and Interactive Techniques}}

@article{kerbl2023gaussian,
  title   = {{3D} {Gaussian} Splatting for Real-Time Radiance Field Rendering},
  author  = {Kerbl, Bernhard and Kopanas, Georgios and Leimk{\"u}hler, Thomas and Drettakis, George},
  journal = TOG,
  volume  = {42},
  number  = {4},
  pages   = {1--14},
  year    = {2023}
}

@inproceedings{fridovich2022plenoxels,
  title     = {Plenoxels: Radiance Fields without Neural Networks},
  author    = {Fridovich-Keil, Sara and Yu, Alex and Tancik, Matthew and Chen, Qinhong and Recht, Benjamin and Kanazawa, Angjoo},
  booktitle = CVPR,
  pages     = {5501--5510},
  year      = {2022}
}

@article{muller2022instant,
  title   = {Instant Neural Graphics Primitives with a Multiresolution Hash Encoding},
  author  = {M{\"u}ller, Thomas and Evans, Alex and Schied, Christoph and Keller, Alexander},
  journal = TOG,
  volume  = {41},
  number  = {4},
  pages   = {102:1--102:15},
  year    = {2022}
}

@inproceedings{mildenhall2020nerf,
  title     = {{NeRF}: Representing Scenes as Neural Radiance Fields for View Synthesis},
  author    = {Mildenhall, Ben and Srinivasan, Pratul P. and Tancik, Matthew and Barron, Jonathan T. and Ramamoorthi, Ravi and Ng, Ren},
  booktitle = ECCV,
  year      = {2020}
}

@inproceedings{schonberger2016colmap,
  title     = {Structure-from-Motion Revisited},
  author    = {Sch{\"o}nberger, Johannes L. and Frahm, Jan-Michael},
  booktitle = CVPR,
  pages     = {4104--4113},
  year      = {2016}
}

@inproceedings{sarlin2019hloc,
  title     = {From Coarse to Fine: Robust Hierarchical Localization at Large Scale},
  author    = {Sarlin, Paul-Edouard and Cadena, Cesar and Siegwart, Roland and Dymczyk, Marcin},
  booktitle = CVPR,
  year      = {2019}
}

@article{wang2004ssim,
  title   = {Image Quality Assessment: From Error Visibility to Structural Similarity},
  author  = {Wang, Zhou and Bovik, Alan C. and Sheikh, Hamid R. and Simoncelli, Eero P.},
  journal = TIP,
  volume  = {13},
  number  = {4},
  pages   = {600--612},
  year    = {2004}
}

@inproceedings{zhang2018lpips,
  title     = {The Unreasonable Effectiveness of Deep Features as a Perceptual Metric},
  author    = {Zhang, Richard and Isola, Phillip and Efros, Alexei A. and Shechtman, Eli and Wang, Oliver},
  booktitle = CVPR,
  year      = {2018}
}

@article{nayar1994focus,
  title   = {Shape from Focus},
  author  = {Nayar, Shree K. and Nakagawa, Yasuo},
  journal = TPAMI,
  volume  = {16},
  number  = {8},
  pages   = {824--831},
  year    = {1994}
}

@article{fischler1981ransac,
  title   = {Random Sample Consensus: A Paradigm for Model Fitting with Applications to Image Analysis and Automated Cartography},
  author  = {Fischler, Martin A. and Bolles, Robert C.},
  journal = CACM,
  volume  = {24},
  number  = {6},
  pages   = {381--395},
  year    = {1981}
}

@inproceedings{rusu2009fpfh,
  title     = {Fast Point Feature Histograms ({FPFH}) for {3D} Registration},
  author    = {Rusu, Radu Bogdan and Blodow, Nico and Beetz, Michael},
  booktitle = ICRA,
  pages     = {3212--3217},
  year      = {2009}
}

@article{chen1992pointtoplane,
  title   = {Object Modelling by Registration of Multiple Range Images},
  author  = {Chen, Yang and Medioni, G{\'e}rard},
  journal = IVC,
  volume  = {10},
  number  = {3},
  pages   = {145--155},
  year    = {1992}
}

@inproceedings{cicconet2017mirror,
  title     = {Finding Mirror Symmetry via Registration and Optimal Symmetric Pairwise Assignment of Curves},
  author    = {Cicconet, Marcelo and Hildebrand, David G. C. and Elliott, Hunter},
  booktitle = ICCVW,
  pages     = {1749--1758},
  year      = {2017}
}

@article{ecc2008,
  title   = {Parametric Image Alignment Using Enhanced Correlation Coefficient Maximization},
  author  = {Evangelidis, Georgios D. and Psarakis, Emmanouil Z.},
  journal = TPAMI,
  volume  = {30},
  number  = {10},
  pages   = {1858--1865},
  year    = {2008}
}

@article{burt1983pyramid,
  title   = {A Multiresolution Spline with Application to Image Mosaics},
  author  = {Burt, Peter J. and Adelson, Edward H.},
  journal = TOG,
  volume  = {2},
  number  = {4},
  pages   = {217--236},
  year    = {1983}
}

@misc{heliconfocus,
  title        = {Helicon Focus},
  author       = {{Helicon Soft Ltd.}},
  howpublished = {\url{https://www.heliconsoft.com/heliconsoft-products/helicon-focus/}},
  year         = {2026},
  note         = {Version 8}
}

@inproceedings{ma2022deblur,
  title     = {{Deblur-NeRF}: Neural Radiance Fields from Blurry Images},
  author    = {Ma, Li and Li, Xiaoyu and Liao, Jing and Zhang, Qi and Wang, Xuan and Wang, Jue and Sander, Pedro V.},
  booktitle = CVPR,
  pages     = {12861--12870},
  year      = {2022}
}

@inproceedings{liu2024mirrorgaussian,
  title     = {{MirrorGaussian}: Reflecting {3D} {Gaussians} for Reconstructing Mirror Reflections},
  author    = {Liu, Jiayue and Tang, Xiao and Cheng, Freeman and Yang, Roy and Li, Zhihao and Liu, Jianzhuang and Huang, Yi and Lin, Jiaqi and Liu, Shiyong and Wu, Xiaofei and Xu, Songcen and Yuan, Chun},
  booktitle = ECCV,
  pages     = {377--393},
  year      = {2024}
}

@inproceedings{meng2024mirror,
  title     = {{Mirror-3DGS}: Incorporating Mirror Reflections into {3D} {Gaussian} Splatting},
  author    = {Meng, Jiarui and Li, Haijie and Wu, Yanmin and Gao, Qiankun and Yang, Shuzhou and Zhang, Jian and Ma, Siwei},
  booktitle = VCIP,
  year      = {2024}
}

@inproceedings{shen2025dof,
  title     = {{DoF-Gaussian}: Controllable Depth-of-Field for {3D} {Gaussian} Splatting},
  author    = {Shen, Liao and Liu, Tianqi and Sun, Huiqiang and Li, Jiaqi and Cao, Zhiguo and Li, Wei and Loy, Chen Change},
  booktitle = CVPR,
  pages     = {26462--26471},
  year      = {2025}
}

@inproceedings{wang2024dofgs,
  title     = {{DOF-GS}: Adjustable Depth-of-Field {3D} {Gaussian} Splatting for Post-Capture Refocusing, Defocus Rendering and Blur Removal},
  author    = {Wang, Yujie and Chakravarthula, Praneeth and Chen, Baoquan},
  booktitle = CVPR,
  year      = {2025}
}

@article{nelson2019dissco,
  title     = {{DiSSCo}, {iDigBio} and the future of global collaboration},
  author    = {Nelson, Gil and Paul, Deborah L},
  journal   = {Biodiversity Information Science and Standards},
  year      = {2019},
  publisher = {Pensoft Publishers}
}

@article{smith2019synthesys+,
  title     = {{SYNTHESYS+} Abridged Grant Proposal},
  author    = {Smith, Vincent Stuart and Gorman, Kristina and Addink, Wouter and Arvanitidis, Christos and Casino, Ana and Dixey, Katherine and Dr{\"o}ge, Gabriele and Groom, Quentin and Haston, Elspeth Margaret and Hobern, Donald and others},
  journal   = {Research Ideas and Outcomes},
  volume    = {5},
  pages     = {e46404},
  year      = {2019},
  publisher = {Pensoft Publishers}
}

@article{Nguyen2014NaturalColour3D,
  title     = {Capturing natural-colour {3D} models of insects for species discovery and diagnostics},
  author    = {Nguyen, Chuong V and Lovell, David R and Adcock, Matt and La Salle, John},
  journal   = {{PLoS} {ONE}},
  volume    = {9},
  number    = {4},
  pages     = {e94346},
  year      = {2014}
}

@article{Strobel2018DISC3D,
  title   = {An automated device for the digitization and {3D} modelling of insects, combining extended-depth-of-field and all-side multi-view imaging},
  author  = {Str{\"o}bel, Bernhard and Schmelzle, Sebastian and Bl{\"u}thgen, Nico and Heethoff, Michael},
  journal = {ZooKeys},
  number  = {759},
  pages   = {1},
  year    = {2018}
}

@article{Plum2021scAnt,
  title     = {{scAnt}---an open-source platform for the creation of {3D} models of arthropods (and other small objects)},
  author    = {Plum, Fabian and Labonte, David},
  journal   = {PeerJ},
  volume    = {9},
  pages     = {e11155},
  year      = {2021}
}

@inproceedings{Li2019PerspectiveConsistent,
  title     = {Perspective-Consistent Multifocus Multiview {3D} Reconstruction of Small Objects},
  author    = {Li, Hengjia and Nguyen, Chuong V.},
  booktitle = DICTA,
  pages     = {1--8},
  year      = {2019}
}

@inproceedings{Xu2021InsectReconstruction,
  title     = {{3D} Reconstruction of Insects: An Improved Multifocus Stacking and an Evaluation of Learning-Based {MVS} Approaches},
  author    = {Xu, Chang and Liu, Jiayuan and Nguyen, Chuong and Casten, Fabien and Maujean, Benoit and Gasparini, Simone},
  booktitle = ThreeDV,
  pages     = {1411--1419},
  year      = {2021}
}

@inproceedings{Amrani2025DNASC,
  title     = {Dynamic Adaptive Sampling for Accurate Image-Based {3D} Insect Reconstruction Using Neural Implicit Surfaces},
  author    = {Amrani, Abderraouf and Laga, Hamid and Framenau, Volker and Thomas, Melissa L.},
  booktitle = DICTA,
  pages     = {1--8},
  year      = {2025}
}

@article{ye2025gsplat,
  title   = {gsplat: An Open-Source Library for {Gaussian} Splatting},
  author  = {Ye, Vickie and Li, Ruilong and Kerr, Justin and Turkulainen, Matias and Yi, Brent and Pan, Zhuoyang and Seiskari, Otto and Ye, Jianbo and Hu, Jeffrey and Tancik, Matthew and Kanazawa, Angjoo},
  journal = JMLR,
  volume  = {26},
  number  = {34},
  year    = {2025}
}

@inproceedings{zhang2021ners,
  title     = {{NeRS}: Neural Reflectance Surfaces for Sparse-View {3D} Reconstruction in the Wild},
  author    = {Zhang, Jason Y. and Yang, Gengshan and Tulsiani, Shubham and Ramanan, Deva},
  booktitle = NeurIPS,
  year      = {2021}
}

@inproceedings{sarlin2020superglue,
  title     = {{SuperGlue}: Learning Feature Matching with Graph Neural Networks},
  author    = {Sarlin, Paul-Edouard and DeTone, Daniel and Malisiewicz, Tomasz and Rabinovich, Andrew},
  booktitle = CVPR,
  pages     = {4938--4947},
  year      = {2020}
}

@inproceedings{mei2022dmr,
  title     = {Unsupervised Severely Deformed Mesh Reconstruction ({DMR}) from a Single-View Image for Longline Fishing},
  author    = {Mei, Jie and Yu, Jingxi and Romain, Suzanne and Rose, Craig S. and Magrane, Kelsey and LeeSon, Graeme and Hwang, Jenq-Neng},
  booktitle = ICME,
  year      = {2022}
}

@article{andersson2020flip,
  title   = {{FLIP}: A Difference Evaluator for Alternating Images},
  author  = {Andersson, Pontus and Nilsson, Jim and Akenine-M{\"o}ller, Tomas and Oskarsson, Magnus and {\AA}str{\"o}m, Kalle and Fairchild, Mark D.},
  journal = PACMCGIT,
  volume  = {3},
  number  = {2},
  pages   = {15:1--15:23},
  year    = {2020}
}

@inproceedings{huang2024_2dgs,
  title     = {{2D} {Gaussian} Splatting for Geometrically Accurate Radiance Fields},
  author    = {Huang, Binbin and Yu, Zehao and Chen, Anpei and Geiger, Andreas and Gao, Shenghua},
  booktitle = SIGGRAPH,
  year      = {2024}
}
}

\clearpage
\appendix

\section{Dataset specimens}
\label{sec:supp_specimens}

\Cref{tab:specimens} lists the nine mounted specimens and describes one diagnostic visual feature of each; \cref{fig:supp_specimens} shows one captured image of every specimen.

\begin{table*}[htbp]
  \centering
  \caption{\textbf{Dataset specimens.} The nine mounted specimens. We describe one diagnostic visual feature each. Two are not butterflies (a tiger moth and an owlfly), kept for pattern and material diversity.}
  \label{tab:specimens}
  \small
  \setlength{\tabcolsep}{4pt}
  \begin{tabular}{@{}lll@{}}
    \toprule
    Common name & Species & Distinctive feature \\
    \midrule
    Old World Swallowtail & \emph{Papilio machaon} & Hindwing tails; red anal eyespots \\
    Common Blue (female) & \emph{Polyommatus icarus} & Brown dorsal; orange ventral lunules \\
    Purple Emperor (male) & \emph{Apatura iris} & Angle-dependent purple iridescence \\
    Common Blue (male) & \emph{Polyommatus icarus} & Uniform violet-blue dorsal \\
    Large White & \emph{Pieris brassicae} & White wings; large black apex patch \\
    Painted Lady & \emph{Vanessa cardui} & Orange dorsal; marbled ventral web \\
    Peak White & \emph{Pontia callidice} & Green-grey chequered ventral \\
    Wood Tiger (moth) & \emph{Arctia plantaginis} & Black-on-yellow banded wings \\
    Owlfly (lacewing rel.) & \emph{Libelloides coccajus} & Clubbed antennae; net-veined wings \\
    \bottomrule
  \end{tabular}
\end{table*}

\ifdefined\dsw\else\newlength{\dsw}\fi\setlength{\dsw}{0.47\linewidth}
\begin{figure}[htbp]
  \centering
  \includegraphics[width=\dsw]{media/dataset/butterfly1.jpg}\hfill\includegraphics[width=\dsw]{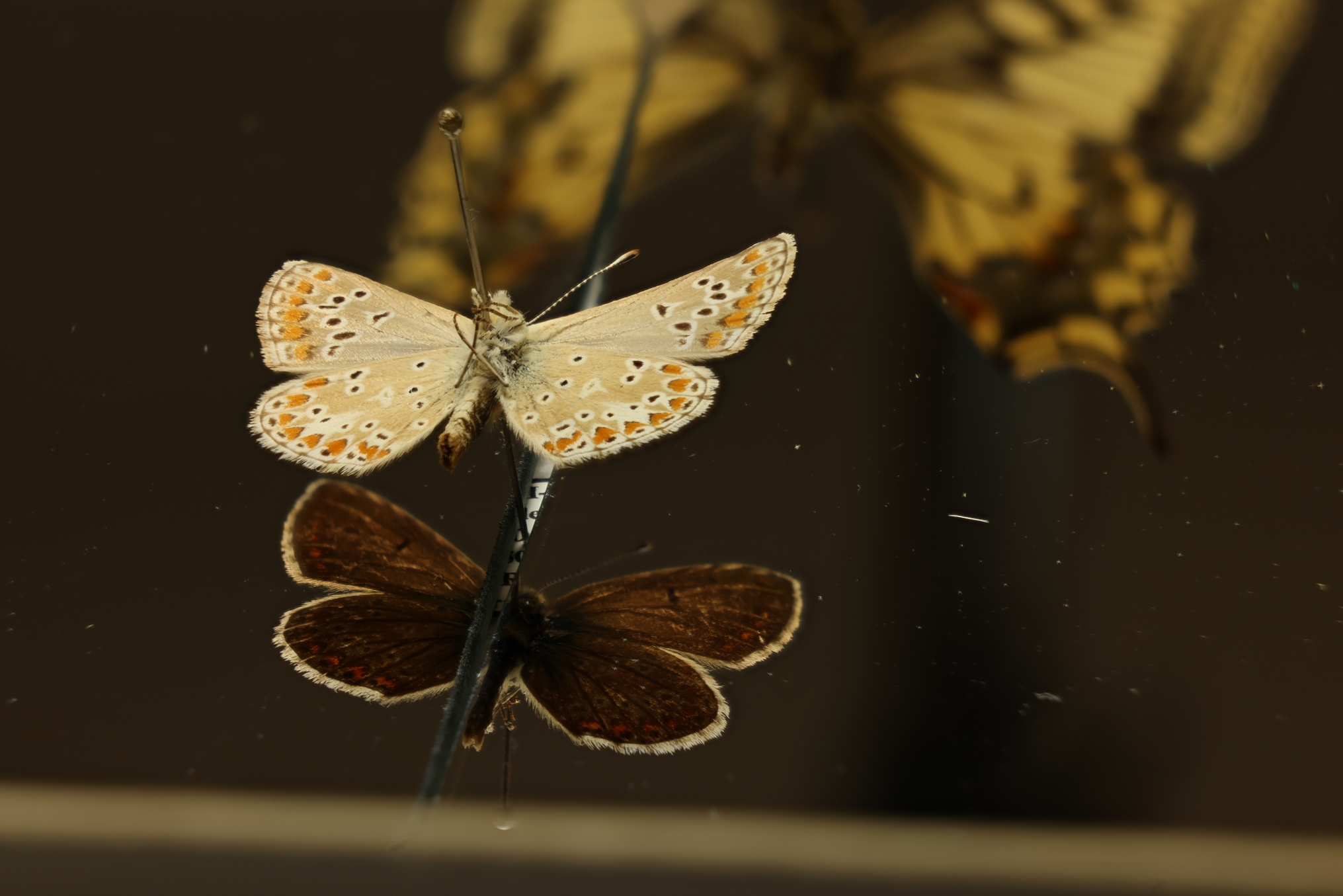}
  \includegraphics[width=\dsw]{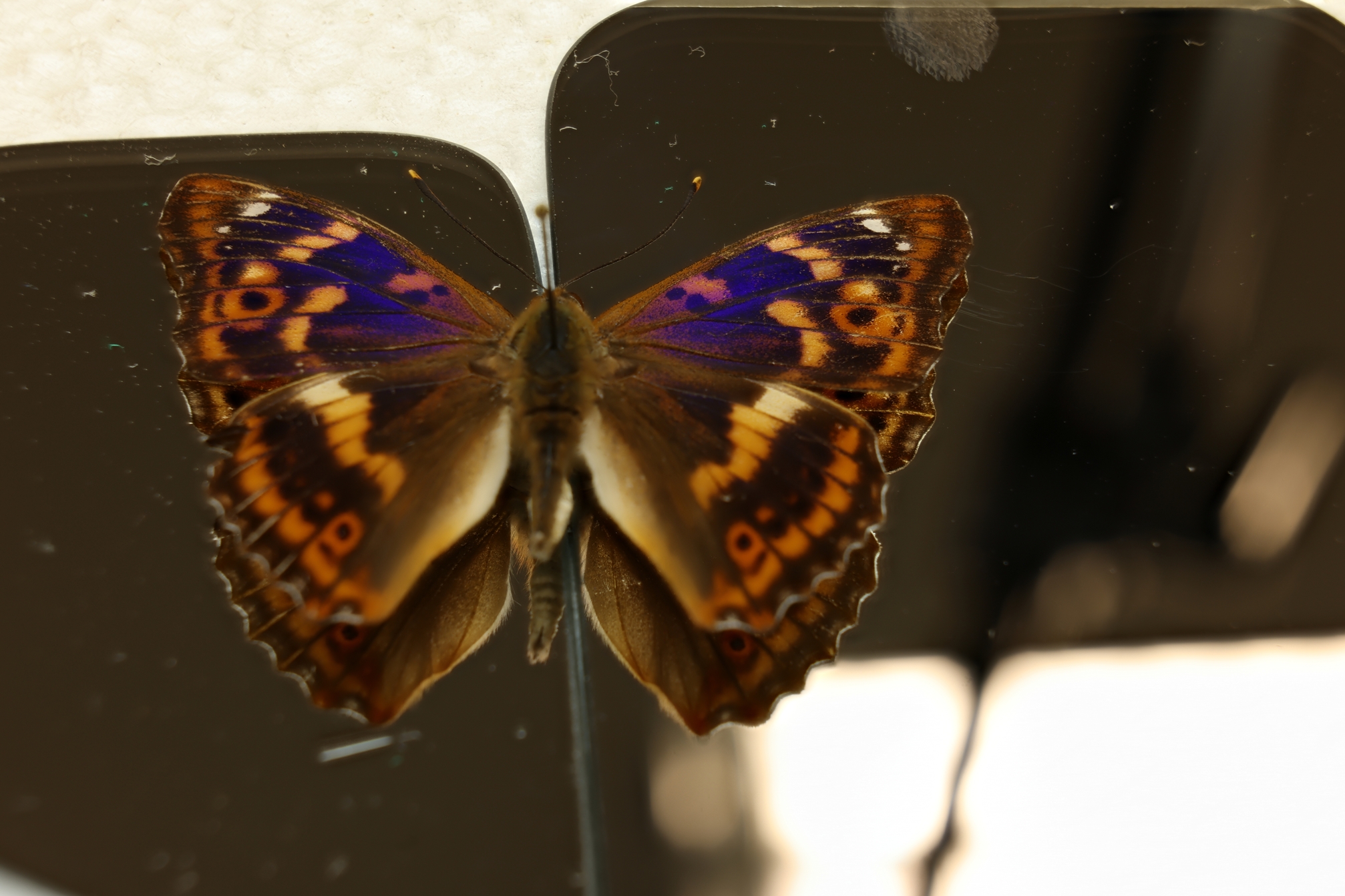}\hfill\includegraphics[width=\dsw]{media/dataset/butterfly4.jpg}
  \includegraphics[width=\dsw]{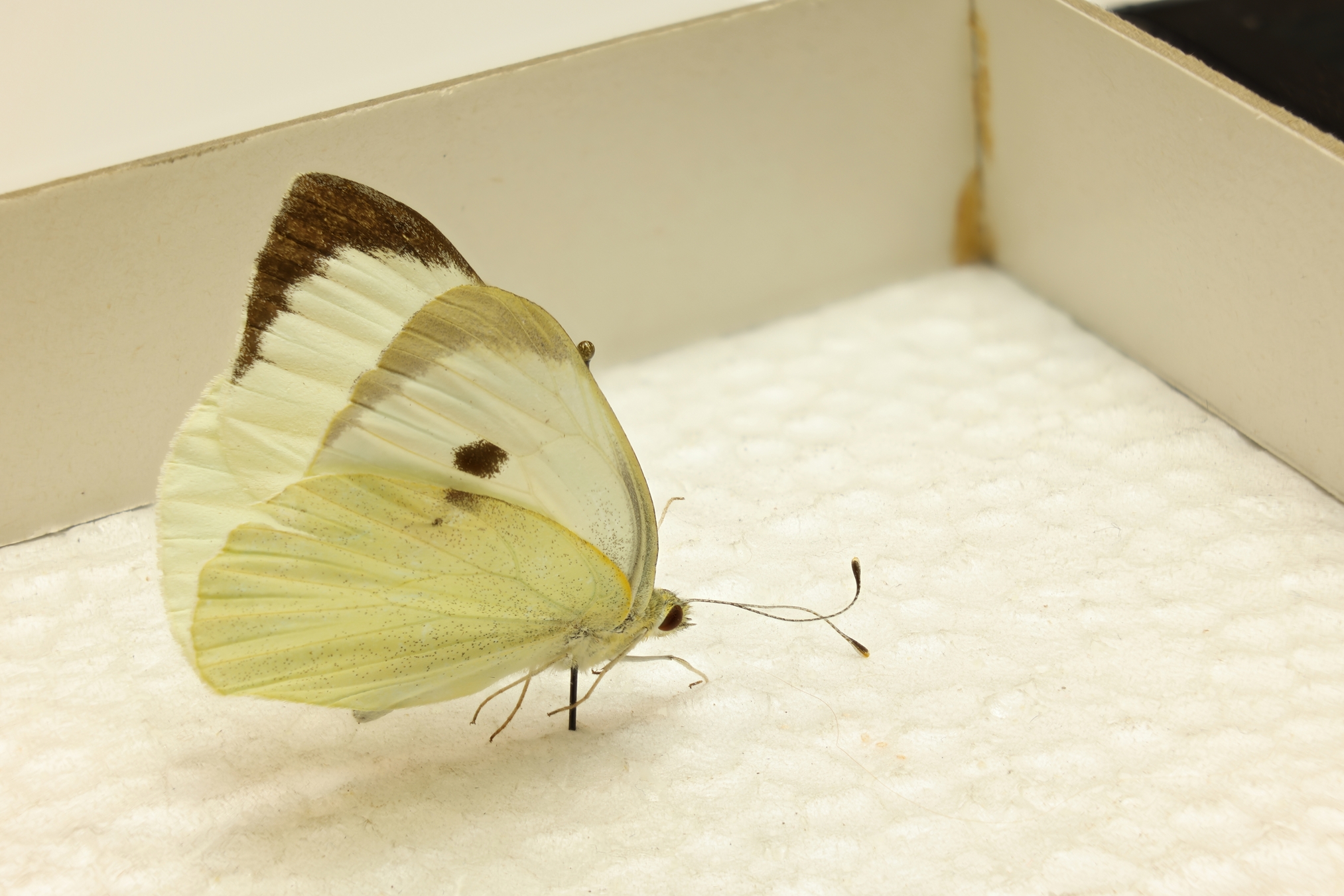}\hfill\includegraphics[width=\dsw]{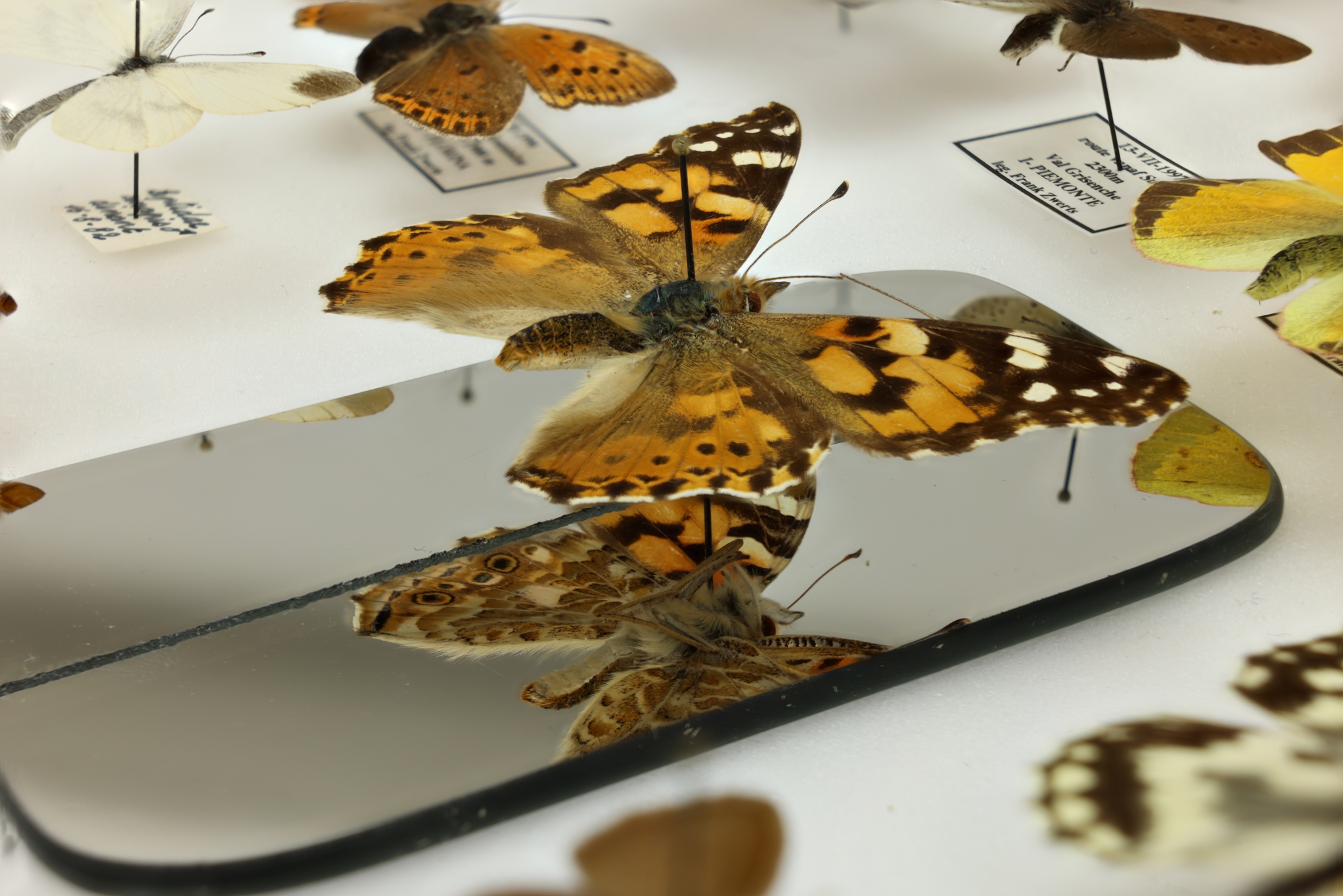}
  \includegraphics[width=\dsw]{media/dataset/butterfly8_and_9.jpg}\hfill\includegraphics[width=\dsw]{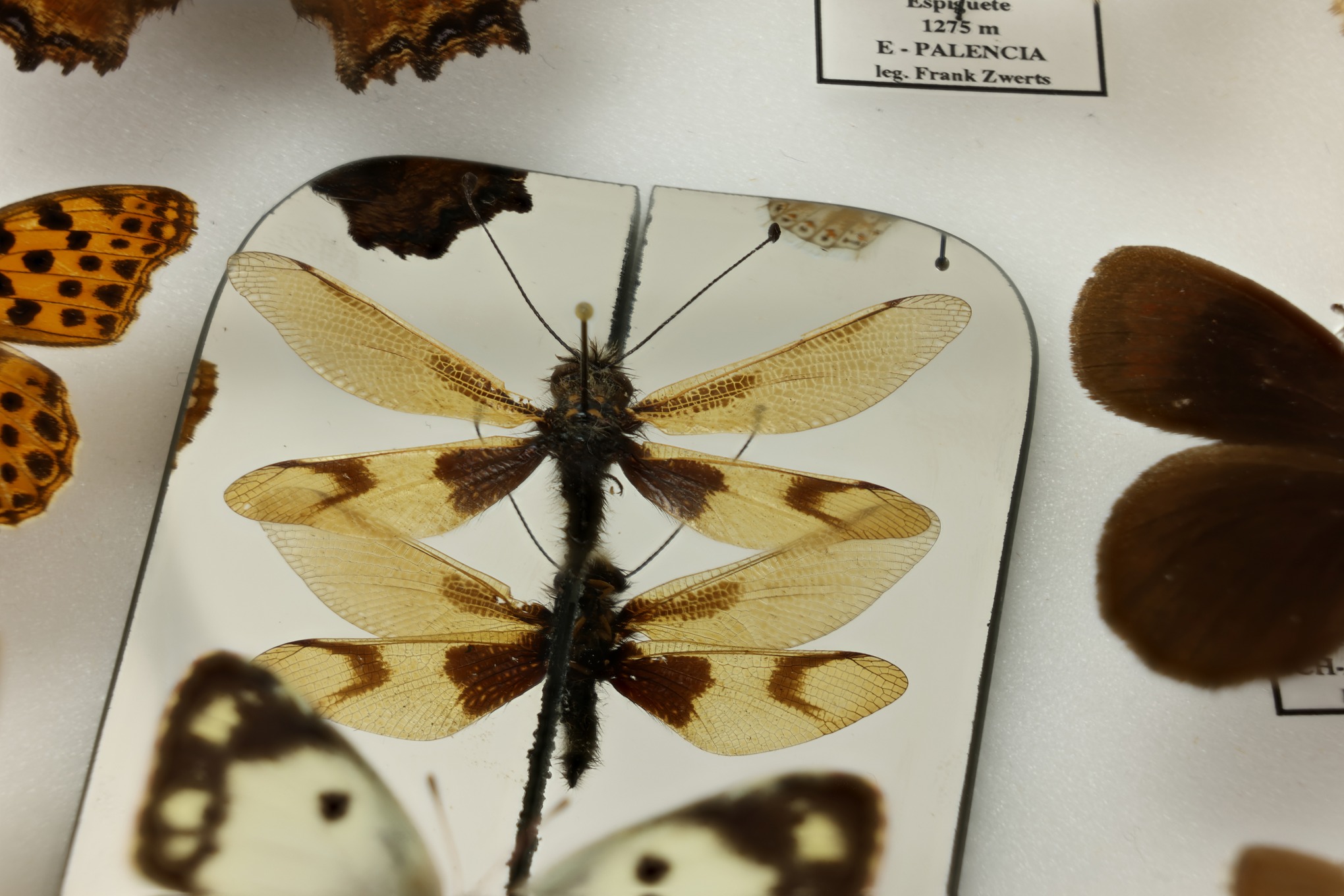}
  \caption{ \textbf{Dataset specimens.} One captured image per specimen, in reading order: Specimens 1--6 (rows 1--3), Specimens 7 and 8 (bottom left), and Specimen 9 (bottom right). Specimens 7 and 8 share a single capture and reconstruction and are evaluated with separate 2D masks. Specimen 5 (row 3, left) is mounted with its wings folded upright, so both wing surfaces are visible directly, and no mirrors are used.}
  \label{fig:supp_specimens}
\end{figure}

\section{Workflow cost}
\label{sec:supp_cost}

Because our target application is practical, museum-scale digitization, \cref{tab:cost} breaks down the end-to-end cost of producing a single specimen's reconstruction on consumer hardware, from photography through focus stacking to training a model. End-to-end, a specimen requires roughly half an hour of hands-on photography and about $1.5$ hours of largely unattended compute---camera calibration, mirror-plane detection, and 3DGS training---on a single RTX~4090.

\begin{table}[htbp]
  \centering
  \caption{ \textbf{Workflow cost.} End-to-end cost of digitizing one specimen, from capture to a trained model, on commodity hardware. Times are per specimen unless noted otherwise; the Gaussian count is for the final chosen configuration. All runs were conducted on a single NVIDIA RTX~4090 (24\,GB). }
  \label{tab:cost}
  \small
  \begin{tabular}{@{}ll@{}}
    \toprule
    Stage / quantity & Value \\
    \midrule
    Capture (photography) & avg. 28 min for 68 views \\
    Image resolution & $8192{\times}5464$\,px \\
    Image alignment & 0.9\,s per image \\
    Focus-stack fusion & 2.9\,s per stack (20--40 images) \\
    Camera calibration (pose + stereo) & 27\,min per scene \\
    Plane identification & avg.\ 4\,min\,56\,s per scene \\
    3DGS training per model               & $\sim$60\,min \\
    Final Gaussian count                  & $\sim$0.38\,M \\
    \bottomrule
  \end{tabular}
\end{table}

\section{Mirror-plane accuracy: per-specimen results}
\label{sec:supp_plane}

\cref{tab:plane_acc} reports the per-specimen mirror-plane accuracy summarized in the main paper. For each specimen, the ground-truth plane is obtained by robustly fitting a plane to points manually selected along the visible edges of the mirror in the dense point cloud (total-least-squares with iterative median-absolute-deviation outlier rejection). $\Delta\theta = \arccos\lvert \mathbf{n}_\text{gt}^{\top} \mathbf{n}_\text{auto}\rvert$ is the angle between the ground-truth and detected plane normals, $\Delta d$ is the offset between the two planes, and the point--plane RMS is the root-mean-square distance of the manually selected mirror points to the detected plane. Offsets and distances are expressed as a percentage of each specimen's spatial extent.

\begin{table}[h]
  \centering
  \caption{\textbf{Per-specimen mirror-plane accuracy.} Ground-truth planes are fit to manually selected mirror-edge points. Offsets are relative to each specimen's spatial extent. Specimens 7 and 8 share a single capture and hence a single mirror plane.}
  \label{tab:plane_acc}
  \small
  \begin{tabular}{@{}lcccc@{}}
    \toprule
    Specimen & \#pts & $\Delta\theta$ (deg) & $\Delta d$ (\% ext.) & RMS (\% ext.) \\
    \midrule
    Specimen 1 & 13 & 1.05 & 0.56 & 0.77 \\
    Specimen 2 &  9 & 0.65 & 0.27 & 0.56 \\
    Specimen 3 & 13 & 1.11 & 0.57 & 1.32 \\
    Specimen 4 &  9 & 0.23 & 0.63 & 0.76 \\
    Specimen 6 & 12 & 0.21 & 0.06 & 0.16 \\
    Specimen 7 &  9 & 0.43 & 0.05 & 0.31 \\
    Specimen 8 &  9 & 0.43 & 0.05 & 0.31 \\
    Specimen 9 & 11 & 0.97 & 1.04 & 0.41 \\
    \midrule
    Mean       &    & 0.64 & 0.40 & 0.57 \\
    \bottomrule
  \end{tabular}
\end{table}

\section{Mirror-plane detection: implementation details}
\label{sec:supp_planedet}

Both point clouds are voxel-downsampled with spacing $\delta = 0.01$ and described by Fast Point Feature Histograms (FPFH), 33-dimensional local geometric descriptors computed at a radius of $5\delta$. Feature-matching RANSAC uses a distance threshold of $1.5\delta$, correspondence checkers for edge length (ratio $0.9$) and point-to-point distance, and up to $10^5$ iterations at $99.9\%$ confidence; the resulting coarse transform $\mathbf{T}_\text{ransac} \in SE(3)$ is refined with point-to-plane ICP at threshold $0.4\delta$ for up to $100$ iterations. Point-to-plane ICP minimizes the squared distance from each source point to the tangent plane of its target correspondence, which is more accurate than point-to-point ICP on curved surfaces such as wing membranes. The plane $\mathbf{n}^\top \mathbf{x} + d = 0$ follows in closed form from the composite reflection $\mathbf{S} = \mathbf{T}_\text{icp}\,\mathbf{M}_X$ with translation part $\mathbf{t}_s$:
\begin{equation}
  \mathbf{n} = \operatorname{eigvec}(\mathbf{S}_{3\times3},\;\lambda{=}{-1}), \qquad
  d = -\mathbf{n}^\top \tfrac{\mathbf{t}_s}{2}.
\end{equation}

\section{Ventral validation: protocol details}
\label{sec:supp_ventral}

\noindent\textbf{Bridging the two calibrations.} The training and ventral acquisitions live in two independent COLMAP coordinate frames. We bridge them with a manual similarity alignment---a coarse point-picking transform followed by an ICP refinement---that maps the ventral point cloud into the training frame; composed, they place each ventral camera in the coordinate system of the trained model.

\noindent\textbf{Refining the held-out poses.} The manual alignment is only approximate, so each rendered ventral view is offset by a few pixels from its photograph, a shift that pixel metrics penalize heavily at the silhouette even when the reconstruction is faithful. Following NeRS~\cite{zhang2021ners}, we therefore refine each held-out camera individually, freezing the reconstruction and optimizing only a per-camera pose delta against its own target view; the scene is never fitted to the held-out views, only the pose moves. The question thus becomes not ``does the model render the exact held-out pose'' but ``does it render \emph{some} pose near it''. All reported ventral metrics use this refinement.

\noindent\textbf{Crops.} All ventral metrics are computed inside a square, $10\%$-padded box around the held-out-view mask, Lanczos-resized to $256{\times}256$; the ground truth is composited over white through the mask, and the render is left unmasked.

\section{Design choice: 3DGS vs.\ 2DGS}
\label{sec:supp_3d2d}

We compare the two splatting primitives---3D Gaussians (3DGS) and 2D surfels (2DGS)---on focus-stacked captures without mirror handling, as in the macro-defocus comparison in the main paper, while holding all other components fixed. For each specimen, we run leave-one-out evaluation over the masked held-out views ($46$ across nine specimens, images downscaled by a factor of four) and report both full-frame metrics and \emph{masked} metrics computed only within the specimen segmentation (the frame is composited to white outside the mask). \cref{tab:design_3d2d} gives per-specimen means, together with the mean number of Gaussian primitives as a proxy for model compactness.

In the specimen region, the two variants are very close, with 2DGS marginally ahead: mean masked PSNR is $30.30$\,dB for 2DGS against $29.71$\,dB for 3DGS, and masked SSIM and LPIPS also slightly favor 2DGS. 2DGS additionally holds a $0.48$\,dB full-frame PSNR advantage. The two primitives use essentially the same number of Gaussians on average ($1.33$M for 2DGS vs.\ $1.37$M for 3DGS including the background). 

\begin{table*}[htbp]
  \centering
  \caption{\textbf{Per-specimen 3DGS vs.\ 2DGS.} Leave-one-out means over the masked held-out views. Masked metrics are computed only within the specimen segmentation; full-frame metrics cover the whole image. \#GS is the mean number of Gaussian primitives (millions). Best of the two per metric in \textbf{bold}.}
  \label{tab:design_3d2d}
  \small
  \begin{tabular}{@{}llccccccc@{}}
    \toprule
    & & \multicolumn{3}{c}{Masked (specimen)} & \multicolumn{3}{c}{Full frame} & \\
    \cmidrule(lr){3-5}\cmidrule(lr){6-8}
    Specimen & Method & PSNR\,$\uparrow$ & SSIM\,$\uparrow$ & LPIPS\,$\downarrow$ & PSNR\,$\uparrow$ & SSIM\,$\uparrow$ & LPIPS\,$\downarrow$ & \#GS\,(M)\,$\downarrow$ \\
    \midrule
    Specimen 1  & 3DGS & 28.98 & 0.933 & 0.0444 & 18.84 & 0.810 & 0.295 & 1.28 \\
                & 2DGS & \textbf{29.24} & \textbf{0.934} & \textbf{0.0430} & \textbf{19.47} & \textbf{0.812} & \textbf{0.289} & \textbf{1.26} \\
    \midrule
    Specimen 2  & 3DGS & 24.97 & 0.952 & 0.0418 & 16.56 & 0.782 & 0.378 & 0.75 \\
                & 2DGS & \textbf{28.05} & \textbf{0.961} & \textbf{0.0319} & \textbf{17.40} & \textbf{0.791} & \textbf{0.368} & \textbf{0.72} \\
    \midrule
    Specimen 3  & 3DGS & 25.77 & 0.912 & 0.0800 & 13.96 & 0.733 & 0.419 & 1.02 \\
                & 2DGS & \textbf{28.74} & \textbf{0.923} & \textbf{0.0678} & \textbf{15.33} & \textbf{0.748} & \textbf{0.398} & \textbf{0.95} \\
    \midrule
    Specimen 4  & 3DGS & \textbf{26.34} & \textbf{0.909} & 0.0773 & 14.67 & 0.700 & 0.420 & \textbf{0.80} \\
                & 2DGS & 25.87 & 0.907 & \textbf{0.0737} & \textbf{15.77} & \textbf{0.729} & \textbf{0.402} & 0.99 \\
    \midrule
    Specimen 5  & 3DGS & 33.32 & 0.976 & 0.0323 & 20.49 & 0.881 & 0.253 & 1.23 \\
                & 2DGS & \textbf{34.53} & \textbf{0.977} & \textbf{0.0300} & \textbf{21.94} & \textbf{0.884} & \textbf{0.235} & \textbf{1.20} \\
    \midrule
    Specimen 6  & 3DGS & \textbf{31.88} & \textbf{0.944} & \textbf{0.0432} & \textbf{22.27} & \textbf{0.833} & \textbf{0.256} & 1.69 \\
                & 2DGS & 30.04 & 0.940 & 0.0485 & 21.12 & 0.821 & 0.275 & \textbf{1.65} \\
    \midrule
    Specimen 7  & 3DGS & 32.72 & \textbf{0.979} & 0.0156 & 20.09 & \textbf{0.815} & 0.278 & 2.23 \\
                & 2DGS & \textbf{32.85} & 0.979 & \textbf{0.0154} & \textbf{20.35} & 0.813 & \textbf{0.277} & \textbf{2.10} \\
    \midrule
    Specimen 8  & 3DGS & 34.16 & \textbf{0.970} & 0.0239 & \textbf{22.41} & \textbf{0.850} & \textbf{0.204} & 2.22 \\
                & 2DGS & \textbf{34.24} & 0.970 & \textbf{0.0233} & 22.23 & 0.846 & 0.209 & \textbf{2.07} \\
    \midrule
    Specimen 9  & 3DGS & \textbf{29.21} & 0.950 & 0.0355 & 20.37 & \textbf{0.809} & 0.297 & 1.09 \\
                & 2DGS & 29.17 & \textbf{0.950} & \textbf{0.0335} & \textbf{20.37} & 0.809 & \textbf{0.292} & \textbf{1.07} \\
    \midrule
    \multirow{2}{*}{Overall ($46$)}
                & 3DGS & 29.71 & 0.947 & 0.0438 & 18.85 & 0.802 & 0.311 & 1.37 \\
                & 2DGS & \textbf{30.30} & \textbf{0.949} & \textbf{0.0408} & \textbf{19.33} & \textbf{0.806} & \textbf{0.305} & \textbf{1.33} \\
    \bottomrule
  \end{tabular}
\end{table*}

\section{Mirror-unaware baseline: effect of densification}
\label{sec:supp_nomirror}

The main paper reports that regular 2DGS, trained without any mirror handling, achieves higher performance on held-out dorsal views when adaptive density control (ADC) is disabled, and the dense multi-view-stereo initialization is kept fixed. \cref{tab:nomirror_full} gives the per-specimen numbers underlying that comparison for the eight mirror specimens (leave-one-out over the same five held-out views per specimen as in \cref{tab:reflect_full}, with images downscaled by a factor of four), including masked and full-frame metrics and the mean number of Gaussians. Disabling densification improves masked PSNR on seven of the eight specimens and every full-frame metric across all specimens, with a third fewer Gaussian per specimen on average. The ADC rows are the 2DGS rows of \cref{tab:design_3d2d} restricted to the mirror specimens.

\begin{table*}[htbp]
  \centering
  \caption{\textbf{Per-specimen mirror-unaware baseline with and without densification.} Regular 2DGS without mirror handling on the eight mirror specimens; leave-one-out means over the masked held-out views. \emph{ADC}: adaptive density control; \emph{none}: Gaussian count frozen at the dense initialization. Masked metrics are computed only within the specimen segmentation, with the frame composited to white outside the mask; full-frame metrics cover the whole image. \#GS is the mean number of Gaussian primitives (millions). Best per metric within each specimen in \textbf{bold}.}
  \label{tab:nomirror_full}
  \small
  \begin{tabular}{@{}llccccccc@{}}
    \toprule
    & & \multicolumn{3}{c}{Masked (specimen)} & \multicolumn{3}{c}{Full frame} & \\
    \cmidrule(lr){3-5}\cmidrule(lr){6-8}
    Specimen & Densify & PSNR\,$\uparrow$ & SSIM\,$\uparrow$ & LPIPS\,$\downarrow$ & PSNR\,$\uparrow$ & SSIM\,$\uparrow$ & LPIPS\,$\downarrow$ & \#GS\,(M)\,$\downarrow$ \\
    \midrule
      \multirow{2}{*}{Specimen 1}
       & ADC & \textbf{29.24} & \textbf{0.934} & \textbf{0.0430} & 19.47 & 0.812 & 0.289 & 1.26 \\
       & none & 28.62 & 0.928 & 0.0469 & \textbf{21.79} & \textbf{0.825} & \textbf{0.278} & \textbf{0.74} \\
    \midrule
      \multirow{2}{*}{Specimen 2}
       & ADC & 28.05 & 0.961 & 0.0319 & 17.40 & 0.791 & 0.368 & 0.72 \\
       & none & \textbf{30.84} & \textbf{0.966} & \textbf{0.0278} & \textbf{18.28} & \textbf{0.821} & \textbf{0.286} & \textbf{0.35} \\
    \midrule
      \multirow{2}{*}{Specimen 3}
       & ADC & 28.74 & 0.923 & 0.0678 & 15.33 & 0.748 & 0.398 & 0.95 \\
       & none & \textbf{29.52} & \textbf{0.927} & \textbf{0.0651} & \textbf{17.51} & \textbf{0.786} & \textbf{0.336} & \textbf{0.73} \\
    \midrule
      \multirow{2}{*}{Specimen 4}
       & ADC & 25.87 & 0.907 & 0.0737 & 15.77 & 0.729 & 0.402 & 0.99 \\
       & none & \textbf{27.88} & \textbf{0.912} & \textbf{0.0689} & \textbf{19.22} & \textbf{0.759} & \textbf{0.353} & \textbf{0.37} \\
    \midrule
      \multirow{2}{*}{Specimen 6}
       & ADC & 30.04 & 0.940 & 0.0485 & 21.12 & 0.821 & 0.275 & 1.65 \\
       & none & \textbf{31.98} & \textbf{0.942} & \textbf{0.0423} & \textbf{22.13} & \textbf{0.841} & \textbf{0.240} & \textbf{1.29} \\
    \midrule
      \multirow{2}{*}{Specimen 7}
       & ADC & 32.85 & 0.979 & 0.0154 & 20.35 & 0.813 & 0.277 & 2.10 \\
       & none & \textbf{34.91} & \textbf{0.982} & \textbf{0.0125} & \textbf{22.01} & \textbf{0.833} & \textbf{0.243} & \textbf{1.58} \\
    \midrule
      \multirow{2}{*}{Specimen 8}
       & ADC & 34.24 & \textbf{0.970} & \textbf{0.0233} & 22.23 & 0.846 & 0.209 & 2.07 \\
       & none & \textbf{34.41} & 0.968 & 0.0242 & \textbf{23.45} & \textbf{0.856} & \textbf{0.189} & \textbf{1.58} \\
    \midrule
      \multirow{2}{*}{Specimen 9}
       & ADC & 29.17 & \textbf{0.950} & 0.0335 & 20.37 & 0.809 & 0.292 & 1.07 \\
       & none & \textbf{31.08} & 0.950 & \textbf{0.0319} & \textbf{21.41} & \textbf{0.820} & \textbf{0.262} & \textbf{0.63} \\
    \midrule
      \multirow{2}{*}{Overall}
       & ADC & 29.77 & 0.945 & 0.0421 & 19.00 & 0.796 & 0.314 & 1.35 \\
       & none & \textbf{31.16} & \textbf{0.947} & \textbf{0.0400} & \textbf{20.73} & \textbf{0.818} & \textbf{0.274} & \textbf{0.91} \\
    \bottomrule
  \end{tabular}
\end{table*}

\section{Design choice: reflection, mirror plane, and densification}
\label{sec:supp_reflect}

We ablate three design choices of our mirror handling jointly: how the reflection is realized 

\begin{itemize}
    \item \emph{reflect} the camera versus \emph{reflect} the splats; 
    \item whether an explicit mirror-plane Gaussian field is modeled;
    \item how the specimen Gaussians are densified (ADC~\cite{kerbl2023gaussian}, or none, i.e.\ the Gaussian count is frozen at the dense multi-view-stereo initialization and only the existing primitives' attributes are optimized).
\end{itemize}

All variants are \emph{segmentation-free} adaptations of the underlying reflection principles and do not use the per-image mirror masks required by the prior methods~\cite{meng2024mirror,liu2024mirrorgaussian}. For each specimen we run leave-one-out evaluation over the masked held-out views (five per specimen; $40$ views over eight specimens, images downscaled by a factor of four) and report both \emph{masked} metrics (computed only within the specimen segmentation, with the frame composited to white outside the mask) and full-frame metrics, together with the mean number of Gaussian primitives (\#GS). \cref{tab:reflect_full} gives the per-specimen means summarized in the main paper, together with the variant without mirror Gaussians (reflect camera, no plane; no densification) and the final model with the two regularizers.

Reflecting the camera with an explicit mirror plane and no densification is the best configuration across the pooled averages of all metrics, masked and full-frame, and for most individual specimens, while using the fewest primitives overall ($0.38$M on average excluding the mirror Gaussians). Because the specimen is already densely covered at initialization, densification predominantly grows redundant Gaussians and background floaters that overfit the sparse training views. Reflecting the splats instead of the camera, both without an explicit plane, is on par once densification is disabled ($28.35$ vs.\ $28.25$\,dB masked PSNR) at twice the primitive count; with ADC it trails every other variant.

\begin{table*}[htbp]
  \centering
  \caption{\textbf{Per-specimen reflection strategy, mirror plane, and densification.} Leave-one-out means over the masked held-out views. \emph{Reflect cam.}\ renders a mirrored-camera view; \emph{reflect splats} mirrors the Gaussians; \emph{plane} denotes an explicit mirror-plane Gaussian field. Masked metrics are computed only within the specimen segmentation; full-frame metrics cover the whole image. \#GS is the mean number of Gaussian primitives (millions). All variants are segmentation-free adaptations, not the mask-based prior methods~\cite{meng2024mirror,liu2024mirrorgaussian}. The \emph{no plane} row drops the mirror Gaussians at the same densification-free setting; the last row is the final model with the alpha-entropy and distortion regularizers. Best per metric within each specimen in \textbf{bold}.}
  \label{tab:reflect_full}
  \small
  \begin{tabular}{@{}llccccccc@{}}
    \toprule
    & & \multicolumn{3}{c}{Masked (specimen)} & \multicolumn{3}{c}{Full frame} & \\
    \cmidrule(lr){3-5}\cmidrule(lr){6-8}
    Specimen & Variant & PSNR\,$\uparrow$ & SSIM\,$\uparrow$ & LPIPS\,$\downarrow$ & PSNR\,$\uparrow$ & SSIM\,$\uparrow$ & LPIPS\,$\downarrow$ & \#GS\,(M)\,$\downarrow$ \\
    \midrule
      \multirow{6}{*}{Specimen 1}
       & Reflect cam., plane; none & 29.26 & 0.936 & 0.0441 & 21.00 & 0.828 & 0.287 & 0.26 \\
       & Reflect cam., plane; ADC & 28.85 & 0.935 & \textbf{0.0416} & 20.24 & 0.814 & 0.308 & 0.73 \\
       & Reflect cam., no plane; none & 26.47 & 0.924 & 0.0590 & 19.34 & 0.799 & 0.351 & 0.29 \\
       & Reflect splats, no plane; ADC & 25.61 & 0.921 & 0.0602 & 18.69 & 0.782 & 0.370 & 1.03 \\
       & Reflect splats, no plane; none & 26.81 & 0.925 & 0.0572 & 20.46 & 0.811 & 0.324 & 0.46 \\
       & $+$ Regularization (full) & \textbf{29.65} & \textbf{0.937} & 0.0464 & \textbf{21.54} & \textbf{0.831} & \textbf{0.286} & \textbf{0.23} \\
    \midrule
      \multirow{6}{*}{Specimen 2}
       & Reflect cam., plane; none & 31.75 & \textbf{0.968} & \textbf{0.0242} & \textbf{18.84} & \textbf{0.825} & \textbf{0.305} & \textbf{0.15} \\
       & Reflect cam., plane; ADC & 23.17 & 0.944 & 0.0540 & 14.72 & 0.729 & 0.485 & 0.69 \\
       & Reflect cam., no plane; none & 28.73 & 0.959 & 0.0367 & 17.35 & 0.782 & 0.374 & 0.19 \\
       & Reflect splats, no plane; ADC & 21.01 & 0.929 & 0.0793 & 14.48 & 0.713 & 0.514 & 0.77 \\
       & Reflect splats, no plane; none & 29.21 & 0.961 & 0.0374 & 17.65 & 0.802 & 0.352 & 0.31 \\
       & $+$ Regularization (full) & \textbf{32.72} & 0.968 & 0.0248 & 18.73 & 0.821 & 0.311 & 0.16 \\
    \midrule
      \multirow{6}{*}{Specimen 3}
       & Reflect cam., plane; none & 29.79 & \textbf{0.933} & \textbf{0.0604} & \textbf{17.04} & \textbf{0.774} & \textbf{0.355} & \textbf{0.33} \\
       & Reflect cam., plane; ADC & 28.50 & 0.928 & 0.0652 & 15.44 & 0.739 & 0.428 & 0.61 \\
       & Reflect cam., no plane; none & 28.22 & 0.924 & 0.0744 & 15.55 & 0.765 & 0.403 & 0.37 \\
       & Reflect splats, no plane; ADC & 26.15 & 0.911 & 0.0856 & 14.55 & 0.719 & 0.465 & 0.93 \\
       & Reflect splats, no plane; none & 27.98 & 0.922 & 0.0738 & 15.93 & 0.760 & 0.396 & 1.00 \\
       & $+$ Regularization (full) & \textbf{29.83} & 0.931 & 0.0620 & 15.89 & 0.738 & 0.369 & 0.35 \\
    \midrule
      \multirow{6}{*}{Specimen 4}
       & Reflect cam., plane; none & \textbf{29.73} & 0.920 & \textbf{0.0572} & 18.22 & \textbf{0.757} & 0.370 & 0.13 \\
       & Reflect cam., plane; ADC & 19.64 & 0.887 & 0.1360 & 11.94 & 0.630 & 0.580 & 0.52 \\
       & Reflect cam., no plane; none & 24.38 & 0.903 & 0.0891 & 14.81 & 0.703 & 0.446 & 0.16 \\
       & Reflect splats, no plane; ADC & 20.49 & 0.881 & 0.1185 & 14.01 & 0.629 & 0.549 & 1.18 \\
       & Reflect splats, no plane; none & 24.49 & 0.898 & 0.0929 & 16.80 & 0.724 & 0.434 & 0.25 \\
       & $+$ Regularization (full) & 29.73 & \textbf{0.920} & 0.0575 & \textbf{18.35} & 0.757 & \textbf{0.366} & \textbf{0.12} \\
    \bottomrule
  \end{tabular}
\end{table*}

\begin{table*}[htbp]
  \centering
  \ContinuedFloat
  \caption{\textbf{Per-specimen reflection strategy, mirror plane, and densification (continued).} Specimens 6--9 and the overall mean; columns and variants as in the first part.}
  \small
  \begin{tabular}{@{}llccccccc@{}}
    \toprule
    & & \multicolumn{3}{c}{Masked (specimen)} & \multicolumn{3}{c}{Full frame} & \\
    \cmidrule(lr){3-5}\cmidrule(lr){6-8}
    Specimen & Variant & PSNR\,$\uparrow$ & SSIM\,$\uparrow$ & LPIPS\,$\downarrow$ & PSNR\,$\uparrow$ & SSIM\,$\uparrow$ & LPIPS\,$\downarrow$ & \#GS\,(M)\,$\downarrow$ \\
    \midrule
      \multirow{6}{*}{Specimen 6}
       & Reflect cam., plane; none & \textbf{28.77} & \textbf{0.936} & \textbf{0.0552} & \textbf{19.69} & 0.813 & 0.314 & \textbf{0.60} \\
       & Reflect cam., plane; ADC & 26.21 & 0.925 & 0.0761 & 18.88 & 0.790 & 0.363 & 1.24 \\
       & Reflect cam., no plane; none & 26.29 & 0.925 & 0.0653 & 18.55 & 0.801 & 0.345 & 0.63 \\
       & Reflect splats, no plane; ADC & 25.85 & 0.923 & 0.0829 & 19.40 & 0.808 & 0.352 & 1.48 \\
       & Reflect splats, no plane; none & 26.96 & 0.933 & 0.0583 & 19.20 & \textbf{0.821} & \textbf{0.300} & 1.24 \\
       & $+$ Regularization (full) & 28.18 & 0.935 & 0.0613 & 18.99 & 0.802 & 0.337 & 0.61 \\
    \midrule
      \multirow{6}{*}{Specimen 7}
       & Reflect cam., plane; none & \textbf{34.75} & \textbf{0.982} & \textbf{0.0135} & \textbf{19.76} & \textbf{0.815} & \textbf{0.290} & \textbf{0.61} \\
       & Reflect cam., plane; ADC & 29.94 & 0.974 & 0.0218 & 16.21 & 0.750 & 0.386 & 1.65 \\
       & Reflect cam., no plane; none & 31.80 & 0.977 & 0.0189 & 17.72 & 0.789 & 0.351 & 0.66 \\
       & Reflect splats, no plane; ADC & 24.51 & 0.956 & 0.0466 & 13.67 & 0.695 & 0.535 & 1.46 \\
       & Reflect splats, no plane; none & 31.92 & 0.978 & 0.0174 & 18.48 & 0.804 & 0.319 & 1.39 \\
       & $+$ Regularization (full) & 34.69 & 0.982 & 0.0139 & 19.66 & 0.811 & 0.298 & 0.64 \\
    \midrule
      \multirow{6}{*}{Specimen 8}
       & Reflect cam., plane; none & \textbf{34.11} & 0.971 & 0.0258 & \textbf{21.34} & \textbf{0.837} & \textbf{0.242} & \textbf{0.61} \\
       & Reflect cam., plane; ADC & 34.07 & \textbf{0.972} & \textbf{0.0231} & 20.52 & 0.817 & 0.264 & 1.57 \\
       & Reflect cam., no plane; none & 32.59 & 0.968 & 0.0293 & 18.91 & 0.809 & 0.299 & 0.66 \\
       & Reflect splats, no plane; ADC & 26.81 & 0.955 & 0.0535 & 16.08 & 0.748 & 0.454 & 1.50 \\
       & Reflect splats, no plane; none & 32.20 & 0.968 & 0.0291 & 20.07 & 0.828 & 0.261 & 1.39 \\
       & $+$ Regularization (full) & 33.85 & 0.971 & 0.0277 & 21.06 & 0.836 & 0.249 & 0.64 \\
    \midrule
      \multirow{6}{*}{Specimen 9}
       & Reflect cam., plane; none & \textbf{30.84} & \textbf{0.952} & \textbf{0.0333} & \textbf{20.17} & \textbf{0.809} & \textbf{0.311} & 0.34 \\
       & Reflect cam., plane; ADC & 29.09 & 0.951 & 0.0347 & 19.64 & 0.798 & 0.325 & 0.83 \\
       & Reflect cam., no plane; none & 27.51 & 0.942 & 0.0447 & 16.79 & 0.774 & 0.404 & 0.41 \\
       & Reflect splats, no plane; ADC & 19.09 & 0.915 & 0.0983 & 11.21 & 0.635 & 0.665 & 1.04 \\
       & Reflect splats, no plane; none & 27.26 & 0.943 & 0.0439 & 17.45 & 0.784 & 0.358 & 0.68 \\
       & $+$ Regularization (full) & 29.74 & 0.949 & 0.0359 & 19.55 & 0.800 & 0.319 & \textbf{0.33} \\
    \midrule
      \multirow{6}{*}{Overall}
       & Reflect cam., plane; none & \textbf{31.12} & \textbf{0.950} & \textbf{0.0392} & \textbf{19.51} & \textbf{0.807} & \textbf{0.309} & \textbf{0.38} \\
       & Reflect cam., plane; ADC & 27.43 & 0.940 & 0.0566 & 17.20 & 0.758 & 0.392 & 0.98 \\
       & Reflect cam., no plane; none & 28.25 & 0.940 & 0.0522 & 17.38 & 0.778 & 0.372 & 0.42 \\
       & Reflect splats, no plane; ADC & 23.69 & 0.924 & 0.0781 & 15.26 & 0.716 & 0.488 & 1.17 \\
       & Reflect splats, no plane; none & 28.35 & 0.941 & 0.0513 & 18.25 & 0.792 & 0.343 & 0.84 \\
       & $+$ Regularization (full) & 31.05 & 0.949 & 0.0412 & 19.22 & 0.799 & 0.317 & 0.38 \\
    \bottomrule
  \end{tabular}
\end{table*}

\section{Macro-defocus handling: per-specimen results}
\label{sec:supp_dof}

\cref{tab:dof_full} reports the per-specimen breakdown of the macro-defocus comparison summarized in the main paper. We evaluate handheld focus stacking under both splatting primitives---3D Gaussians (3DGS) and 2D surfels (2DGS)---against DoF-Gaussian~\cite{shen2025dof} on the masked held-out views ($46$ across nine specimens, images downscaled by a factor of four). Every method is scored with an identical metric stack on the same held-out image, and we report both \emph{masked} metrics (computed only within the specimen segmentation, with the frame composited to white outside the mask) and full-frame metrics. On the specimen, both focus-stacking variants outperform DoF-Gaussian by a wide margin ($+3.6$--$4.2$\,dB mean masked PSNR and less than half the perceptual error). DoF-Gaussian struggles to recover detailed, high-frequency structures.

\begin{table*}[htbp]
  \centering
  \caption{\textbf{Per-specimen macro-defocus handling.} Handheld focus stacking under both primitives (3DGS, 2DGS) versus DoF-Gaussian~\cite{shen2025dof}, over the masked held-out views. Masked metrics are computed only within the specimen segmentation; full-frame metrics cover the whole image. Best PSNR per specimen in \textbf{bold}.}
  \label{tab:dof_full}
  \small
  \begin{tabular}{@{}llcccccc@{}}
    \toprule
    & & \multicolumn{3}{c}{Masked (specimen)} & \multicolumn{3}{c}{Full frame} \\
    \cmidrule(lr){3-5}\cmidrule(lr){6-8}
    Specimen & Method & PSNR\,$\uparrow$ & SSIM\,$\uparrow$ & LPIPS\,$\downarrow$ & PSNR\,$\uparrow$ & SSIM\,$\uparrow$ & LPIPS\,$\downarrow$ \\
    \midrule
    \multirow{3}{*}{Specimen 1}
      & 3DGS & 28.98 & 0.933 & 0.0444 & 18.84 & 0.810 & 0.295 \\
      & 2DGS & \textbf{29.24} & 0.934 & 0.0430 & \textbf{19.47} & 0.812 & 0.289 \\
      & DoF-Gaussian~\cite{shen2025dof} & 25.06 & 0.896 & 0.0867 & 17.33 & 0.768 & 0.347 \\
    \midrule
    \multirow{3}{*}{Specimen 2}
      & 3DGS & 24.97 & 0.952 & 0.0418 & 16.56 & 0.782 & 0.378 \\
      & 2DGS & \textbf{28.05} & 0.961 & 0.0319 & \textbf{17.40} & 0.791 & 0.368 \\
      & DoF-Gaussian~\cite{shen2025dof} & 27.70 & 0.949 & 0.0464 & 17.07 & 0.784 & 0.376 \\
    \midrule
    \multirow{3}{*}{Specimen 3}
      & 3DGS & 25.77 & 0.912 & 0.0800 & 13.96 & 0.733 & 0.419 \\
      & 2DGS & \textbf{28.74} & 0.923 & 0.0678 & \textbf{15.33} & 0.748 & 0.398 \\
      & DoF-Gaussian~\cite{shen2025dof} & 19.04 & 0.851 & 0.1541 & 10.36 & 0.563 & 0.618 \\
    \midrule
    \multirow{3}{*}{Specimen 4}
      & 3DGS & \textbf{26.34} & 0.909 & 0.0773 & 14.67 & 0.700 & 0.420 \\
      & 2DGS & 25.87 & 0.907 & 0.0737 & \textbf{15.77} & 0.729 & 0.402 \\
      & DoF-Gaussian~\cite{shen2025dof} & 18.22 & 0.882 & 0.1528 & 10.55 & 0.586 & 0.615 \\
    \midrule
    \multirow{3}{*}{Specimen 5}
      & 3DGS & 33.32 & 0.976 & 0.0323 & 20.49 & 0.881 & 0.253 \\
      & 2DGS & \textbf{34.53} & 0.977 & 0.0300 & \textbf{21.94} & 0.884 & 0.235 \\
      & DoF-Gaussian~\cite{shen2025dof} & 22.68 & 0.957 & 0.0966 & 14.54 & 0.810 & 0.522 \\
    \midrule
    \multirow{3}{*}{Specimen 6}
      & 3DGS & \textbf{31.88} & 0.944 & 0.0432 & \textbf{22.27} & 0.833 & 0.256 \\
      & 2DGS & 30.04 & 0.940 & 0.0485 & 21.12 & 0.821 & 0.275 \\
      & DoF-Gaussian~\cite{shen2025dof} & 28.60 & 0.926 & 0.0943 & 20.78 & 0.815 & 0.340 \\
    \midrule
    \multirow{3}{*}{Specimen 7}
      & 3DGS & 32.72 & 0.979 & 0.0156 & 20.09 & 0.815 & 0.278 \\
      & 2DGS & \textbf{32.85} & 0.979 & 0.0154 & 20.35 & 0.813 & 0.277 \\
      & DoF-Gaussian~\cite{shen2025dof} & 32.49 & 0.974 & 0.0290 & \textbf{21.29} & 0.804 & 0.360 \\
    \midrule
    \multirow{3}{*}{Specimen 8}
      & 3DGS & 34.16 & 0.970 & 0.0239 & \textbf{22.41} & 0.850 & 0.204 \\
      & 2DGS & \textbf{34.24} & 0.970 & 0.0233 & 22.23 & 0.846 & 0.209 \\
      & DoF-Gaussian~\cite{shen2025dof} & 32.23 & 0.957 & 0.0643 & 21.57 & 0.812 & 0.354 \\
    \midrule
    \multirow{3}{*}{Specimen 9}
      & 3DGS & \textbf{29.21} & 0.950 & 0.0355 & 20.37 & 0.809 & 0.297 \\
      & 2DGS & 29.17 & 0.950 & 0.0335 & 20.37 & 0.809 & 0.292 \\
      & DoF-Gaussian~\cite{shen2025dof} & 28.50 & 0.939 & 0.0724 & \textbf{20.82} & 0.813 & 0.361 \\
    \midrule
    \multirow{3}{*}{Overall ($46$)}
      & 3DGS & 29.71 & 0.947 & 0.0438 & 18.85 & 0.802 & 0.311 \\
      & 2DGS & \textbf{30.30} & 0.949 & 0.0408 & \textbf{19.33} & 0.806 & 0.305 \\
      & DoF-Gaussian~\cite{shen2025dof} & 26.06 & 0.926 & 0.0885 & 17.14 & 0.750 & 0.433 \\

    \bottomrule
  \end{tabular}
\end{table*}

\section{Iridescence}
\label{sec:supp_iridescence}

A hallmark of many Lepidoptera is \emph{iridescence}---color produced by light interference in nanostructured wing scales rather than by pigment, and therefore strongly dependent on viewing angle. Our reconstructions reproduce it directly, inheriting it from the view-dependent spherical-harmonic color model of 3DGS (\cref{fig:iridescence}).

\newlength{\irw}\setlength{\irw}{0.48\linewidth}
\begin{figure}[htbp]
  \centering
  \includegraphics[width=\irw]{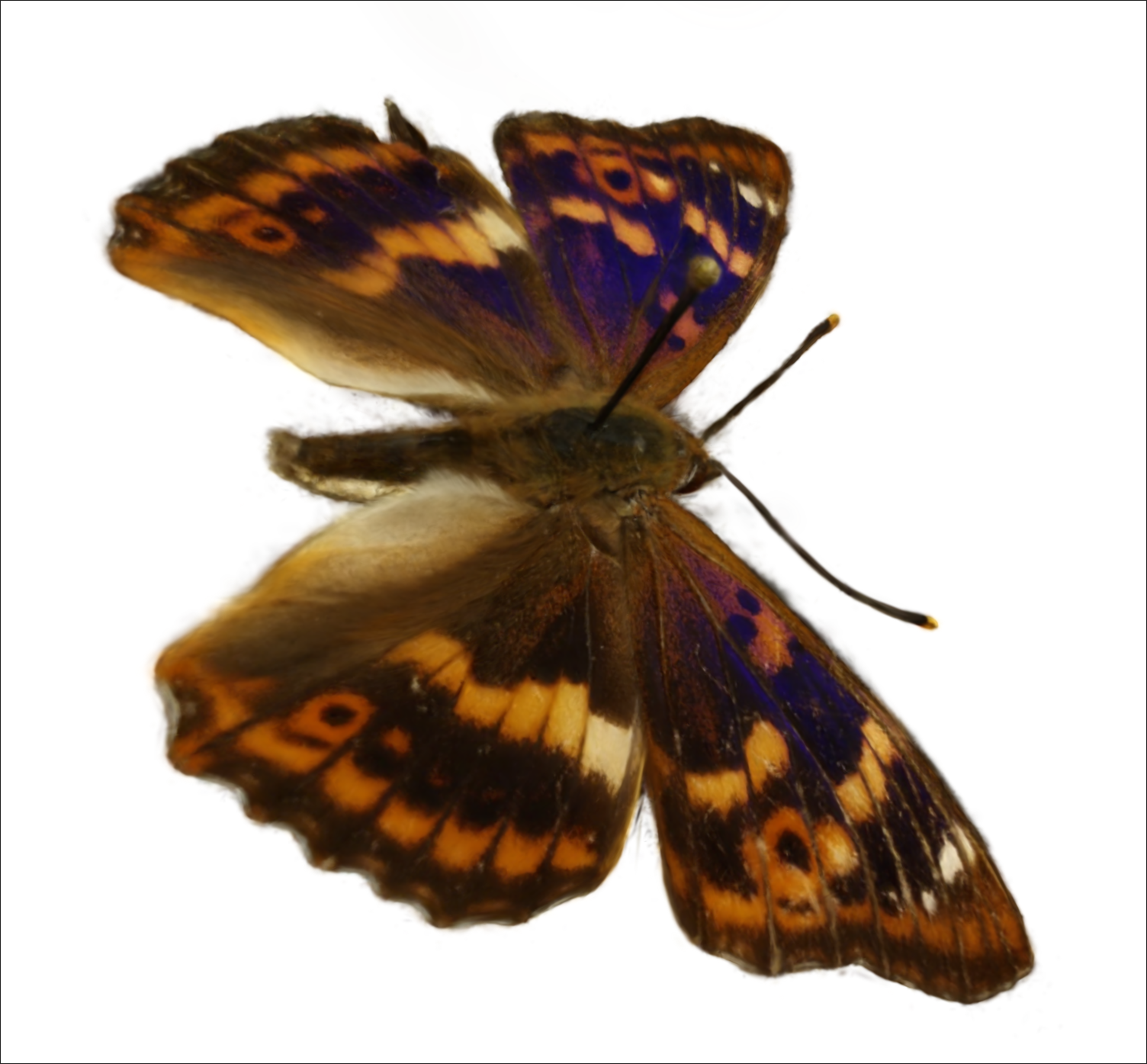}
  \includegraphics[width=\irw]{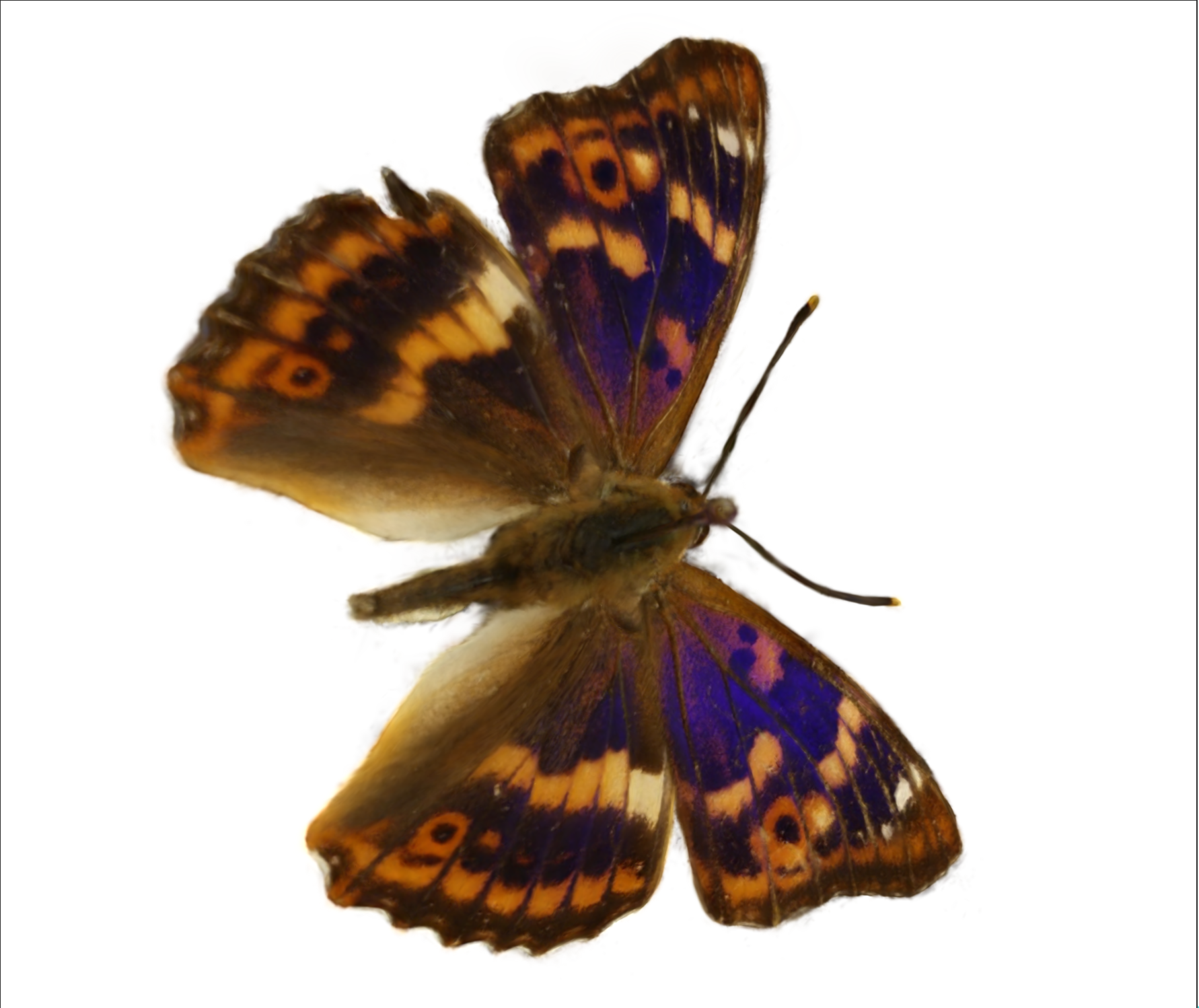}
  \caption{\textbf{Iridescence.} 3DGS faithfully reproduces the angle-dependent \emph{structural color} (iridescence) of butterfly wings: the same reconstruction rendered from two viewing angles shows the characteristic shifting blue--purple sheen. This view-dependent effect is an emergent property of the spherical-harmonic color model in 3DGS rather than a contribution of our pipeline.}
  \label{fig:iridescence}
\end{figure}

\end{document}